\documentclass[a4paper]{article}

\usepackage[utf8]{inputenc}
\usepackage[T1]{fontenc} 
\usepackage{RR,RRthemes}
\usepackage[%
  bookmarks, 
  linkcolor= blue,
  citecolor= blue,
  filecolor= blue,
  urlcolor=  blue,
  colorlinks=true, 
  hyperindex=true,
  pdftitle={Modèles stochastiques du chemostat}, 
  pdfauthor={Fabien Campillo, Marc Joannides and Irène Larramendy-Valverde}
]{hyperref}             
\usepackage[english]{babel}
\usepackage{graphicx}           
\usepackage{url}
\usepackage{makeidx}
\usepackage[plain]{algorithm}
\usepackage{algorithmic} 
\usepackage{color} 
\usepackage[margin=3.3cm]{geometry}
\usepackage{pifont}
\usepackage{tikz}
\usepackage{amsmath,amsfonts,amssymb,latexsym,stmaryrd}

\newcommand{\eqdef}     {\stackrel{{\textrm{\rm\tiny def}}}{=}}
\newtheorem{theorem}      {Theorem}[section]
\newtheorem{theorem*}     {theorem}
\newtheorem{proposition}  [theorem]{Proposition}

\newtheorem{lemma}        [theorem]{Lemma}

\newtheorem{remark}       [theorem]{Remark}

\newcommand{\proof}        {\paragraph{Proof}}
\newcommand{\N}        {\mathbb N}

\newcommand{\E}        {\mathbb E}

\newcommand{\R}      {\mathbb R}

\renewcommand{\P}      {\mathbb P}
\newcommand{\cov}       {\hbox{{\textrm cov}}}
\newcommand{\norm}   [1] {\left\Vert #1 \right\Vert}
\newcommand{\crochet}[1] {\langle #1 \rangle}
\newcommand{\ot}        {\leftarrow}
\newcommand{\carre}     {\hfill$\Box$}
\newcommand{\AAA}  {{\mathcal A}}

\newcommand{\DD}   {{\mathcal D}}

\newcommand{\NN}   {{\mathcal N}}

\newcommand{\PP}   {{\mathcal P}}

\newcommand{\RR}   {{\mathcal R}}

\newcommand{\rmd}   {{{\textrm{\upshape d}}}}

\newcommand{\indic}{{\mathrm\mathbf1}}
\renewcommand{\epsilon}{\varepsilon}

\def\dobm{
    \copy1\kern-\wd1\kern0.05ex\copy1\kern-\wd1\kern0.05ex\box1}

\newcommand{\fenumi}  {\textrm{\rm({\textit{i}}\/)}}
\newcommand{\fenumii} {\textrm{\rm({\textit{ii}}\/)}}

\newcommand{\Tmax}{T^{\textrm{\tiny max}}}
\newcommand{\bnu}{\bar\nu}
\newcommand{\bbm}{\boldsymbol{m}}

\newcommand{\bv}{\boldsymbol{v}}
\newcommand{\wb}{w^{\textrm{\tiny b}}}
\newcommand{\ws}{w^{\textrm{\tiny s}}}
\newcommand{\Wb}{W^{\textrm{\tiny b}}}
\newcommand{\Ws}{W^{\textrm{\tiny s}}}
\newcommand{\cb}{c^{\textrm{\tiny b}}}
\newcommand{\cs}{c^{\textrm{\tiny s}}}
\newcommand{\mb}{m_{\textrm{\tiny b}}}
\newcommand{\ks}{k_{\textrm{\tiny s}}}
\newcommand{\ki}{k_{\textrm{\tiny i}}}
\newcommand{\ms}{m_{\textrm{\tiny s}}}
\newcommand{\ssin}{s^{\textrm{\tiny in}}}
\newcommand{\Btrue}   {{\mathcal B}}
\newcommand{\Btruebio}{{\mathcal B}^{\textrm{\tiny bio}}}
\newcommand{\Btrueout}{{\mathcal B}^{\textrm{\tiny out}}}
\newcommand{\Strue}{{\mathcal S}}
\newcommand{\Struebio}{{\mathcal S}^{\textrm{\tiny bio}}}
\newcommand{\Struein} {{\mathcal S}^{\textrm{\tiny in}}}
\newcommand{\Strueout}{{\mathcal S}^{\textrm{\tiny out}}}

\newcommand{\Sin} {S^{\textrm{\tiny in}}}

\newcommand{\Xbb}{X^{\textrm{\tiny b,bio}}}
\newcommand{\Xbo}{X^{\textrm{\tiny b,out}}}
\newcommand{\Xsb}{X^{\textrm{\tiny s,bio}}}
\newcommand{\Xsi}{X^{\textrm{\tiny s,in}}}

\newcommand{\sigmamin}{\sigma_{\textrm{\tiny min}}}
\newcommand{\mumax}{\mu_{\textrm{\tiny max}}}
\newcommand*\circled[1]{%
  \tikz[baseline=(C.base)]\node[draw,circle,inner sep=0.5pt](C) {#1};\!
}
\newcommand{\circledi}{{\scriptsize $\protect\circled{\it i\/}$}}
\newcommand{\circledip}{{\scriptsize $\protect\circled{\it i\/}\;'$}}
\newenvironment{psmallmatrix}
{\left(\begin{smallmatrix}}
{\end{smallmatrix}\right)}

\RRdate{November 2010}

\RRauthor{
Fabien Campillo\thanks{\protect\url{Fabien.Campillo@inria.fr} --- 
           Project--Team MERE, INRIA/INRA, UMR MISTEA, b\^at. 29, 2 place Viala, 34060 Montpellier cedex 06, France.}%
  \and
Marc Joannides\thanks{\protect\url{marc.joannides@univ-montp2.fr} ---
   Universit\'e Montpellier 2 / I3M, case courrier 51,
        place Eug\`ene Bataillon, 34095 Montpellier cedex 5; this author is associate researcher for Project--Team MERE, INRIA/INRA, UMR MISTEA.}%
  \and
Irène Larramendy-Valverde\thanks{\protect\url{larra@math.univ-montp2.fr} ---
   Universit\'e Montpellier 2 / I3M, case courrier 51,
        place Eug\`ene Bataillon, 34095 Montpellier cedex 5.}%
}
\authorhead{Campillo, Joannides, Laramendy-Valdere}

\RRtitle{Modèles stochastiques du chemostat}
\RRetitle{Stochastic models of the chemostat}
\titlehead{Stochastic models of the chemostat}

\RRabstract{We consider the modeling of the dynamics of the chemostat at its very source. The chemostat is classically represented as a system of ordinary differential  equations. Our goal is to establish a stochastic model that is valid at the scale immediately preceding the one corresponding to the deterministic model. At a microscopic scale we present a pure jump stochastic model that gives rise, at the macroscopic scale, to the ordinary differential equation model. At an intermediate scale, an approximation diffusion allows us to propose a model in the form of a system of stochastic differential equations. We expound the mechanism to switch from one model to another, together with the associated simulation procedures. We also describe the domain of validity of the different models.}

\RRkeyword{stochastic differential equations, chemostat, pure jump process,
diffusion approximation, tau-leap method, Monte Carlo method, Gillespie algorithm}

\RRresume{Nous reprenons la modélisation de la dynamique du chemostat à sa source. Le chemostat est classiquement représenté par un système d’équations différentielles. Notre objectif est d'établir un modèle stochastique qui est valable à l'échelle qui précède immédiatement celle qui correspond au modèle déterministe. Partant d'une échelle microscopique, nous présentons un modèle stochastique de sauts purs qui conduit, à l'échelle macroscopique, au modèle d'équation différentielle. À une échelle intermédiaire, une approximation diffusion nous permet de proposer un modèle sous la forme d'un système d'équations différentielles stochastiques. Nous détaillons les techniques qui permettent de passer d'une échelle à une autre ainsi que de simuler ces différents modèles. Nous décrivons également les domaines de validité des différents modèles.}

\RRmotcle{équations différentielles stochastiques, chemostat, processus de saut, approximation diffusion, méthode ``tau-leap'', méthode de Monte Carlo, algorithme de Gillespie}

\RRprojet{MERE}
\RRdomaine{1} 
\RRthemeProj{mere} 
\RRdomaineProjBis{mere} 
\RCSophia 

\begin{document}
 
\RRNo{7458}
 
\makeRR   

\clearpage
\tableofcontents
\mbox{}
\cleardoublepage
\mbox{}
\cleardoublepage

\section{Introduction}

The evolution of the state of a single species/single substrate chemostat is 
usually described by a set of ordinary differential equations (ODE) derived from a 
mass balance principle, see~\cite{smith1995a}.  More precisely, if $s(t)$ denotes 
the concentration of nutrient (substrate) and $b(t)$ the concentration of the 
organism (biomass) at time $t$(expressed in \textit{g/L}), then the couple $x(t)=
(b(t),s(t))^{*}$ is the solution of the following ODE \cite{smith1995a}:  
\begin{subequations}
\label{eq.chemostat}
\begin{align}
\label{eq.chemostat.b}
  \dot b(t)
  &=
  [\mu(s(t))-D]\,b(t)\,,
\\
\label{eq.chemostat.s}
  \dot s(t)
  &=
  -k\,\mu(s(t))\,b(t)+D\,[\ssin-s(t)]
\end{align}
\end{subequations}
where $D>0$ is the dilution rate, $\ssin>0$ the substrate concentration
in the influent, and $k>0$ the stoichiometric coefficient. The initial
condition lies in the positive orthant, that is $b(0)\geq 0$ and
$s(0)\geq 0$. Equation \eqref{eq.chemostat} will also be denoted:
\[
  \dot x(t) = f(x(t))\,.
\]
 
The specific growth rate function  $\mu(s)$ is non-negative; we
suppose that $\mu(0)=0$, $\mu(s)>0$ for $s>0$, $\mu(s)\leq \mumax
<\infty$ and that it is continuous at 0. Commonly used models are the Monod
model (uninhibited growth) and the Haldane model (inhibited growth)
that reads respectively: 
\begin{align}
\label{eq.specific.growth.rate}
 \mu(s)
 &=
 \mumax\,\frac{s}{\ks + s}\,,
 &
 \mu(s)
 &=
 \mumax\,\frac{s}{\ks + s +\frac{s^{2}}{\ki}}\,.
\end{align}

This approach relies on the fact that the stochastic effects can be neglected, 
thanks to the law of large numbers, or at least can be averaged out. Although this level of 
description is sufficient for a number of applications of interest, it could be 
a valuable way of accounting for the stochastic nature of the system. Indeed, 
at small population sizes the chemostat could present stochastic behaviors, 
also the accumulation of small perturbations in the context of multi-species 
could not be neglected. Moreover, whereas the experimental results observed in 
well mastered laboratory conditions match closely the ODE theoretical behavior, 
a noticeable difference may occur in operational conditions. In these cases,
stochastic features may not be neglected. We aim to build a model that 
still relies on a mass balance principle and that encompasses the useful 
stochastic information. 

\medskip

Many works \cite
{stephanopoulos1979a,gard1999a,imhof2005a} propose to superpose a stochastic term 
on Equation~\eqref{eq.chemostat} in order to model the uncertainty on 
the phenomenon, principally due to imprecise experimental conditions. 
Paradoxically, this amounts to the addition of an \emph{ad hoc} perturbation to a model that 
has been obtained by neglecting these perturbations. We propose instead to consider 
the stochastic aspect at the very beginning of the modeling process, and to 
determine the conditions under which it is insignificant. This approach is not 
individual-based per se, as it starts from the macroscopic model~\eqref
{eq.chemostat}. However, the first stochastic model proposed will be described at 
the individual level. This method will allow for a justification of the specific structure 
of the stochastic perturbation that affects the mean behavior. More generally, we 
will outline a modeling strategy based on many available tools, either stochastic 
or deterministic, depending on the regularity of the phenomenon to be modeled. In 
this paper we focus on the modeling and simulation process rather than on the 
mathematical developments; moreover we make use of known mathematical results. Our goal is to establish a stochastic model that is valid at the 
scale immediately preceding the one corresponding to the deterministic model \eqref{eq.chemostat}.

\medskip

The paper is organized as follows: in Section~\ref{sec.time.scale}, we recall the 
origin of model~\eqref{eq.chemostat} and the assumptions ensuring its validity. We 
show that since different timescales naturally appear in the problem, these 
assumptions need to be checked at each scale. Section~\ref{sec.models} is devoted 
to the different models: the pure jump description that will be considered as the 
reference model is introduced in Section \ref{subsec.jump.model}; the discrete 
time approximation, Poisson and normal, are presented in Section \ref
{subsec.discrete.time.approximations}; the discrete-time normal approximation 
appeared to be a time discretization of a diffusion process given by a stochastic 
differential equation presented in Section \ref{subsec.diffusion.model}. In 
Section \ref{subsec.asymptotic.analysis} we describe the asymptotic results that 
bridge these different models. Section \ref{sec.simulation.algorithms} is devoted 
to the associated simulation algorithm, Section \ref{sec.simulation} to numerical 
tests.

\section{Scale and geometry issues in ODE model}
\label{sec.time.scale}

An individual-based model should keep track of the position in space of each cell, 
together with their current biological states, it should also account for discrete events such as the division of a cell. Such a description of the system at the finest level could be of interest but unnecessary in view of our goals, namely to set a macroscopic model that gives account for stochastic phenomena.  At this 
scale, the system is reduced to a $\R^2$--vector and its dynamics. 

Model~\eqref{eq.chemostat} is obtained according to the
classical approach, by choosing a small time interval $\Delta t$ on
which a mass balance principle is applied to the state. However,
$\Delta t$ should be large enough as we do not describe the dynamic at the timescale of jumps of one unit of substrate or bacteria but rather at the timescale of jumps of packet units. Such an interval
could be called {\em macroscopically   infinitesimal} \cite{gillespie2000a}.

\subsubsection*{Mass balance}

Let $(\Btrue_t,\Strue_t)$ denote the true concentrations at time $t$, 
assumed to be constant throughout the medium. The mass balance on interval 
$[t, t +  \Delta t)$  reads 
\begin{subequations}
\label{eq.mass.balance}
\begin{align} 
\label{eq.mass.balance.b}
  \Btrue_{t + \Delta t} - \Btrue_t 
  &= 
  \Delta \Btruebio_t+ \Delta \Btrueout_t \,,
\\
\label{eq.mass.balance.s}
  \Strue_{t + \Delta t} - \Strue_t 
  &= \Delta \Struebio_t 
  + \Delta \Struein_t + \Delta \Strueout_t
\end{align}
\end{subequations}  
where
\begin{itemize} 
\item 
  $\Delta \Btruebio_t$ and $ \Delta \Btrueout_t$ are
  the increments of biomass due to natural growth and to the
  outflow respectively, within $[t, t + \Delta t)$,
\item  
  $ \Delta \Struebio_t,\, \Delta \Struein_t$ and $\Delta
  \Strueout_t$ are the increments of substrate due to the consumption
  by  the biomass, the inflow and the outflow respectively, within
  $[t, t + \Delta t)$.
\end{itemize} 
Since we want to obtain an ODE, we now assume that the stochastic
fluctuations are negligible relative to the increments. Again this
requires $\Delta t$ to be large enough, so that sufficiently many
discrete events have occurred. Moreover, $\Delta t$ should be taken even
larger in case of inhomogeneity of the dynamics.

We denote by $(\bar  b(t), \bar s(t))$
for $t = 0,\, \Delta t,\, 2\, \Delta t, \dots$ the deterministic sequence
constructed by using the mean increments of $(\Btrue_t,\Strue_t)$:
\begin{align*} 
  \bar b(t + \Delta t) - \bar b(t) 
      &= \E [\Delta \Btruebio_t + \Delta \Btrueout_t]\,, 
  \\
  \bar s(t + \Delta t) - \bar s(t) 
      &= \E [\Delta \Struebio_t  + \Delta \Struein_t + \Delta \Strueout_t] \,.
\end{align*} 
Next, using the mass action law for the biomass we have
\begin{align} 
\label{eq.delta.b}
  \E[\Delta \Btruebio_t ] 
  &\simeq \mu(\bar s(t))\, \bar b(t)\, \Delta t\,,
  & 
  \E[\Delta \Struebio_t] 
  &\simeq - k \, \mu(\bar s(t))\, \bar b(t) \,
  \Delta t
\end{align}  
where $\mu(s)$ is the specific growth rate and $k>0$ the stoichiometric
coefficient. Note that we again require $\Delta t$ to be large enough,
since $\mu(s)$ and $k$ make sense only for a sufficiently large
population of bacteria. Now, since we have assumed perfect
homogeneity of the medium, we get:
\begin{align} 
\label{eq.delta.in.out}
  \E[\Delta \Btrueout_t] 
  &\simeq - D \, \bar b(t)\, \Delta t\,,
  & 
  \E[\Delta \Struein_t] 
  &\simeq - D \, \ssin\, \Delta t\,,
  & 
  \E[\Delta \Strueout_t] 
  &\simeq - D \, \bar s(t)\, \Delta t \,.
\end{align} 
Note that~\eqref{eq.delta.b} and~\eqref{eq.delta.in.out} are
approximations because we have used a constant value for  $\bar b(t)$
and $\bar s(t)$  within $[t, t + \Delta t)$. For this approximations to
be correct, none of the quantities involved should vary significantly
within $[t, t + \Delta t)$. We finally obtain the construction of the
sequence $(\bar b(t), \bar s(t))$ by 
\begin{subequations}
\label{eq.chemostat.approxdet}
\begin{align}
\label{eq.chemostat.approxdet.b}
 \bar b(t + \Delta t) - \bar b(t) 
  &=
  [\mu(\bar s(t))-D]\,\bar b(t)\,\Delta t,
\\
\label{eq.chemostat.approxdet.s}
\bar s(t + \Delta t) - \bar s(t) 
  &=
  (-k\,\mu(\bar s(t))\,\bar b(t) + D\,[\ssin-\bar s(t)])\,\Delta t,
\end{align}
\end{subequations}
Model~\eqref{eq.chemostat} is obtained by letting $\Delta t\to 0$ in System~\eqref{eq.chemostat.approxdet}. However, since $\Delta t$ is
bounded from below, some care should be taken when this
limit is achieved. System~\eqref{eq.chemostat.approxdet} can be understood as the
discretization of~\eqref{eq.chemostat} using an explicit Euler scheme
with time-step $\Delta t$. Whenever there exists $\Delta t$
sufficiently small, the deterministic sequence $(\bar b(t), \bar s(t))$
will be close to model~\eqref{eq.chemostat}, sampled at time $0,
\Delta t,\, 2\,\, \Delta t, \dots$

\subsubsection*{Geometry and scales}

The mass balance established in~\eqref{eq.mass.balance} features five
terms that can be gathered according to the three sources of
variations. This gives rise to a geometric structure that can be
emphasized by writing~\eqref{eq.chemostat} under the form: 
\begin{align}
\label{eq.geometry}
\frac{\rmd }{\rmd t}
\begin{psmallmatrix}
  b(t) \\ s(t)
\end{psmallmatrix}
=
	\underbrace{
	\mu(s(t))\,b(t) \, 
	\begin{psmallmatrix}
	  1\\ 
	  -k \\ 
	\end{psmallmatrix}
	\vphantom{\begin{psmallmatrix}  b(t) \\ s(t) \end{psmallmatrix}}
	}_{\textrm{biology}}
	\underbrace{
	+
	D\,
	\begin{psmallmatrix}
	  0 \\ \ssin
	\end{psmallmatrix}
	\vphantom{\begin{psmallmatrix}  b(t) \\ s(t) \end{psmallmatrix}}
	}_{\textrm{inflow}}
	\underbrace{
	-
	D\,
	\begin{psmallmatrix}  b(t) \\ s(t) \end{psmallmatrix}
	}_{\textrm{outflow}} 
\end{align}
However, whereas the geometry is well captured, the timescale of the
five original terms is not readable in~\eqref{eq.chemostat} nor
\eqref{eq.geometry}. Indeed, the fact that the approximations
in~\eqref{eq.delta.b} and~\eqref{eq.delta.in.out} may be of different
quality for each individual term is not exploited at all.

\section{Models at different scales}
\label{sec.models}

In the previous section, we mentioned that the lower bound for $\Delta t$ is
related to the size of the population and to the regularity of the
phenomenon. Often, the experimental conditions are such that this
bound is low enough, so that System~\eqref{eq.chemostat.approxdet}
is  correctly  approximated by~\eqref{eq.chemostat} sampled with
period $\Delta t$. If a smaller period is to be considered, then the
conditions under which~\eqref{eq.chemostat.approxdet} has been
obtained are not fulfilled. Particularly, the stochastic fluctuations
should be accounted for. 

We now  introduce a stochastic process built on
the same premise, that is a  mean mass balance principle at a given
$\Delta t$. This model will have~\eqref{eq.chemostat} as a fluid
limit as $\Delta t$ goes to~0. This latter model suitably
features the geometry of the chemostat but, as a limit model, cannot
feature all its natural scales. The proposed stochastic models will
respect both the geometry and the natural scales of the chemostat. We
first establish a pure jump process representation of the chemostat at
a microscopic scale, then we derive a diffusion process representation
which will be valid at mesoscopic and macroscopic scales.

\subsection{Pure jump model $X_{t}=(B_{t},S_{t})^*$}
\label{subsec.jump.model}

Even if do not aim at deriving an individual-based model, we try to
preserve the discrete feature in the dynamics. We achieve this by
considering only \emph{aggregated jumps} obtained by adding up small
and frequent jumps resulting from individual events. The resulting
stochastic process will be a pure jump process
$X_{t}=(B_{t},S_{t})^{*}$, fully determined by its jumps and the
corresponding jump rates; the state variable will be denoted $x=(b,s)^{*}$. 

\begin{figure}
\begin{center}
\includegraphics[width=7cm]{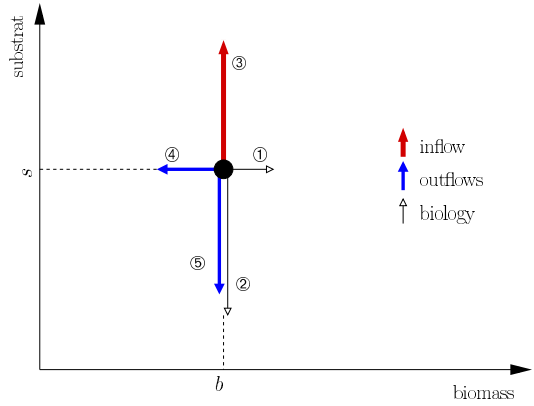} 
\end{center}
\vspace{-1em}
\caption{\it In this model, from a position $x=(b,s)^*$ the process could jump according to 5 mechanisms (2 due to the biology, 1 inflow, and 2 outflows), the basic jump \circledi\ has a length $\frac{1}{K_{i}}$ for $i=1,\dots, 5$.}
\label{fig.jumps}
\end{figure}

In view of~\eqref{eq.mass.balance}, we are led to  consider five jumps:
\begin{itemize}
\item[\ding{192}] biology term: biomass increase of size $\nu_1(x)$ at
  rate $\lambda_{1}(x)$;
\item[\ding{193}] biology term: substrate decrease of size $\nu_2(x)$ at
  rate $\lambda_{2}(x)$;
\item[\ding{194}] inflow term: substrate inflow of size $\nu_3(x)$ at
  rate $\lambda_{3}(x)$;
\item[\ding{195}] outflow term: biomass outflow of size $\nu_4(x)$ at
  rate $\lambda_{4}(x)$;
\item[\ding{196}] outflow term: substrate outflow of size $\nu_5(x)$ at
  rate $\lambda_{5}(x)$;
\end{itemize}
(see Figure \ref{fig.jumps}). It remains to set the jump size rates
so  as to comply with the mass balance principle and the stochastic
mass action law.

For a macroscopically infinitesimal $\Delta t$,
denote by $\Delta \Xbb_{t}$,
$\Delta \Xsb_{t}$,
$\Delta \Xsi_{t}$,
$\Delta \Xbo_{t}$,
$\Delta \Xsb_{t}$ the cumulated jump of type~\ding{192}, \ding{193}, \ding{194}, \ding{195}, \ding{196} respectively, on state process $X_{t}$ within the time interval $[t,
  t + \Delta   t)$. 
  
We first focus on the first two expressions.  The stochastic mass action law
\cite{wilkinson2006a} requires
\begin{align*} 
  \E[\Delta \Xbb_t\vert X_{t} = x] 
  &\simeq
  \begin{psmallmatrix}\mu(s)\, b\, \Delta t \\ 0 \end{psmallmatrix} \,,
\\
  \E[\Delta \Xsb_t  \vert X_{t} = x] 
  &\simeq
  \begin{psmallmatrix}0 \\ -k \, \mu(s)\, b \,  \Delta t \end{psmallmatrix}
  \, .
\end{align*} 
Now notice that, for small $\Delta t$, the number of jumps of
type~\ding{192} (resp.~\ding{193}) within $[t, t+ \Delta t)$ is
approximately $\mathcal{P}(\lambda_1(x)\, \Delta t)$ (resp.
$\mathcal{P}(\lambda_2(x)\, \Delta t)$), so that
\begin{align*} 
  \E[\Delta \Xbb_t | X_{t} = x] 
  &\simeq
  \lambda_1(x)\, \Delta t\, \nu_1(x)\,,
  \\
  \E[\Delta \Xsb_t | X_{t} = x] 
  &\simeq
  \lambda_2(x)\, \Delta t\, \nu_2(x) \,.
\end{align*} 
So we are looking for $(\lambda_i(x), \nu_i(x))$ satisfying:
\begin{align}
\label{eq.mass.action}
  \lambda_1(x)\, \nu_1(x)  
  = 
  \begin{psmallmatrix} \mu(s)\, b \\  0  \end{psmallmatrix} 
  \quad \textrm{ and } \quad 
  \lambda_2(x)\,\nu_2(x) 
  = 
  \begin{psmallmatrix} 0\\  -k\, \mu(s)\, b   \end{psmallmatrix} \,.
\end{align}
We therefore introduce the \emph{scale} parameters $K_1$ and $K_2$ and
we choose:
\begin{align*} 
  \lambda _1(x) 
  & \eqdef  K_1\, \mu(s)\, b\,, 
  & 
  \nu_{1} 
  &\eqdef  
    \begin{psmallmatrix} \frac{1}{K_{1}}\\ 0 \end{psmallmatrix}\,,
\\
  \lambda _2(x) 
  &\eqdef  
  K_2\, k\, \mu(s)\, b 
  &
  \nu_{2} 
  &\eqdef
 -\begin{psmallmatrix}0\\ \frac{1}{K_{2}}\end{psmallmatrix} \,.
\end{align*} 
this choice is not unique and will be explain later in Section \ref{sec.other.models}.

Here by ``scale'' we mean that jumps due to \circledi\ will be of
magnitude $\frac{1}{K_i}$ and the corresponding rates will be of
magnitude $K_i$. Large $K_i$ yields frequent and small jumps. Using
the Poisson argument mentioned above, we see that these scale
parameters $K_{i}$ do not act on the mean values of the increments but
on their variances (large $K_{i}$ will correspond to small
variances). The $K_i$'s can thus be regarded as tuning parameters
quantifying the uncertainty or regularity of the corresponding source
of variation. 

\medskip

Reproducing this discussion with the three other types of jumps, and
considering only admissible jumps (in the positive orthant), we obtain
a pure jump Markov process with rate coefficients $\lambda_{i}(x)$ and
associated jumps $\nu_{i}(x)$ defined in Table~\ref{table:rates:jumps}.

\begin{table}[ht] 
\centering \small
\begin{tabular}{|l|ccccc|}
\hline 
   & \ding{192} & \ding{193} & \ding{194} & \ding{195} & \ding{196} \\ 
   & biomass & substrate & substrate & biomass & substrate \\
   & increase & decrease & inflow & outflow & outflow \\ \cline{2-6}
   &  \multicolumn{2}{c|}{biology} & inflow & \multicolumn{2}{|c|}{outflow}
   \\
\hline\hline 
rate $\lambda_{i}(x)$ 
   & $K_{1}\,\mu(s)\,b$ 
   & $K_{2}\,k\,\mu(s)\,b$ 
   & $K_{3}\,D\,\ssin$ 
   & $K_{4}\,D\,b$ 
   & $K_{5}\,D\,s$ 
    \vphantom{$\displaystyle\int$}
   \\ \hline 
jump $\nu_{i}(x)$  
   & $\begin{pmatrix}\frac{1}{K_{1}}\\ 0\end{pmatrix}$
   & $-\begin{pmatrix}0\\ \frac{1\wedge K_{2}\,s}{K_{2}}\end{pmatrix}$
   & $\begin{pmatrix}0\\ \frac{1}{K_{3}}\end{pmatrix}$
   & $-\begin{pmatrix}\frac{1\wedge K_{4}\,b}{K_{4}}\\ 0\end{pmatrix}$
   & $-\begin{pmatrix}0\\ \frac{1\wedge K_{5}\,s}{K_{5}} \end{pmatrix}$
    \vphantom{$\begin{pmatrix}0\\ -1 \\ 0\end{pmatrix}$}
    \\ [2ex] 
\hline 
\end{tabular} 
\caption{\it Rates and jumps of the five basic mechanisms of the pure jump process. Note that the jumps $\nu_{i}(x)$ essentially do not depend on $x$ except for the negative jumps near the border $\{x=(b,s)^{*}\in\R^{2}_{+};b=0 \textrm{ or }s=0\}$.} 
\label{table:rates:jumps} 
\end{table}

\subsubsection*{About scales parameters}

Let $\mb$ and $\ms$ denote the  representative masses of a single bacteria and of a single molecule of substrate. Typically $\mb\gg \ms$ (e.g. $\mb\simeq 10^{6}\,\ms$). Hence:
\begin{align*}
  0<K_{i}\leq \frac{1}{\mb}\,,\ \textrm{for }i=1,4
  \quad\textrm{and}\quad
  0<K_{i}\leq \frac{1}{\ms}\,,\ \textrm{for }i=2,3,5\,.
\end{align*}

In most cases $K_{i}\ll K_{j}$ for $i=1,4$ and $j=2,3,5$, but it is
possible to adjust the coefficients $K$'s to the specific application
considered. For example $K_{2}$ will be large in laboratory
experimental conditions, but for a real implementation the substrate
inflow concentration could have a large variance. Also for the
outflow, in regular conditions $K_{4}$ and $K_{5}$ could be large, but
in bad mixing conditions they could be smaller. Finally $K_{1}$ could
be smaller than $K_{4}$, as the biomass concentration increase
presents more variance than the substrate decrease (which is more
regular as it is related to the diffusion of substrate across cell
membranes). 



As proved later in Lemma \ref{lemma.f.fK}, the jumps $\nu_{i}(x)$ are
essentially constant and equal to:
\begin{align}
\label{eq.nui}
 \nu_{1}&\eqdef
 \begin{pmatrix}\frac{1}{K_{1}}\\ 0\end{pmatrix}\,,
 &
 \nu_{2}&\eqdef
 -\begin{pmatrix}0\\ \frac{1}{K_{2}}\end{pmatrix}\,,
 &
 \nu_{3}&\eqdef
 \begin{pmatrix}0\\ \frac{1}{K_{3}}\end{pmatrix}\,,
 &
 \nu_{4}&\eqdef
 -\begin{pmatrix}\frac{1}{K_{4}}\\ 0\end{pmatrix}\,,
 &
 \nu_{5}&\eqdef
  -\begin{pmatrix}0\\ \frac{1}{K_{5}} \end{pmatrix}\,.
\end{align}

\subsubsection*{Representation of $X_{t}$}

The constructive description of the process $X_t$ that has been just presented would be used for simulation purposes, see Section \ref{subsec.simu.pure.jump}. Nevertheless, it should be completed by a more comprehensible and synthetic representation. This will require some mathematical developments which we summarize now and that are detailed in Appendix \ref{sec.representation.of.X}.

First we should notice that the jump process $X_{t}$ can be represented as the following (jump) SDE:
\begin{align}
\label{eq.jump.representation}
  X_{t}
  =
  X_{0}
  +
  \sum_{i=1}^{5} \int_{(0,t]\times [0,\infty)} 
          \nu_{i}(X_{u^{-}})\,\indic_{\{v\leq \lambda_{i}(X_{u^{-}})\}}
          \,N^{i}(\rmd u\times\rmd v)
\end{align}
where $N^{i}$ are independent random Poisson measures with intensity measure $\rmd u\times \rmd v$ (Lebesgue measure).

The process $X_{t}$ can be described as a Markov process with infinitesimal generator   $\AAA \phi(x)
   =
   \lim_{t\to 0}\frac1t\,[\E\phi(X^{x}_{t})-\phi(x)]$
where $X^{x}_{t}$ is the process $X_{t}$ starting from $x$. This operator completely characterizes the law of the process $X_{t}$. In Appendix \ref{sec.representation.of.X} we prove that this process is non explosive, i.e. it is defined for all $t\geq 0$; that it admits moments as soon as $X_{0}$ does; and that it is solution of \eqref{eq.jump.representation} see Proposition \ref{proposition.jump.representation}.

Still representation \eqref{eq.jump.representation} is opaque. We can establish that the process $X_{t}$ essentially admits the following representation:
\begin{subequations}
\label{eq.representation.pure.jump.simple}
\begin{align}
\label{eq.representation.pure.jump.simple.B}
  \rmd B_{t}
  &=
  \big(
    \mu(S_{t})\,B_{t}
    -
    D\,B_{t}
  \big)\,\rmd t 
  \textstyle
  +
    \frac{\rmd\bar m^{1}_{t}}{\sqrt{K_{1}}}
    +
    \frac{\rmd\bar m^{4}_{t}}{\sqrt{K_{4}}}
    \,,
\\
\label{eq.representation.pure.jump.simple.S}
  \rmd S_{t}
  &=
  \big(
    -k\,\mu(S_{t})\,B_{t}
    + D\,(\ssin-S_{t})
  \big)\,\rmd t
  \textstyle
  +
    \frac{\rmd \bar m^{2}_{t}}{\sqrt{K_{2}}}
    +
    \frac{\rmd \bar m^{3}_{t}}{\sqrt{K_{3}}} 
    +
    \frac{\rmd \bar m^{5}_{t}}{\sqrt{K_{5}}}
\end{align}
\end{subequations}
where $\bar m^{i}_{t}$ are independent square integrable martingales with zero mean. The exact representation \eqref{eq.representation.pure.jump} differs from \eqref{eq.representation.pure.jump.simple} only through terms $(1\wedge K_{i} \,s)$ and $(1\wedge K_{i} \,b)$ that are equal to 1 except on a very limited neighborhood of the axes.

The martingales $\bar m^{i}_{t}$  are of mean 0 and they are explicitly known, see  \eqref{eq.bar.mi}, as well as their quadratic variation, see \eqref{eq.quadratic.variation}. From Equation \eqref{eq.representation.pure.jump.simple}, the deterministic part of the process $X_{t}$, its drift coefficient, appears to be essentially the classical ODE \eqref{eq.chemostat}; and the stochastic part of this dynamics, the martingale terms, are of order $1/\sqrt{K_{i}}$.

\subsection{Discrete time approximations}
\label{subsec.discrete.time.approximations}

\subsubsection*{Poisson approximation  $\tilde X_{t_n}=(\tilde B_{t_n},\tilde S_{t_n})^*$}

For any small $\Delta t>0$ given, let $t_{n}=n\,\Delta t$. We propose a discrete time Poisson approximation $(\tilde X_{t_{n}})_{n\geq 0}$ of $(X_{t})_{t\geq 0}$: on the interval $[t_n,t_{n+1})$ we froze the rate functions $\lambda_{i}(X_{t})$ to $\lambda_{i}(X_{t_n})$ so that we get a Poisson distribution. The jumps $\nu_{i}(X_{t})$ are also frozen to $\nu_{i}(X_{t_n})$. Let $\tilde X_{0}=X_{0}$, the approximation is defined by:
\begin{align}
\label{eq.poisson.approximation}
  \tilde X_{t_{n+1}}
  =
  \tilde X_{t_{n}}
  +\sum_{i=1}^{5} \nu_{i}(\tilde X_{t_{n}})\,
  \PP_n^{i}(\Delta t\,\lambda_{i}(\tilde X_{t_n}))
\end{align}
where $(\PP^{i}_n(\rho))_{n\in\N,i=1\cdots 5}$ are independent Poisson variables with intensities $\rho$. 

We have:
\begin{align}
\nonumber
  \E[\tilde X_{t_{n+1}}|\tilde X_{t_n}=x]
  &=
  x+\sum_{i=1}^{5} \nu_{i}(x)\,\E[\PP_n^{i}(\Delta t\,\lambda_{i}(\tilde X_{t_n}))|\tilde X_{t_n}=x]
\\  
\label{eq.poisson.approximation.mean}
  &=
  x+\Delta t\,\sum_{i=1}^{5} \nu_{i}(x)\,\lambda_{i}(x)
\end{align}
and let
\begin{align}
\label{eq.fK1}
 f_{K}(x) \eqdef \sum_{i=1}^{5} \nu_{i}(x)\,\lambda_{i}(x)\,,
\end{align}
$f_{K}(x)$ is ``essentially'' the r.h.s. function $f(x)$ of the O.D.E. \eqref{eq.chemostat}, more precisely $f_{k}(x)=f(x)$ except near the axes (see. Lemma \ref{lemma.f.fK}). In other words, the infinitesimal increments of the conditional mean follow the O.D.E. \eqref{eq.chemostat}.

Also:
\begin{subequations}
\label{eq.poisson.approximation.cov}
\begin{align}
\label{eq.poisson.approximation.cov.1}
  \cov[\tilde X_{t_{n+1}}|\tilde X_{t_n}=x]
  &=
  \sum_{i=1}^{5} \cov[\nu_{i}(x)\,\PP_n^{i}(\Delta t\,\lambda_{i}(\tilde X_{t_n}))|\tilde X_{t_n}=x]
  =
  \begin{pmatrix}
    \tilde\Sigma^{2}_{1}& 0\\ 0 & \tilde\Sigma^{2}_{2}
  \end{pmatrix}
\end{align}
with
\begin{align}
\nonumber
  \tilde\Sigma^{2}_{1}
  &=
  \textstyle
  \frac{1}{K_{1}^{2}}\,\cov[\PP_n^1(\Delta t\,\lambda_1(x))]
  +
  \frac{1}{K_{4}^{2}}\,
  (1\wedge K_{4}\,b)^{2}\,
  \cov[\PP_n^4(\Delta t\,\lambda_4(x))]
\\
\label{eq.poisson.approximation.cov.2}
  &
  \textstyle
  =
  \Delta t\,\Big\{\frac{1}{K_{1}}\,\mu(s)\,b
  +
  \frac{1}{K_{4}}\,(1\wedge K_{4}\,b)^2\,D\,b
  \Big\}\,,
\\[0.5em]
\nonumber
  \tilde\Sigma^{2}_{2}
  &=
  \textstyle
  \frac{1}{K_{2}^{2}}\,(1\wedge K_{2}\,s)^{2}\,
         \cov[\PP_n^2(\Delta t\,\lambda_2(x))]
  +
  \frac{1}{K_{3}^{2}}\,
         \cov[\PP_n^3(\Delta t\,\lambda_3(x))]
\\
\nonumber
  & \qquad\qquad\qquad\qquad \qquad \qquad
  \textstyle
  +
  \frac{1}{K_{5}^{2}}\,(1\wedge K_{5}\,s)^{2}\,
         \cov[\PP_n^5(\Delta t\,\lambda_5(x))]
\\
\label{eq.poisson.approximation.cov.3}
  &
  \textstyle
  =
  \Delta t\,\Big\{
  \frac{1}{K_{2}}\,(1\wedge K_{2}\,s)^2\,k\,\mu(s)\,b
  +
  \frac{1}{K_{3}}\,D\,\ssin
  +
  \frac{1}{K_{5}}\,(1\wedge K_{5}\,s)^2\,D\,s
  \Big\}\,.
\end{align}
\end{subequations}

\subsubsection*{Diffusion approximation $\tilde \xi_{t_n}=(\tilde \beta_{t_n},\tilde \sigma_{t_n})^*$}

In \eqref{eq.poisson.approximation}, the variable $\PP_n^{i}(\Delta t\,\lambda_{i}(x))$ is Poisson distributed with parameter $\Delta t\,\lambda_{i}(x)$. When this parameter is large (greater than 10 or 20) then this last distribution is very close to the normal distribution of mean $\Delta t\,\lambda_{i}(x)$ and variance $\Delta t\,\lambda_{i}(x)$.  Hence, we get a (discrete time) normal approximation $(\tilde \xi_{t_{n}})_{n\geq 0}$ of $(X_{t})_{t\geq 0}$ by letting $\tilde \xi_{0}=X_{0}$ and, conditionally on $\tilde \xi_{t_{n-1}}=x$:
\begin{align*}
  \tilde \xi_{t_{n+1}}
  =
  x
  +
  \sum_{i=1}^{5} \nu_{i}(x)\,\NN_{n}^{i}
\end{align*}
where $\NN^{i}_{n}$ are 5 independent Gaussian random variables :
\[
   \NN^{i}_{n} \sim 
   \NN\big(
     \lambda_{i}(x)\,\Delta t
     \,,\,
     \lambda_{i}(x)\,\Delta t
   \big)
\]
So conditionally on $\tilde \xi_{t_n}=x$, $\tilde \xi_{t_{n+1}}$ is normal with mean 
\eqref{eq.poisson.approximation.mean} and covariance matrix \eqref{eq.poisson.approximation.cov}.

Let $\tilde\xi_{t_n}=(\tilde\beta_{t_{n}},\tilde\sigma_{t_{n}})^{*}$, given $\tilde\beta_{t_n}=b$ and $\tilde\sigma_{t_n}=s$:
\begin{subequations}
\label{eq.normal.approximation}
\begin{align}
\label{eq.normal.approximation.1}
  \tilde \beta_{t_{n+1}}
  &=
  b+
  \big[\mu(s) - (1\wedge K_{4}\,b)\,D\big]\,b\,\Delta t
  \textstyle
  +
  \sqrt{\Delta t\,\frac{\mu(s)\,b}{K_{1}}}\;w_n^{1}
  +
  \sqrt{\Delta t\,\frac{(1\wedge K_{4}\,b)^{2}\,D\,b}{K_{4}}}\;w_n^{4}
\\[0.5em]
\nonumber
  \tilde \sigma_{t_{n+1}}
  &=
  s+
  \big[ -  (1\wedge K_{2} \,s)\,k\,\mu(s)\,b + D\,\ssin - (1\wedge K_{5} \,s)\,D\,s\big]
  \,\Delta t
\\
\label{eq.normal.approximation.2}
  &
  \quad\ \;\, 
  +
  \textstyle
  \sqrt{\Delta t\,\frac{(1\wedge K_{2}\,s)^{2}\,k\,\mu(s)\,b}{K_{2}}}\;w_n^{2}
  +
  \sqrt{\Delta t\,\frac{D\,\ssin}{K_{3}}}\;w_n^{3}
  +
  \sqrt{\Delta t\,\frac{(1\wedge K_{5}\,s)^{2}\,D\,s}{K_{5}}}\;w_n^{5}
\end{align}
\end{subequations}
where $w^{i}_{n}$ are i.i.d. $\NN(0,1)$ random variables.

\subsubsection*{Boundary conditions}

In both approximations \eqref{eq.poisson.approximation} and \eqref{eq.normal.approximation}, no mechanism prevents the processes $\tilde X_{t_{n}}$ or $\tilde\xi_{t_{n}}$ from staying within the positive orthant $\R^{2}_{+}$.
An ad hoc solution is to set the concentration to 0 whenever it becomes negative, see Section \ref{sec.simulation.algorithms}.

\subsection{Diffusion model $\xi_{t}=(\beta_{t},\sigma_{t})^*$}
\label{subsec.diffusion.model}

\subsubsection*{A stochastic differential equation model}

System \eqref{eq.normal.approximation} is the Euler-Maruyama time discretization of the diffusion process $\xi_t =(\beta_t ,\sigma_t )^*$ solution of the following SDE:
\begin{align*}
  \rmd \beta_{t}
  &=
  \big[\mu(\sigma_t) - (1\wedge K_4\,\beta_t)\, D\big]\,\beta_t\,\rmd t
  \textstyle
  +
  \sqrt{\frac{\mu(\sigma_t)\,\beta_t}{K_1}}\;\rmd W_t^1
  +
  \sqrt{\frac{(1\wedge K_4\,\beta_t)^2\,D\,\beta_t}{K_4}}\;\rmd W_{t}^{4}
\\[0.5em]
  \rmd \sigma_t
  &=
  \big[ -  (1\wedge K_{2} \,\sigma_t)\,k\,\mu(\sigma_t)\,\beta_t + D\,\ssin - (1\wedge K_{5} \,\sigma_t)\,D\,\sigma_t\big]\,\rmd t
\\
  &
  \qquad 
  +
  \textstyle
  \sqrt{\frac{(1\wedge K_{2}\,\sigma_t)^2\,k\,\mu(\sigma_t)\,\beta_t}{K_2} }\;\rmd W_t^{2}
  +
  \sqrt{\frac{D\,\ssin}{K_3}}\;\rmd W_t^{3}
  +
  \sqrt{\frac{(1\wedge K_{5}\,\sigma_t)^2\,D\,\sigma_t}{K_5}}\;\rmd W_t^{5}
\end{align*}
where $W^{i}_{t}$ are independent standard Wiener processes. Note that this result can be obtained directly from the process $X_t$ without the help of the discrete-time approximation. Indeed the infinitesimal generator of process $X_{t}$ given by \eqref{eq.GI.jump} is a difference operator, and by Taylor development, it can be approximated by a second order differential operator corresponding to a diffusion process \cite{ethier1986a}.

For small $K_{i}$'s the last system is equivalent considering:
\begin{align*}
  \rmd \beta_{t}
  &=
  \big[\mu(\sigma_t)\,\beta_t - D\,\beta_t\big]\,\rmd t
  \textstyle
  +
  \sqrt{\frac{\mu(\sigma_t)\,\beta_t}{K_1}}\;\rmd W_t^1
  +
  \sqrt{\frac{D\,\beta_t}{K_4}}\;\rmd W_{t}^{4}
\\[0.5em]
\nonumber
  \rmd \sigma_t
  &=
  \big[ - k\,\mu(\sigma_t)\,\beta_t + D\,\ssin - D\,\sigma_t\big]\,\rmd t
\\
  &
  \qquad 
  \qquad 
  \qquad 
  +
  \textstyle
  \sqrt{\frac{k\,\mu(\sigma_t)\,\beta_t}{K_{2}} }\;\rmd W_t^{2}
  +
  \sqrt{\frac{D\,\ssin}{K_{3}}}\;\rmd W_t^{3}
  +
  \sqrt{\frac{D\,\sigma_t}{K_{5}}}\;\rmd W_t^{5}
\end{align*}
then we can group the Brownian motions in the following way:
\begin{subequations}
\label{eq.diffusion.approximation}
\begin{align}
\label{eq.diffusion.approximation.beta}
  \rmd \beta_{t}
  &=
  \big[\mu(\sigma_t)- D\big]\,\beta_t\,\rmd t
  \textstyle
  +
  \sqrt{
    \frac{\mu(\sigma_t)\,\beta_t}{K_1}
    +
    \frac{D\,\beta_t}{K_4}}\;\rmd \Wb_t
\\[0.5em]
\label{eq.diffusion.approximation.sigma}
  \rmd \sigma_t
  &=
  \big[ - k\,\mu(\sigma_t)\,\beta_t + D\,(\ssin - \sigma_t)\big]\,\rmd t
  +
  \textstyle
  \sqrt{
    \frac{k\,\mu(\sigma_t)\,\beta_t}{K_{2}}
    +
    \frac{D\,\ssin}{K_{3}}
    +
    \frac{D\,\sigma_t}{K_{5}}
    }\;\rmd \Ws_t
\end{align}
\end{subequations}
where $\Wb$ and $\Ws$ are independent standard Wiener processes.

\subsubsection*{Behavior of the system of SDE's near the axes}

System \eqref{eq.normal.approximation} is the Euler-Maruyama time discretization of the SDE \eqref{eq.diffusion.approximation} (for large $K_{i}$'s). Even if the diffusion approximation  is only valid for large values of the biomass and the substrate, we can study the behavior of \eqref{eq.diffusion.approximation} near the axes. 

As for the discrete-time normal approximation, we should clarify the boundary conditions. As we well see, the component $\beta_{t}$ given by \eqref{eq.diffusion.approximation.beta} will remain positive, but the component $\sigma_{t}$ given by \eqref{eq.diffusion.approximation.sigma} could become negative. We must first require that $\mu(s)=0$ for $s<0$. Then, note that each equation of \eqref{eq.diffusion.approximation} is related to the following well-known CIR model for interest rates:
\begin{remark}[Cox-Ingersoll-Ross model]
\label{remark.CIR}
Consider the one--dimensional SDE:
\begin{align}
\label{eq.cir}
  \rmd X_{t}
  =
  (a+b\,X_{t})\,\rmd t
  +
  \sigma\,\sqrt{X_{t}}\,\rmd W_{t}
  \,,\quad X_{0}=x_{0}\geq 0\,.
\end{align}
with $a\geq 0$, $b\in\R$, $\sigma>0$. According to \cite[Prop. 6.2.4]{lamberton1996a}, for all $x_{0}\geq 0$, $X$ is a continuous process taking values in $\R^{+}$, and let $\tau=\inf\{t\geq 0,\, X_{t}=0\}$, then:
\begin{enumerate}
\item
If $a\geq \sigma^2/2$, then $\tau=\infty$ $\P_{x}$--a.s.;
\item
if $0\leq a< \sigma^2/2$ and $b\leq 0$ then $\tau<\infty$ $\P_{x}$--a.s.;
\item
if $0\leq a< \sigma^2/2$ and $b> 0$ then $\P_{x}(\tau<\infty)\in(0,1)$.
\end{enumerate}
In the first case, $X$ never reaches 0. In the second case $X$ a.s. reaches the state 0, in the third case it may reach 0. If $a=0$ then the state 0 is absorbing. 
\end{remark}

It is clear that $\beta=0$ is an absorbing state for \eqref{eq.diffusion.approximation.beta}, and when $\beta=0$, \eqref{eq.diffusion.approximation.sigma} reduces to
\begin{align*}
  \rmd \sigma_t
  &=
  D\,(\ssin - \sigma_t)\,\rmd t
  +
  \textstyle
  \sqrt{
    \frac{D\,\ssin}{K_{3}}
    +
    \frac{D\,\sigma_t}{K_{5}}
    }\;\rmd \Ws_t
\end{align*}
and from Remark \ref{remark.CIR}, the solution of this SDE will stay on the half-line
$[-\frac{K_{5}}{K_{3}}\,\ssin,\infty)$ and:
\begin{enumerate}
\item 
if $\ssin\,(\frac{1}{K_{3}}+\frac{1}{K_{5}}) \geq \frac{1}{2\,K_{5}^{2}}$ then $\sigma_{t}$ never reaches $-\frac{K_{5}}{K_{3}}\,\ssin$;
\item
if $\ssin\,(\frac{1}{K_{3}}+\frac{1}{K_{5}}) < \frac{1}{2\,K_{5}^{2}}$ then $\sigma_{t}$ reaches $-\frac{K_{5}}{K_{3}}\,\ssin$ in finite time and is reflected.
\end{enumerate}
Indeed, it is enough to apply Itô formula to $\tilde \sigma_{t} = \frac{D\,\ssin}{K_{3}} +\frac{D\,\sigma_t}{K_{5}} $ and to use the Remark \ref{remark.CIR}. Note that, as $K_{5}$ is large, condition \fenumi\ is more realistic than condition \fenumii.

To extend the definition of  \eqref{eq.diffusion.approximation} for negative value of $\sigma$, let suppose that $\mu(\sigma)=0$ for $\sigma\leq 0$. As we seen, $\beta_{t}$ will stay non-negative and $\beta=0$ is an absorbing state. Also $\sigma_{t}\geq -\frac{K_{5}}{K_{3}}\,\ssin$ and for large $K_{5}$ this state will be repulsive. Note that for small values of $\sigma_{t}$, as the $K_{i}$ are large, the diffusion term in \eqref{eq.diffusion.approximation.sigma} will be small and the drift part will be dominated by $D\,\ssin$ so that $\sigma_{t}$ will increase fast and its probability to be negative will be small. 

The fact that the substrate concentration could be ``negative''  is due to the normal approximation. This approximation is valid for large values of concentration and the validity of the diffusion system \eqref{eq.diffusion.approximation} is questionable for small concentration. Nonetheless we can study its properties.

A possibility to get an SDE with positive solution is to consider an SDE with boundary condition  \cite[\S\,IV-7]{ikeda1981a} by adding a local time in $\{\sigma=0\}$ to the Equation \eqref{eq.diffusion.approximation.sigma}. This solution is rather artificial and will not be retained.

\medskip

So the solution of the system \eqref{eq.diffusion.approximation} remains in the domain
$\DD=[0,\infty)\times [-\frac{K_{5}}{K_{3}}\,\ssin,\infty)$. The proof that this system admits a strong solution with pathwise uniqueness is presented in Appendix \ref{sec.existence.uniqueness}.

\subsection{Asymptotic analysis}
\label{subsec.asymptotic.analysis}

The convergence of the pure jump model \eqref{eq.jump.representation} or of the 
diffusion approximation \eqref{eq.diffusion.approximation} to the deterministic  model \eqref{eq.chemostat} as all the $K_{i}\to\infty$ can be rigorously established. 

Let $X^K_t$ be the pure jump model defined at the beginning of Section \ref{subsec.jump.model}, or as the solution of the Equation \eqref{eq.jump.representation} for a given $K \eqdef (K_{1},K_{2},K_{3},K_{4},K_{5})$. Let $\xi^K_t$ be the solution of the SDE \eqref{eq.diffusion.approximation}. Let $x(t)$ be the EDO model solution of Equation \eqref{eq.chemostat}. Then $X^K_{t}$ converges toward $x(t)$ in the following way: for all $T>0$ and all $\delta>0$,
\begin{align}
\label{eq.cv.X.to.x}
  \P\Big(
     \sup_{0\leq t \leq T}\norm{X^K_{t} - x(t)}\geq \delta
  \Big)
  \longrightarrow 0
\end{align}
as $K_{i}\to\infty$ for all $i=1\cdots 5$. 
This result is not surprising if we consider the representation \eqref{eq.representation.pure.jump} of  $(X_{t})_{t\geq 0}$; it was obtained in a context of martingale convergence theorems in \cite{kurtz1970a,kurtz1971b} or in a more general context of convergence of sequences of infinitesimal generators in \cite{ethier1986a}.

We can also prove the same type of convergence for the process $\xi^K_{t}$. Indeed, in Equation \eqref{eq.diffusion.approximation} the scale coefficients appears as $1/\sqrt{K_{i}}$ in the diffusion part of the SDE, and the convergence clearly holds as all the $K_{i}$ tends to infinity.

\subsection{Other models}
\label{sec.other.models}

\begin{figure}
\begin{center}
\includegraphics[width=7cm]{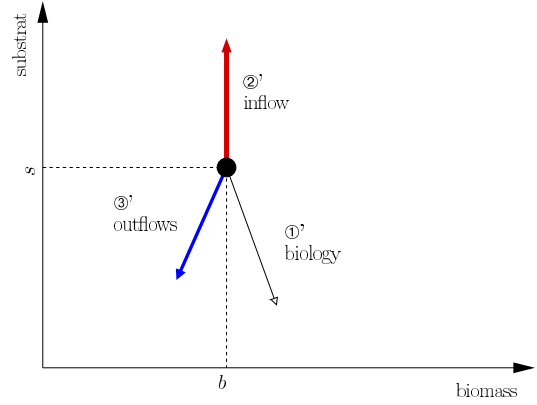} 
\end{center}
\vspace{-1em}
\caption{\it In this simplified model, from a position $x=(b,s)^*$ the process could jump according to 3 mechanisms (biology, inflow, and outflow), the basic jump  \circledip\ has a length $\frac{1}{K_{i}'}$ for $i=1,\dots, 3$}
\label{fig.jumps.simplified}
\end{figure}

As already noticed, the choice of $(\lambda_i(x), \nu_i(x))$ satisfying 
\eqref{eq.mass.action} is not unique. We choose not to make the jump sizes depend on the state value $x$ (except for the boundary conditions), only the jump rates depend on $x$. Another possibility is to choose jump sizes that depend on the state value $x$. For example instead of the choice of Table \ref{table:rates:jumps}, we can choose:
{\small
\begin{align*}
\lambda_{1}(x) &= K_{1}\,\mu(s)\,,
&
\lambda_{2}(x) &= K_{2}\,k\,\mu(s)\,,
&
\lambda_{3}(x) &= K_{3}\,D\,,
&
\lambda_{4}(x) &= K_{4}\,D\,,
&
\lambda_{5}(x) &= K_{5}\,D
\end{align*}
}
and
{\small
\begin{align*}
\nu_{1}(x) 
& = \begin{psmallmatrix}\frac{b}{K_{1}}\\ 0\end{psmallmatrix}\,,
&
\nu_{2}(x) 
& = -\begin{psmallmatrix}0\\ \frac{b}{K_{2}}\end{psmallmatrix}\,,
&
\nu_{3}(x) 
& = \begin{psmallmatrix}0\\ \frac{\ssin}{K_{3}}\end{psmallmatrix}\,,
&
\nu_{4}(x) 
& = -\begin{psmallmatrix}\frac{b}{K_{4}}\\ 0\end{psmallmatrix}\,,
&
\nu_{5}(x) 
& = -\begin{psmallmatrix}0\\ \frac{s}{K_{5}} \end{psmallmatrix}
\end{align*}}
\noindent(if we neglect the boundary condition). Then in place of \eqref{eq.diffusion.approximation} we have the following set of equations:
\begin{subequations}
\label{eq.diffusion.approximation.other}
\begin{align}
\label{eq.diffusion.approximation.other.beta}
  \rmd \beta_{t}
  &=
  \big[\mu(\sigma_t)\,\beta_t - D\,\beta_t\big]\,\rmd t
  \textstyle
  +
  \sqrt{\frac{\mu(\sigma_t)}{K_1}}\,\beta_t\;\rmd W_t^1
  +
  \sqrt{\frac{D}{K_4}}\,\beta_t\;\rmd W_{t}^{4}
\\[0.5em]
\nonumber
  \rmd \sigma_t
  &=
  \big[ - k\,\mu(\sigma_t)\,\beta_t + D\,\ssin - D\,\sigma_t\big]\,\rmd t
\\
\label{eq.diffusion.approximation.other.sigma}
  &
  \qquad 
  \qquad 
  \qquad 
  +
  \textstyle
  \sqrt{\frac{k\,\mu(\sigma_t)}{K_{2}} }\,\beta_t\;\rmd W_t^{2}
  +
  \sqrt{\frac{D}{K_{3}}}\,\ssin\;\rmd W_t^{3}
  +
  \sqrt{\frac{D}{K_{5}}}\,\sigma_t\;\rmd W_t^{5}
\end{align}
\end{subequations}
where $W^i$ are independent standard Wiener processes.

\subsubsection*{A three components model}

Instead of the five components \ding{192} to \ding{196}, we can consider a case with three independent sources of jump variation. This example strictly preserves the geometry \eqref{eq.geometry} by considering
three independent sources of jump variation:
\begin{itemize}
\item[\ding{192}$'$] biology term: biomass increase and substrate
  decrease  at scale $K'_{1}$; 
\item[\ding{193}$'$] inflow term: substrate inflow at scale $K'_{2}$;
\item[\ding{194}$'$] outflow term: biomass and substrate  outflow at
  scale $K'_{3}$ 
\end{itemize}
see Figure \ref{fig.jumps.simplified}. 

Again the jump sizes and rates should be chosen so as to satisfy the
mass balance principle and the stochastic mass action, with no
canonical choice. An ad hoc choice is given in the following table:
\begin{table}[ht] 
\centering \small
\begin{tabular}{|l|ccc|}
\hline 
   & \ding{192}$'$ & \ding{193}$'$ & \ding{194}$'$ \\ \cline{2-4}
   &  biology & inflow & outflow \\
\hline\hline 
rate $\lambda_{i}'(x)$ 
   & $K'_{1}\,\mu(s)\,b$ 
   & $K'_{2}\,D$ 
   & $K'_{3}\,D$ \vphantom{$\displaystyle\int$} \\ \hline 
jump $\nu'_{i}(x)$  
   & $\begin{pmatrix}\frac{1}{K'_{1}}\\[0.4em]
                     - \frac{k \wedge K'_1\, s}{K'_1}
     \end{pmatrix}$
   & $\begin{pmatrix}0\\ \frac{\ssin}{K'_{2}}\end{pmatrix}$
   & $\begin{pmatrix}
      -  \frac{1 \wedge K'_{3}\, \Vert x \Vert}{K'_{3}} \, b
      \\[0.4em] 
      - \frac{1 \wedge K'_{3}\, \Vert x \Vert}{K'_{3}} \, s
      \end{pmatrix}$
    \vphantom{$\begin{pmatrix}0\\ -1 \\ 0 \\ 0\end{pmatrix}$}
    \\ [2ex] 
\hline 
\end{tabular} 
\caption{\it Rates and jumps of an ad hoc choice for three mechanisms of the pure jump process. Note that the third jump $\nu'_{3}(x)$ is radial.} 
\label{table:rates:jumps:adhoc} 
\end{table}

These jumps are now essentially equal to 
\begin{align*}
 \nu'_{1}&\eqdef \frac{1}{K'_{1}}\, 
 \begin{pmatrix}1\\-k\end{pmatrix}\,,
 &
 \nu'_{2}&\eqdef \frac{1}{K_{2}}
 \begin{pmatrix}0\\ \ssin \end{pmatrix}\,,
 &
 \nu'_{3}&\eqdef\frac{1}{K'_{3}}
 \begin{pmatrix}-b\\ -s\end{pmatrix}.
\end{align*} 
This setting forces the jumps to be directed along the corresponding
vector field, which is a strong constraint. In particular, the
stoichiometry is strictly respected:  the production of 1~unit of
biomass requires \emph{exactly} $k$~units of substrate. Moreover the
outflow jump is always radial, so that the increments of biomass and
substrate are again strongly linked. Notice that for this particular
choice of $\lambda'_3$ and $\nu'_3$, the jump rate is constant but the
jump size is not. In other words, the jump carries information both in
the direction and the intensity of the variation. This will affect the
qualitative behavior of the process and of its diffusion
approximation, regarding extinction for example. 

As for our canonical model, we obtain a SDE for the diffusion
approximation of the jump process~:
\begin{align*}
  \rmd \beta_{t}
  &=
  \big[\mu(\sigma_t) - (1\wedge K'_3\,\beta_t)\, D\big]\,\beta_t\,\rmd t
  \textstyle
  +
  \sqrt{\frac{\mu(\sigma_t)\,\beta_t}{K'_1}}\;\rmd W_t^1
  +
  \sqrt{\frac{D}{K'_3}}(1\wedge K'_3\,\Vert
  \xi_t\Vert)\,\,\beta_t\;\rmd W_{t}^{3} 
\\[0.5em]
  \rmd \sigma_t
  &=
  \big[ -  (k \wedge K'_{1} \,\sigma_t)\,\mu(\sigma_t)\,\beta_t 
        + D\,\ssin - 
        (1\wedge K'_{3} \,\Vert \xi_t\Vert)\,D\,\sigma_t\big]\,\rmd t
\\
  &
  \qquad 
  +
  \textstyle
  \sqrt{\frac{\mu(\sigma_t)\,\beta_t}{K_2} }\, 
  (k\wedge K'_{1}\,\sigma_t)\;\rmd W_t^{1}
  +
  \sqrt{\frac{D}{K'_2}}\, \ssin\;\rmd W_t^{2}
  +
  \sqrt{\frac{D}{K'_3}}\,(1\wedge K'_3\,\Vert
  \xi_t\Vert)\,\,\sigma_t \;\rmd W_t^{5} \ .
\end{align*}
Notice that since $W^1$ and $W^3$ affect both components of $\xi_t$,
the quadratic variation process $\langle \xi \rangle _t$ will not be a
diagonal matrix. In comparison with~\eqref{eq.geometry}, we write the
vector form of the SDE for large $K_i$'s: 
\begin{multline*} 
  \rmd \begin{psmallmatrix}  \beta_t \\ \sigma_t\end{psmallmatrix}
  =
  \Big [
	\underbrace{
	\mu(\sigma_t)\,\beta_t \, 
	\begin{psmallmatrix}
	  1\\ 
	  -k \\ 
	\end{psmallmatrix}
	}_{\textrm{biology}}
	\underbrace{
	+
	D\,
	\begin{psmallmatrix}
	  0 \\ \ssin
	\end{psmallmatrix}
	}_{\textrm{inflow}}
	\underbrace{
	-
	D\,
	\begin{psmallmatrix}
	  \beta_t \\ \sigma_t
	\end{psmallmatrix}
	}_{\textrm{outflow}} 
  \Big  ] 
  \;\rmd t 
 \\
 + 
 \underbrace{\textstyle
	\sqrt{\frac{\mu(\sigma_t)\,\beta_t }{K'_1}}
	\begin{psmallmatrix}
	  1\\ 
	  -k 
	\end{psmallmatrix} \;\rmd W_t^1
	}_{\textrm{biology}}
	+
	\underbrace{
	\textstyle 
	\sqrt{\frac{D}{K'_2}}\,
	\begin{psmallmatrix}
	  0 \\ \ssin
	\end{psmallmatrix}\;\rmd W_t^2
	}_{\textrm{inflow}}
	+
	\underbrace{
	\textstyle
	\sqrt{\frac{D}{K'_3}}\,
	\begin{psmallmatrix} \beta_t \\ \sigma_t \end{psmallmatrix}\;\rmd W_t^3
	}_{\textrm{outflow}} \ .
\end{multline*} 
The diffusion term appears as the conjunction of three perturbations
acting along the three vector fields determined by the sources of
variation. Moreover, the intensity of the noise could be different for
each type of perturbation. Considering this model could therefore be
of interest, if the geometric interpretation of the noise is
meaningful,~see \cite{joannides2010a}.

\subsubsection*{Comparison with the Imhof-Walcher model \cite{imhof2005a}}

We finally mention that the diffusion model appearing in~\cite{imhof2005a},
is obtained from  
\eqref{eq.diffusion.approximation.other} by letting $K_{1}=K_{2}=K_{3}=0$ which leads to:
\begin{subequations}
\label{eq.diffusion.approximation.adhoc}
\begin{align}
\label{eq.diffusion.approximation.adhoc.beta}
  \rmd \beta_{t}
  &=
  \big[\mu(\sigma_t)- D\big]\,\beta_t\,\rmd t
  \textstyle
  +
  \cb\,\beta_t\;\rmd \Wb_t
\\[0.5em]
\label{eq.diffusion.approximation.adhoc.sigma}
  \rmd \sigma_t
  &=
  \big[ - k\,\mu(\sigma_t)\,\beta_t + D\,(\ssin - \sigma_t)\big]\,\rmd t
  +
  \cs\,\sigma_t\;\rmd \Ws_t
\end{align}
\end{subequations}
The choice of these coefficients is justified in~\cite{imhof2005a} by
constructing an approximating Markov chain, and then taking the limit
as the sampling rate goes to~0. This model will be compared to the
diffusion approximation \eqref{eq.diffusion.approximation} model on a
simulation test in Section \ref{sec.simulation.comp.adhoc}.

\section{Simulation algorithms}
\label{sec.simulation.algorithms}

We presented several models for the chemostat system: the pure jump model $(X_{t})_{t\geq 0}$ could be considered as a detailed model at the microscopic scale. The Poisson approximation 
$(\tilde X_{t_{n}})_{n\in\N}$  given by \eqref{eq.poisson.approximation} and the normal approximation $(\tilde \xi_{t_{n}})_{n\in\N}$ given \eqref{eq.normal.approximation} are constant time step approximation of the pure jump process. Finally the diffusion process $(\xi_t)_{t\geq 0}$  solution of the SDE \eqref{eq.diffusion.approximation} is a continuous time approximation of the pure jump process.

The now present the three associated simulation algorithms that will be valid at different scales.

\subsection{Pure jump model}
\label{subsec.simu.pure.jump}

The pure jump model in continuous time described in Section \ref{subsec.jump.model} can be exactly simulated thanks to the Gillespie algorithm, also called stochastic simulation algorithm, described in Algorithm \ref{algo.gillespie}.

\begin{algorithm}
\begin{center}
\begin{minipage}{13cm} 
\hrule
\vspace{0.3em}
\begin{algorithmic}
\STATE $t\ot 0$, $x\ot x_0$
\STATE save $(t,x)$
\WHILE{$t\leq\Tmax$}
  \STATE compute $\lambda_{i}(x)$
    \quad\texttt{\% see Table \ref{table:rates:jumps}}
  \STATE $\lambda = \sum_{i=1}^{5}\lambda_{i}(x)$
  \STATE $\Delta t \sim \textrm{\rm Exp}(\lambda)$
    \quad\texttt{\% exponential distribution}
  \STATE $u \sim \textrm{\rm U}[0,1]$
    \quad\texttt{\% uniform distribution}
  \STATE $t \ot t+\Delta t$
  \IF{$u\leq \lambda_1(x)/\lambda$}
    \STATE $x\ot x+\bnu_{1}$ 
    \quad\texttt{\% biomass reproduction}
  \ELSIF{$u\leq \{\lambda_1(x)+\lambda_2(x)\}/\lambda$}
    \STATE $x\ot [x-\bnu_{2}]_{+} $ 
    \quad\texttt{\% consumption}
  \ELSIF{$u\leq \{\lambda_1(x)+\lambda_2(x)+\lambda_3(x)\}/\lambda$}
    \STATE $x\ot x+\bnu_{3} $ 
    \quad\texttt{\% substrate inflow}
  \ELSIF{$u\leq \{\lambda_1(x)+\lambda_2(x)+\lambda_3(x)+\lambda_4(x)\}/\lambda$}
    \STATE $x\ot [x-\bnu_{4}]_{+} $ 
    \quad\texttt{\% biomass outflow}
  \ELSE
    \STATE $x\ot [x-\bnu_{5}]_{+} $ 
    \quad\texttt{\% substrate outflow}
  \ENDIF
  \STATE save $(t,x)$
\ENDWHILE
\end{algorithmic}
\vspace{0.3em}
\hrule
\vspace{0.7em}
Here 
  $\bnu_1 = \big(\begin{smallmatrix}1/K_1\\ 0\end{smallmatrix}\big),$
  $\bnu_2 = \big(\begin{smallmatrix}0\\ 1/K_2\end{smallmatrix}\big),$
  $\bnu_3 = \big(\begin{smallmatrix}0\\ 1/K_3\end{smallmatrix}\big),$
  $\bnu_4 = \big(\begin{smallmatrix}1/K_4\\ 0\end{smallmatrix}\big),$
  $\bnu_5 = \big(\begin{smallmatrix}0\\ 1/K_5\end{smallmatrix}\big)$
and 
$[x]_{+}$ is the projection on the positive quadrant: $[x]_{+} =
  \left[\left(\begin{smallmatrix}\beta\\ \sigma\end{smallmatrix}\right)\right]_{+}
  =
  \left(\begin{smallmatrix}\beta\vee 0\\ \sigma\vee 0\end{smallmatrix}\right)$.
\end{minipage} 
\end{center}
\vskip-1em
\caption{\it Gillespie algorithm (or stochastic simulation algorithm).}
\label{algo.gillespie}
\end{algorithm}

When the rate coefficients $\lambda_{i}(x)$ are large the time increment will be small and the Gillespie algorithm is impractical. As the scale coefficients $K_{i}$ are large, the $\lambda_{i}(x)$, $i\neq 3$, are large only when $\beta$ and $\sigma$ are small; $\lambda_3(x)$ will remain large as it does not depend on $x$.

\subsection{Poisson approximation}
\label{subsec.simu.poisson}

\begin{algorithm}
\begin{center}
\begin{minipage}{13cm} 
\hrule
\vspace{0.3em}
\begin{algorithmic}
\STATE $t\ot 0$, $x\ot x_0$
\STATE save $(t,x)$
\WHILE{$t\leq\Tmax$}
  \STATE compute $\lambda_{i}(x)$
    \quad\texttt{\% see Table \ref{table:rates:jumps}}
  \STATE $\lambda = \sum_{i=1}^{5}\lambda_{i}(x)$
  \STATE compute $\bbm_{i}(x)$, $\bv_{i}(x)$
    \quad\texttt{\% see \eqref{eq.tau.leap.functions}}
  \STATE $\Delta t\ot \min_{i=1\cdots 5}\big\{
	    \epsilon\,\lambda/|\bbm_i(x)|
	    \,,\,
	    \epsilon^2\,\lambda^2/\bv_i(x)
  \big\}$
  \STATE $t \ot t+\Delta t$
  \STATE $\PP_i \sim \textrm{Poisson}(\lambda_{i}(x)\,\Delta t)$ for $i=1\cdots 5$
  \STATE $x \ot [x+\bnu_1\,\PP_1-\bnu_2\,\PP_2+\bnu_3\,\PP_3-\bnu_4\,\PP_4-\bnu_5\,\PP_5]_{+}$
  \STATE save $(t,x)$
\ENDWHILE
\end{algorithmic}
\vspace{0.3em}
\hrule
\vspace{0.7em}
Here 
  $\bnu_1 = \big(\begin{smallmatrix}1/K_1\\ 0\end{smallmatrix}\big),$
  $\bnu_2 = \big(\begin{smallmatrix}0\\ 1/K_2\end{smallmatrix}\big),$
  $\bnu_3 = \big(\begin{smallmatrix}0\\ 1/K_3\end{smallmatrix}\big),$
  $\bnu_4 = \big(\begin{smallmatrix}1/K_4\\ 0\end{smallmatrix}\big),$
  $\bnu_5 = \big(\begin{smallmatrix}0\\ 1/K_5\end{smallmatrix}\big)$
and 
$[x]_{+}$ is the projection on the positive quadrant: $[x]_{+} =
  \left[\left(\begin{smallmatrix}\beta\\ \sigma\end{smallmatrix}\right)\right]_{+}
  =
  \left(\begin{smallmatrix}\beta\vee 0\\ \sigma\vee 0\end{smallmatrix}\right)$.
\end{minipage} 
\end{center}
\vskip-1em
\caption{\it Poisson approximation or tau-leap method.}
\label{algo.poisson.approximation}
\end{algorithm}

The simulation of the previous mo\-del could be cumbersome for very high rates of event. In this case it is desirable to use the fixed time step Poisson approximation method \eqref{eq.poisson.approximation} also called tau-leap \cite{gillespie2001a}. Recently many papers have addressed the numerical analysis of this approximation scheme \cite{rathinam2005a,tiejun-li2007a,anderson-david2009a}. In this method the time step should be small enough so that it fulfills the following ``leap condition'': the state change in any leap should be small enough that no rate function $\lambda_{i}(x)$ will experience a macroscopically significant change in its value, that is:
\begin{align}
\label{eq.leap.condition}
\textstyle
  \Big|
    \lambda_{i}\big(x+\sum_{i'}\nu_{i'}(x)\,\PP^{i'}_{n}(\Delta t\,\lambda_{i'}(x))\big)
    -
    \lambda_{i}(x)
  \Big|
  \leq
  \epsilon\,\lambda(x)
\end{align}
for $i=1\cdots 5$, where $0<\epsilon\ll 1$ is an error control parameter.

For this method to be practicable \cite{gillespie2003a} proposed an automatic and simple way of determining the largest time step $\Delta t$ compatible with the leap condition. Define:
\begin{subequations}
\label{eq.tau.leap.functions}
\begin{align}
\label{eq.tau.leap.functions.mu}
  \bbm_{i}(x)
  &\eqdef
  \sum_{i'=1}^5 \lambda_{i'}(x)\,\big(\nabla\lambda_{i}(x)\cdot\nu_{i'}\big)
  \,,  
\\
\label{eq.tau.leap.functions.sigma2}
  \bv_i(x)
  &\eqdef
  \sum_{i'=1}^5 \lambda_{i'}(x)\,\big(\nabla\lambda_{i}(x)\cdot\nu_{i'}\big)^{2}
\end{align}
\end{subequations}
for $i,i'=1\cdots 5$, and let
\begin{align}
\label{eq.tau.leap.dt}
  \Delta t
  &=
  \min_{i=1\cdots 5}
  \left\{
    \frac{\epsilon\,\lambda(x)}{|\bbm_i(x)|}
    \,,\,
    \frac{\epsilon^{2}\,\lambda^2(x)}{|\bv_i(x)|}
  \right\}
\end{align}
where $\epsilon$ is an error control parameter ($0<\epsilon\ll 1$), see Algorithm \ref{algo.poisson.approximation}. Note that in the original context the jumps $\nu_{i}(x)$ do not depends on $x$, but in our situation they do not essentially depend on $x$, the dependence on $x$ was introduced to handle the jump near the axes in order to avoid negative concentration.

\subsection{Diffusion (normal) approximation}
\label{subsec.simu.diffusion}

\begin{algorithm}
\begin{center}
\begin{minipage}{14cm} 
\hrule
\vspace{0.3em}
\begin{algorithmic}
\STATE $t\ot 0$, $(\beta,\sigma)\ot(\beta_{0},\sigma_{0})$
\STATE save $(t,\beta,\sigma)$
\WHILE{$t\leq\Tmax$}
  \STATE $\wb\sim\NN(0,1)$, $\ws\sim\NN(0,1)$
  \STATE
  $\beta'
  \ot
  \beta
  +
  (\mu(\sigma) - D)\,\beta\,\Delta t
  +
  \sqrt{
      \frac{\mu(\sigma)\,\beta}{K_1}
      +
      \frac{D\,\beta}{K_4}
  }\,\sqrt{\Delta t}\,\wb
  $
  \STATE 
  $\sigma'
  \ot
  \sigma
  +
  ( - k\,\mu(\sigma)\,\beta + D\,(\ssin - \sigma))
  \,\Delta t
  +
  \sqrt{
     \frac{k\,\mu(\sigma)\,\beta}{K_2}
     +
     \frac{D\,\ssin}{K_{3}}
     +
     \frac{D\,\sigma}{K_{5}}
  }
  \,\sqrt{\Delta t}\,\ws$
  \STATE $\beta \ot [\beta']^{+}$
    \quad\texttt{\% 0 is an absorbing state for $\beta$}
  \STATE $\sigma \ot |\sigma'-\sigmamin|+\sigmamin$
    \quad\texttt{\% reflection at $\sigmamin=-\frac{K_{5}}{K_{3}}\,\ssin$ for $\sigma$}
  \STATE $t \ot t+\Delta t$
  \STATE save $(t,\beta,\sigma)$
\ENDWHILE
\end{algorithmic}
\vspace{0.3em}
\hrule
\end{minipage} 
\end{center}
\caption{\it Normal approximation.}
\label{algo.normal.approximation}
\end{algorithm}

The normal approximation \eqref{eq.normal.approximation} can be slightly modified in order to take into account the qualitative behavior of the SDE \eqref{eq.diffusion.approximation} near the axes. We propose the following scheme:
\begin{subequations}
\label{eq.sde.discretization}
\begin{align}
\nonumber
  \tilde \beta_{t_{n+1}}
  &=
  \Big[\tilde\beta_{t_n}
  +
  \big[\mu(\tilde\sigma_{t_n}) - (1\wedge K_{4}\,\tilde \beta_{t_n})\,D\big]\,\tilde \beta_{t_n}\,\Delta t
\\
\label{eq.sde.discretization.1}
  &
  \quad 
  \quad 
  \textstyle
  +
  \sqrt{\Delta t}\;
  \sqrt{
     \frac{\mu(\tilde\sigma_{t_n})\,\tilde \beta_{t_n}}{K_{1}}
     +
     \frac{(1\wedge K_{4}\,\tilde \beta_{t_n})^{2}\,D\,\tilde \beta_{t_n}}{K_{4}}
   }\;\wb_n
   \Big]^+
   \,,
\\
\nonumber
  \tilde \sigma_{t_{n+1}}
  &=
  \Big|
  \tilde\sigma_{t_n}
  +
  \big[ -  (1\wedge K_{2} \,\tilde\sigma_{t_n})\,k\,\mu(\tilde\sigma_{t_n})\,\tilde \beta_{t_n} + D\,\ssin - (1\wedge K_{5} \,\tilde\sigma_{t_n})\,D\,s\big]
  \,\Delta t
\\[0.3em]
\label{eq.sde.discretization.2}
  &
  \quad 
  \quad 
  +
  \textstyle
  \sqrt{\Delta t}\;
  \sqrt{
      \frac{(1\wedge K_{2}\,\tilde\sigma_{t_n})^{2}\,k\,\mu(s)\,\tilde \beta_{t_n}}{K_{2}}
      +
      \frac{D\,\ssin}{K_{3}}
      +
      \frac{(1\wedge K_{5}\,\tilde\sigma_{t_n})^{2}\,D\,s}{K_{5}}
  }\;\ws_n
  -\sigmamin\Big|+\sigmamin
  \,.
\end{align}
\end{subequations}
Indeed as $\beta=0$ is an absorbing state for the component $\beta_{t}$ of the SDE, instead of the standard Euler-Maruyama \eqref{eq.normal.approximation.1}, we can use \eqref{eq.sde.discretization.1} where $[\cdot]^{+}$ is the positive part and $\wb_{n}$ are i.i.d. $\NN(0,1)$ random variables. 

Also, to take into account that the component $\sigma$ is reflected in $\sigmamin=-\frac{K_{5}}{K_{3}}\,\ssin$ we use the scheme \eqref{eq.sde.discretization.2} where $\ws_{n}$ are i.i.d. $\NN(0,1)$ random variables. This discretization scheme was proposed in \cite{diop2003a} in the context of the CIR diffusion process. In order to get a positive substrate concentration we can consider $\tilde\sigma_{t_n}^{+}=\tilde\sigma_{t_n}\vee 0$ or let $\sigmamin =0$ in \eqref{eq.sde.discretization.2}. The simulation procedure is presented in Algorithm \ref{algo.normal.approximation}.

\begin{remark}[Scales and hybrid simulation]
The three algorithms proposed here are valid at different scales. In the Gillespie algorithm all the detailed microscopic jumps of the dynamics are simulated. 

The idea of the Poisson approximation is to consider a time step $\Delta$ that should be small enough so that the different event rates barely evolve in the time interval $[t,t+\Delta t]$, but large enough for the approximation to be worthwhile. Starting in $x$ at $t$, the time step $\Delta t$ is given by  \eqref{eq.tau.leap.dt} but if it is less than a few multiples of $1/\lambda(x)$ then the Gillespie algorithm should be preferred.

Now the Poisson variables $\PP_{i}$ of parameter $\lambda_{i}(x)\,\Delta t$ could be approximated by normal variables $\NN(\lambda_{i}(x)\,\Delta t,\lambda_{i}(x)\,\Delta t)$ as soon as $\lambda_{i}(x)\,\Delta t\geq 20$.

The simulation method can automatically switch from one algorithm to another one according to the scale. We can also imagine that different components of the state vector are simulated with different algorithms.
\end{remark}

\section{Simulation study}
\label{sec.simulation}

We present simulation results of the discretized diffusion model \eqref{eq.sde.discretization} with Monod  and Haldane specific growth rates \eqref{eq.specific.growth.rate}. The ODE \eqref{eq.chemostat} is integrated with a Runge-Kutta\footnote{The routine \texttt{ode45} of \texttt{Matlab}, an explicit Runge-Kutta (4,5) formula.} scheme but the Euler scheme, corresponding to \eqref{eq.sde.discretization} with $K_{i}=\infty$, gives very close results.

In addition to the deterministic case (case 0 with $K_{i}=\infty$ for all $i$), we consider 3 basic cases (see Table \ref{table.simulation.cases}):
\begin{description}
\item 
[\ ``Standard'' scales:] $K_{2,3,5}=100\times K_{1,4}$ corresponds to the ``standard'' case where the substrate concentration dynamics is closer to the deterministic case than the biomass concentration dynamics.
\item 
[\ ``Unstirred inflow/outflows'' scales:] $K_{1,2}=100\times K_{3,4,5}$ corresponds to the case where inflow and outflows are unstirred.
\item 
[\ ``Fluid substrate'' scales:] $K_{2,3,4}=\infty$, in this case the substrate equation \eqref{eq.diffusion.approximation.sigma} is deterministic, i.e. the substrate dynamics is in fluid limit.
\item 
[\ ``Biological only'' scales:] $K_{3,4,5}=\infty$, in this case we consider that the randomness is only due to biological aspects of the system.
\end{description}

\begin{table}
\begin{center}\small
\begin{tabular}{cllccccc}
\hline
cases &  & & $K_1$& $K_2$& $K_3$& $K_4$& $K_5$ 
\\ \hline\hline
0    & deterministic && $\infty$& $\infty$& $\infty$& $\infty$& $\infty$
\\ \hline
1    & ``standard''        
                       & \it case 1.1 & $10^4$   & $10^6$   & $10^6$   & $10^4$   & $10^6$   \\
     & \footnotesize(see Figure \ref{fig.simu.case1})
                       & \it case 1.2 & $10^5$   & $10^7$   & $10^7$   & $10^5$   & $10^7$   \\
     &                 & \it case 1.3 & $10^7$   & $10^9$   & $10^9$   & $10^7$   & $10^9$
\\ \hline
2    & ``unstirred inflow/outflows''        
                       & \it case 2.1 & $10^6$   & $10^6$   & $10^4$   & $10^4$   & $10^4$   \\
     & \footnotesize(see Figure \ref{fig.simu.case2})
                       & \it case 2.2 & $10^7$   & $10^7$   & $10^5$   & $10^5$   & $10^5$   \\
     &                 & \it case 2.3 & $10^9$   & $10^9$   & $10^7$   & $10^7$   & $10^7$
\\ \hline
3    & ``fluid substrate'' 
                       & \it case 3.1 & $10^6$   & $\infty$ & $\infty$ & $10^4$   & $\infty$ \\
     & \footnotesize(see Figure \ref{fig.simu.case3})
                       & \it case 3.2 & $10^7$   & $\infty$ & $\infty$ & $10^5$   & $\infty$ \\
     &                 & \it case 3.3 & $10^9$   & $\infty$ & $\infty$ & $10^7$   & $\infty$ 
\\ \hline
4    & ``biological only'' 
                       & \it case 4.1 & $10^6$   & $10^4$   & $\infty$ & $\infty$ & $\infty$ \\
     & \footnotesize (see Figure \ref{fig.simu.case4})
                       & \it case 4.2 & $10^7$   & $10^5$   & $\infty$ & $\infty$ & $\infty$ \\
     &                 & \it case 4.3 & $10^9$   & $10^7$   & $\infty$ & $\infty$ & $\infty$ 
\\ \hline
\end{tabular}
\\[0.3em]
\hspace{8.6cm}   \tiny (here ``$\infty=10^{20}$'')
\end{center}
\caption{\it Simulation cases.}
\label{table.simulation.cases}
\end{table}

\subsection{A first comparison of trajectories}
\label{sec.first.comparison}

\begin{table}
\begin{center}\small
\begin{tabular}{llll}
\hline
\multicolumn{4}{r}{Monod model}\\
\hline
& set 1 & set 2 \\
\hline
$k$       & $10$    & $10$        & stoichiometric constant \\
$\mumax$  & $3$     & $0.5$       & maximal growth rate ($h^{-1}$)\\
$D$       & $0.12$  & $0.4$       & dilution rate ($h^{-1}$)\\
$\ssin$    & $0.5$   & $10$        & input concentration ($g/l$)\\
$\ks$     & $6$     & $1$         & half saturation constant ($g/l$)\\
\\ \hline
\end{tabular}
\\[0.6cm]
\begin{tabular}{cc}
\includegraphics[width=3cm]{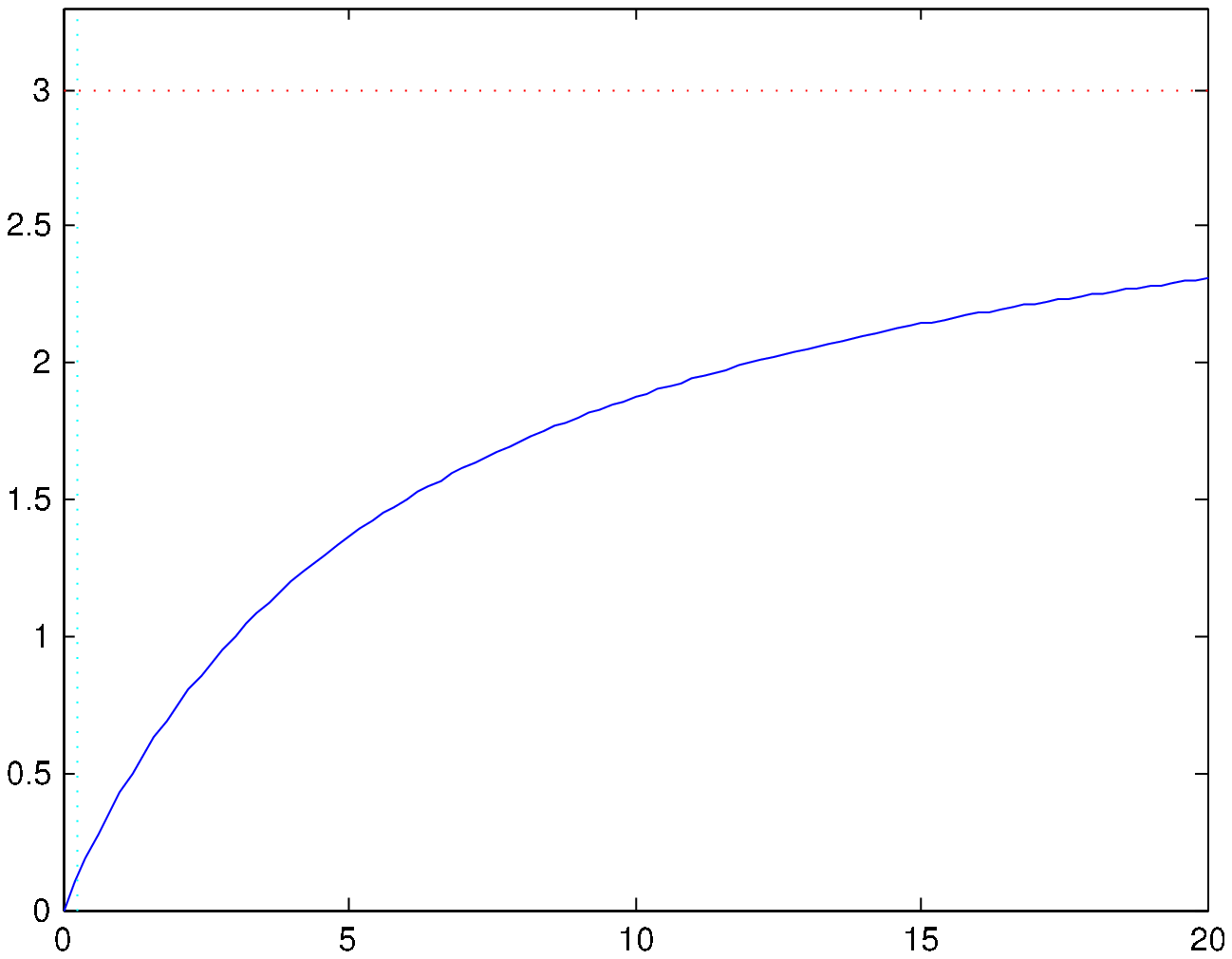}
&
\includegraphics[width=3cm]{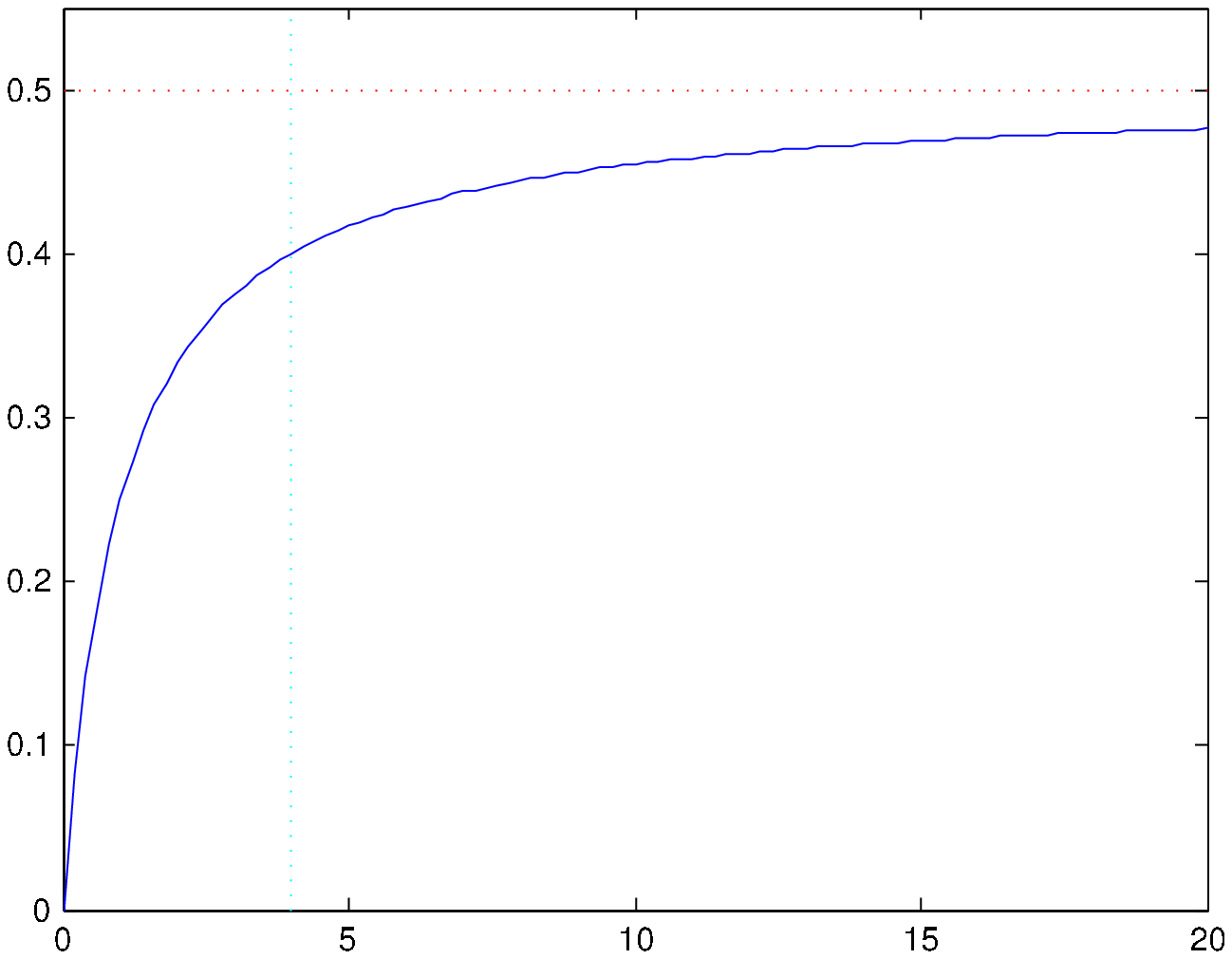}
\\
set 1 & set 2 \end{tabular}
\end{center}
\caption{\it Sets of simulation parameters and the corresponding Monod growth rate functions $s\to\mu(s)$. The horizontal doted line is the maximum capacity $\mumax$ and the vertical doted line the asymptotic substrate concentration of the ODE corresponding to the non-washout case.} 
\label{table.simu.parameter.monod} 
\end{table}

We consider the set of parameters of Table \ref{table.simu.parameter.monod} (set 1) for the Monod case in the ``standard'' scales $K_{1,4}=10^{6}$ and $K_{2,3,5}=10^{8}$. 

\begin{figure}
\begin{center}
\includegraphics[width=7cm]{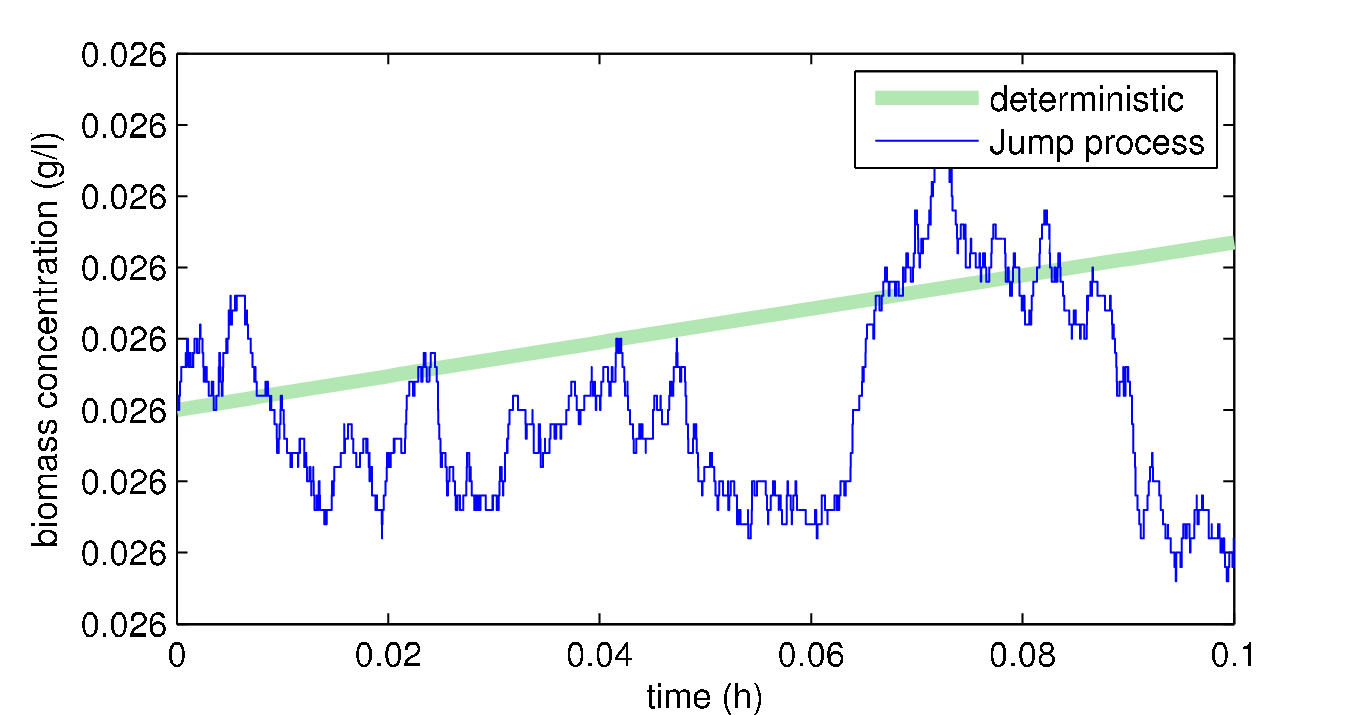}
\includegraphics[width=7cm]{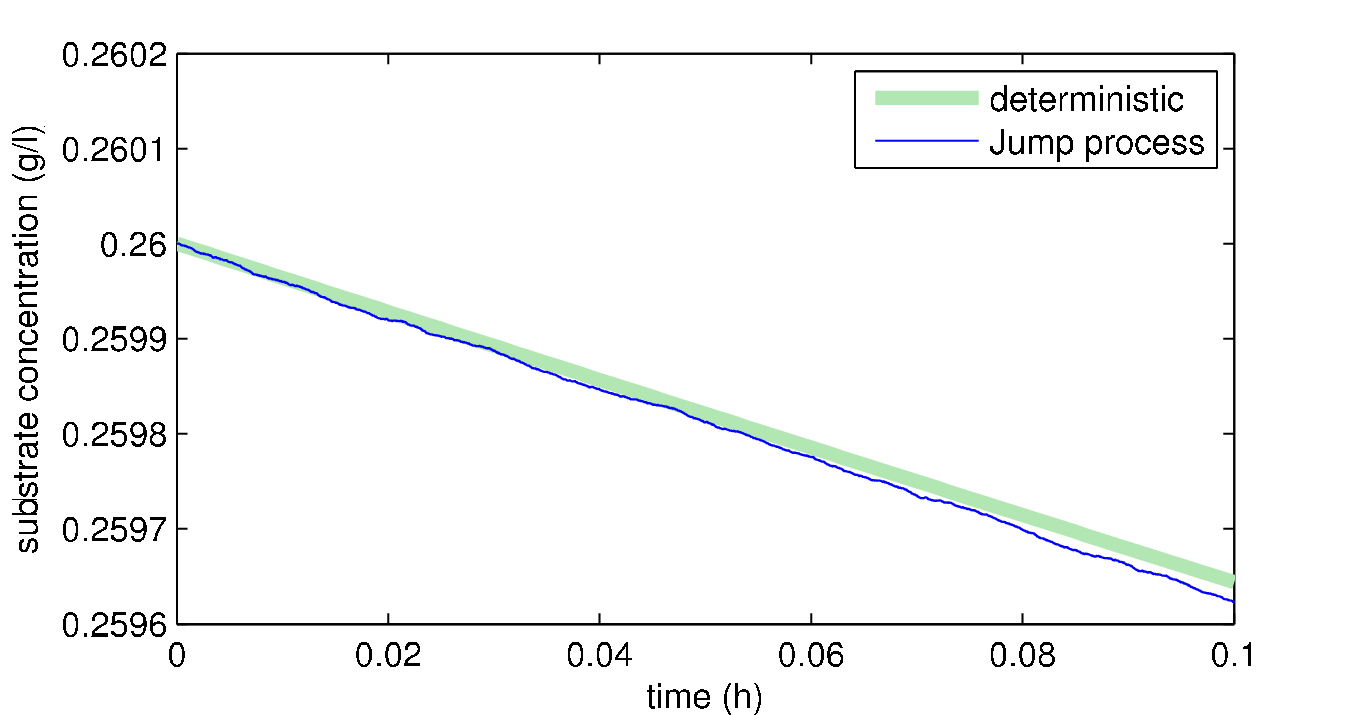}
\end{center}
\caption{\it Simulation of the pure jump process with the Gillespie Algorithm \ref{algo.gillespie} for $K_{1,4}=10^{6}$ and $K_{2,3,5}=10^{8}$ and 1237928 events (blue: a realization of the jump process $(X_{t})_{0\leq t\leq 0.1}$, green: the ODE $(x(t))_{0\leq t\leq 0.1}$) --- the substrate dynamics (RIGHT) presents many small size jumps, the biomass dynamics (LEFT) less jumps but with higher amplitude. The corresponding phase-portrait is plotted in Figure \ref{fig.pur.jump+poisson} (LEFT).}
\label{fig.pur.jump1}
\end{figure}

In Figure \ref{fig.pur.jump1} we present a simulation of the pure jump process with the Gillespie method. As expected, most of the events corresponds to small jumps of the substrate concentration. Before addressing the question of reliability of these algorithms, see next subsection, we first focus on the qualitative nature of the trajectories proposed by the various methods.

\begin{figure}
\begin{center}
\includegraphics[width=7cm]{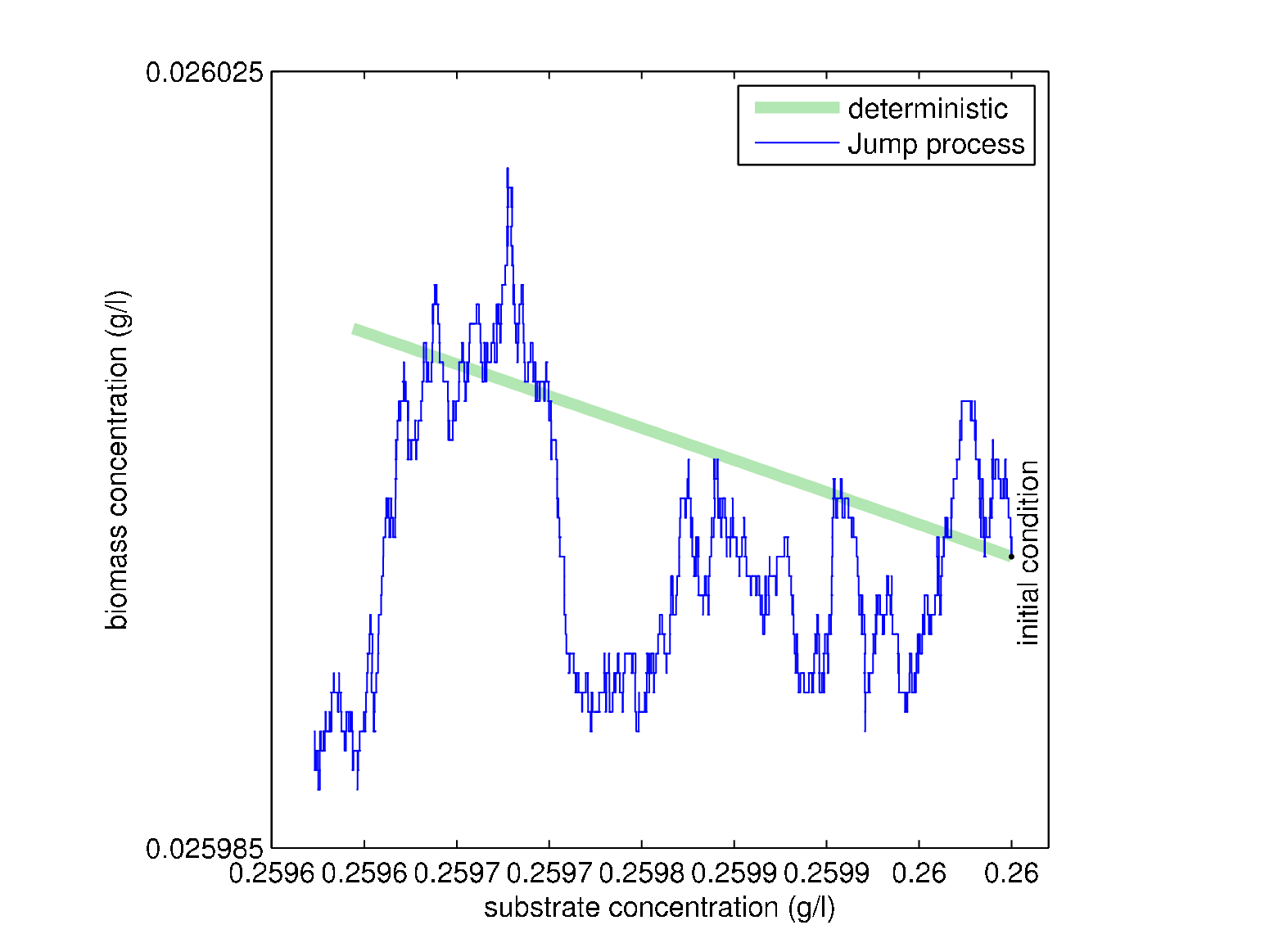}
\includegraphics[width=7cm]{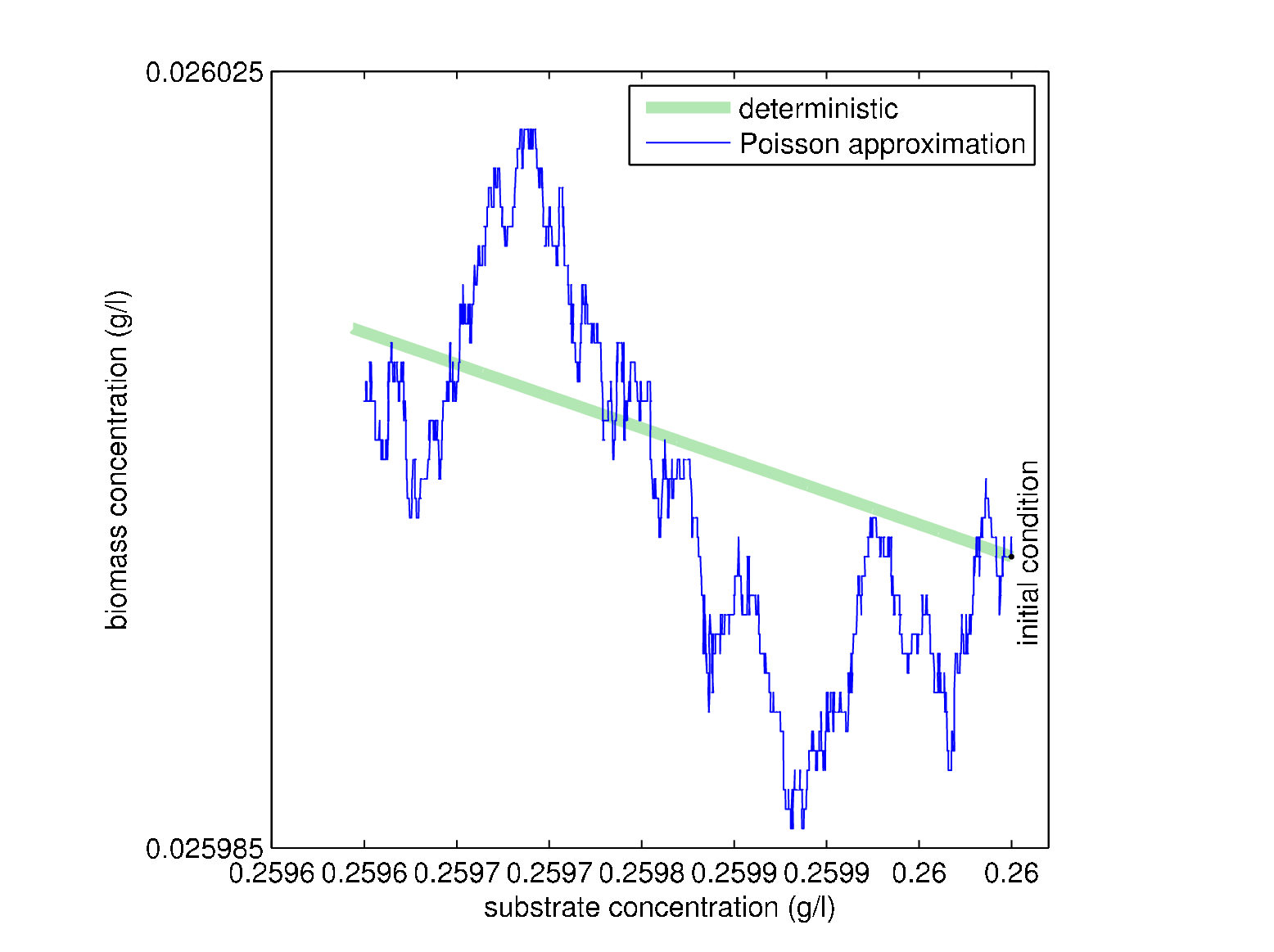}
\end{center}
\caption{\it LEFT : Simulation of the pure jump process with the Gillespie Algorithm \ref{algo.gillespie} for $K_{1,4}=10^{6}$ and $K_{2,3,5}=10^{8}$ and 1237928 events (blue: a realization of the jump process $(X_{t})_{0\leq t\leq 0.1}$, green: the ODE $(x(t))_{0\leq t\leq 0.1}$) --- the substrate dynamics presents many small size jumps, the biomass dynamics less jumps but with higher amplitude --- the final time of simulation is $0.1$ (h). RIGHT: same simulation with the Poisson approximation $(\tilde X_{t_{n}})_{0\leq t_{n}\leq 0.1}$  (tau-leap) with 3456 events for $\epsilon=10^{-6}$ (the time step is $\simeq$ $1.89\times 10^{-5}$).}
\label{fig.pur.jump+poisson}
\end{figure}

In Figure \ref{fig.pur.jump+poisson}, a simulation in a short time horizon of $0.1$ (h) is proposed with Gillespie method (exact simulation) and the Poisson approximation  method (tau-leaping) with a very small error control parameter $\epsilon=10^{-6}$. The corresponding trajectories are very similar though $10^6$ events are needed for the Gillespie method and only 3500 time steps are needed for the tau-leap method. 

\begin{figure}
\begin{center}
\includegraphics[width=7cm]{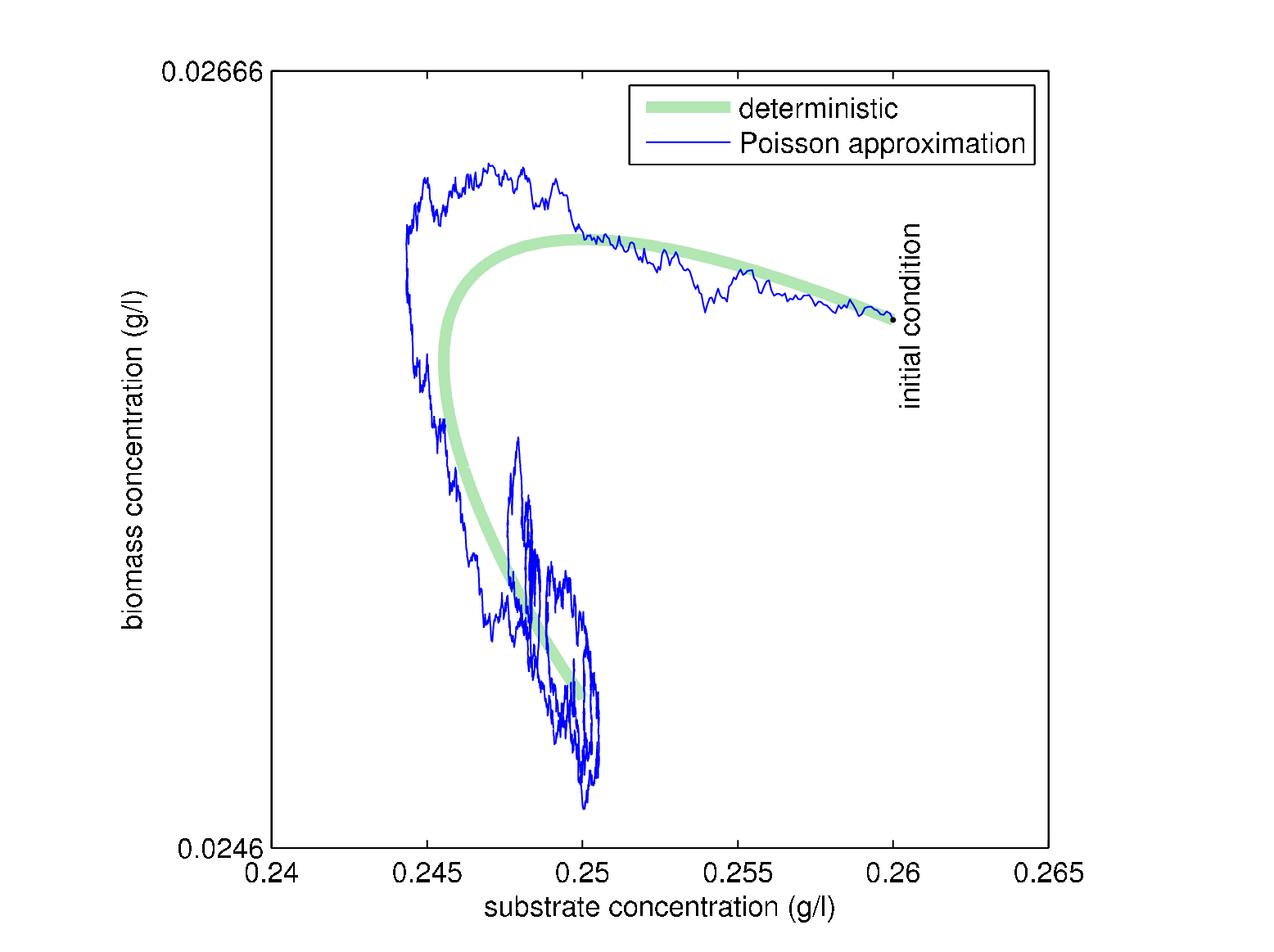}
\includegraphics[width=7cm]{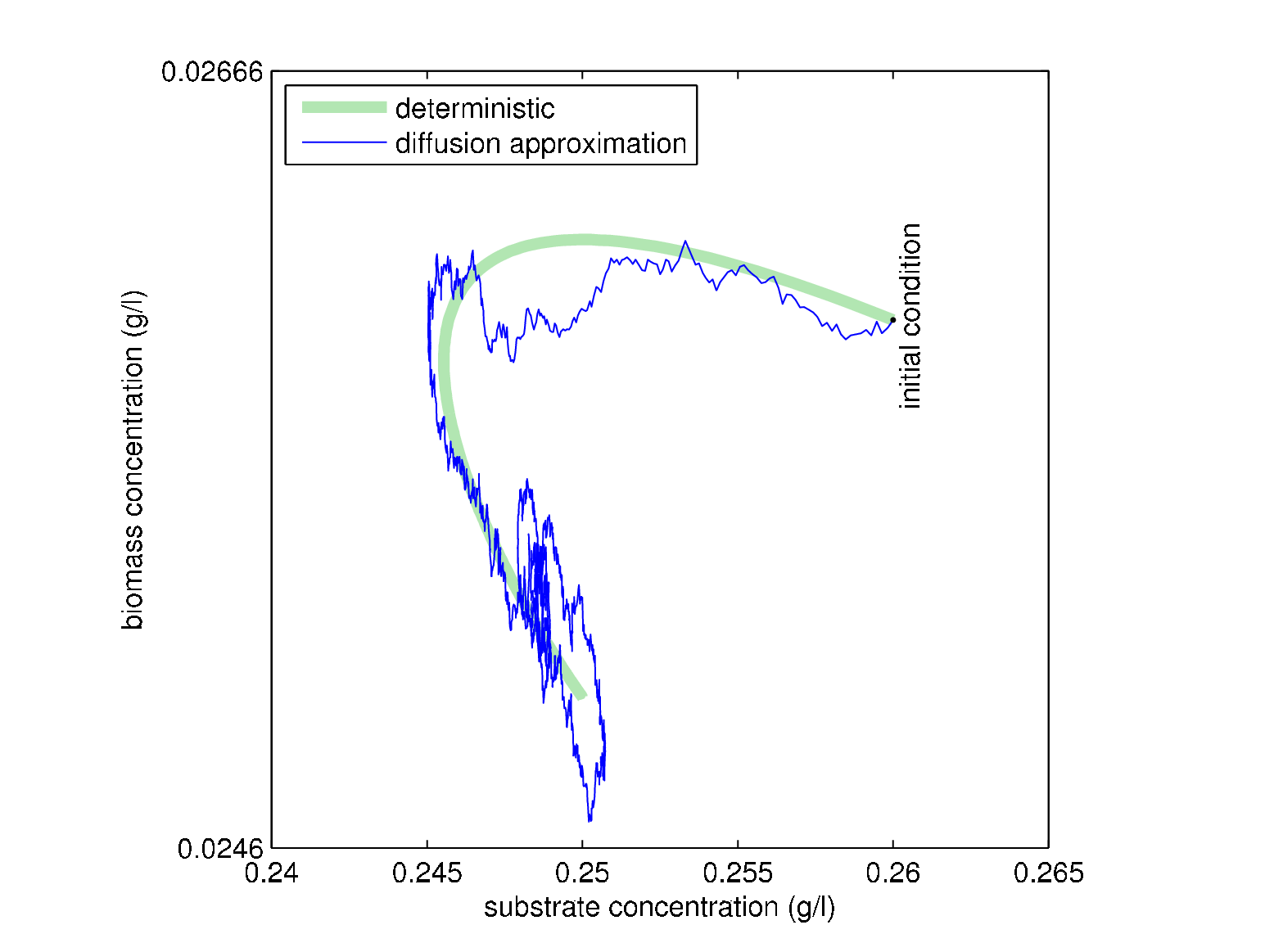}
\end{center}
\caption{\it LEFT : Simulation of the Poisson approximation with the $\tau$-leap Algorithm \ref{algo.gillespie} for $K_{1,4}=10^{6}$ and $K_{2,3,5}=10^{8}$ and 3004 events (blue: a realization of the jump process $(\tilde X_{t_{n}})_{0\leq t_{n}\leq 100}$, green: the ODE $(x(t))_{0\leq t\leq 100}$), $\epsilon=10^{-3}$, $\Delta t$ between $ 0.028 $ and $0.0344 $. RIGHT: same simulation with the normal approximation $(\tilde \xi_{t_{n}})_{0\leq t_{n}\leq 100}$ (diffusion approximation) with $2000$ time steps with $\Delta t=0.05$.}
\label{fig.poisson+normal}
\end{figure}

Figure \ref{fig.poisson+normal} present a simulation on a realistic length of time 100 (h). Here only the Poisson approximation and the normal approximation are reliable. For the Poisson approximation we use $\epsilon=10^{-3}$ (so that $\Delta t$ is between $0.016612$ and $0.034232 $, for 3004 time steps) and for the normal approximation we use $2000$ (corresponding to $\Delta t=0.025$). Again, the associated trajectories are very similar.

\subsection{Law of the concentrations at a given time $t$}
\label{sec.simulation.mc.t.fixe}

\begin{figure}
\begin{center}
\includegraphics[width=15cm]{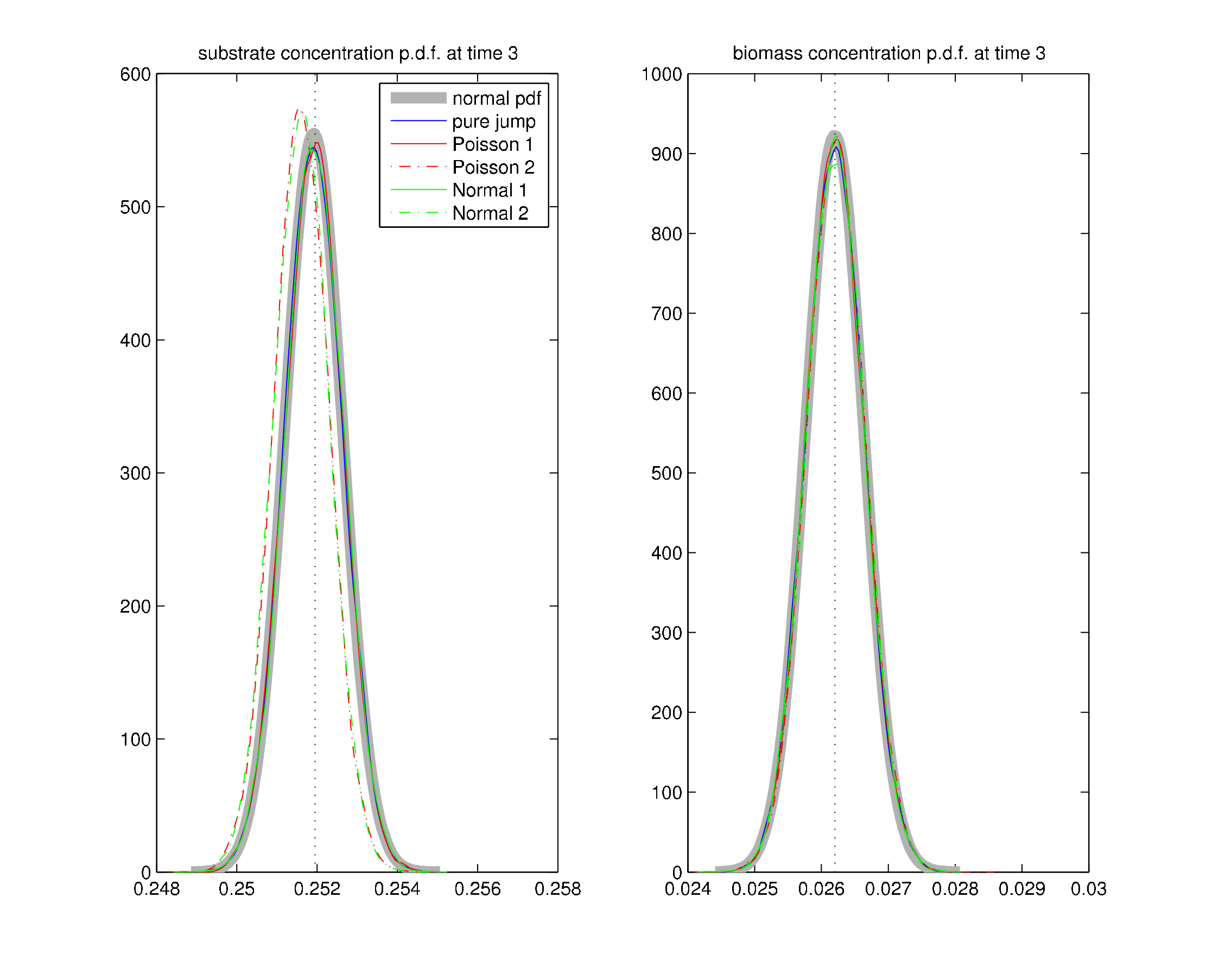}
\end{center}
\caption{\it Empirical densities for the substrate and the biomass concentrations  at time $t=3$ obtained with the exact simulation of the pure jump process $(S_{t},B_{t})$ with the Gillespie method (blue line), with the Poisson approximation $(\tilde S_{t},\tilde B_{t})$ with constant time step (red solid line for a step 0.05, red dash line for a step 0.5), with the normal approximation $(\tilde \sigma_{t},\tilde \beta_{t})$ with constant time step (green solid line for a step 0.05, green dash line for a step 0.5). The corresponding CPU time are respectively: 5 h 45 min 32.6 s, 33.2 s, 4.6 s, 0.7 s, 0.1 s. The simulation parameters are the Set 1 of Table \ref{table.simu.parameter.monod} (Monod growth function) with  ``standard'' scales $K_{1,4}=10^{5}$ and $K_{2,3,5}=10^{7}$. We compute the substrate and the biomass concentrations  at $t=3$ (h) and for $j=1\cdots 20000$ independent Monte Carlo trials. The empirical densities are obtained with a kernel approximation procedure. We also compute the empirical mean and standard deviation from the sample of the pure jump process and plot with a thick grey line the corresponding normal densities: the match is very good. The vertical doted line the value of the substrate and biomass concentration at time $t=3$ given by the ODE; again it matches the mean of all the empirical densities (except the ones corresponding to the time step 0.5).}
\label{fig.huge.mc}
\end{figure}

In Figure \ref{fig.huge.mc}, we propose a Monte Carlo simulation to approximate the marginal densities of the
biomass concentration $B_{t}$ and of the substrate concentration $S_{t}$ at a given time $t$. 
We consider the set of parameters of Table \ref{table.simu.parameter.monod} (set 1) for the Monod case in the ``standard'' scales $K_{1,4}=10^{5}$ and $K_{2,3,5}=10^{7}$. 

We compute $(S_{t}^{(j)},B_{t}^{(j)})$ for $t=3$ (h) for $j=1\cdots 20000$ independent Monte Carlo trials of the pure jump process (with the Gillespie method), with the Poisson approximation (tau-leap method) and with the normal approximation. For the tau-leap method we choose a constant time step. For ``Poisson 1'' and ``Normal 1'' we use a step of $0.05$, for ``Poisson 2'' and ``Normal 2'' we use a step of $0.5$. For each test, we compute the approximate density of $S_{t}$ and $B_{t}$ from the sample with a kernel method. Hence we compare 5 probability density functions for each component $S_{t}$ and $B_{t}$. We also compute the empirical mean and standard deviation associated with the sample obtained from the pure jump process and we plot the associated normal density.

Initial conditions are $B_{0}=0.026$ and $S_{0}=0.26$, corresponding to the case of Figure \ref{fig.poisson+normal}, which is quite far from the equilibrium state.

\begin{figure}
\begin{center}
\includegraphics[width=15cm]{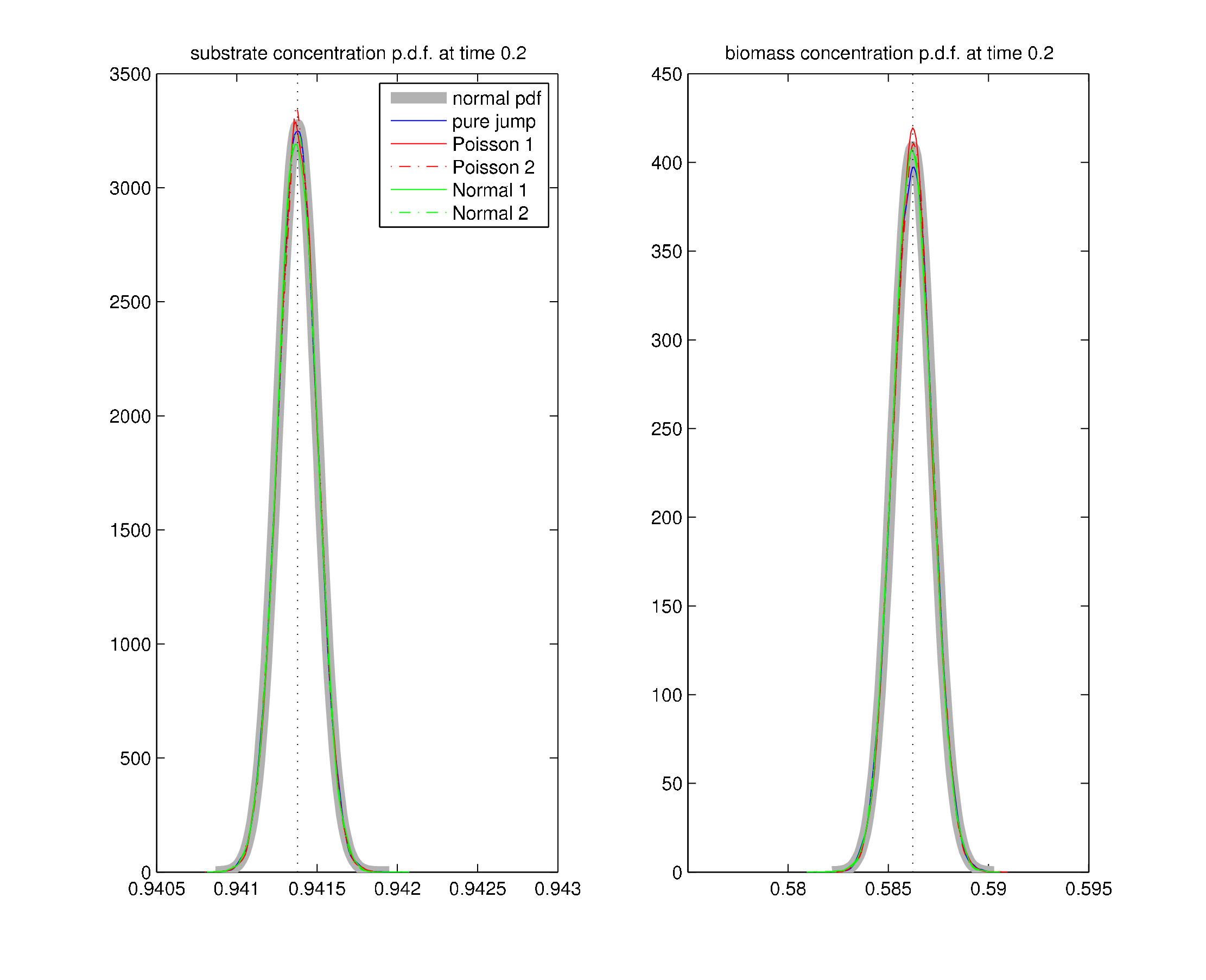}
\end{center}
\caption{\it Empirical densities for the substrate and the biomass concentrations $S_{t}$ and $B_{t}$ at time $t=0.2$ obtained with the exact simulation of the pure jump process $(S_{t},B_{t})$ with the Gillespie method (blue line), with the Poisson approximation $(\tilde S_{t},\tilde B_{t})$ with constant time step (red solid line for a step 0.005, red dash line for a step 0.01), with the normal approximation $(\tilde \sigma_{t},\tilde \beta_{t})$ with constant time step (green solid line for a step 0.005, green dash line for a step 0.01). The corresponding CPU time are respectively: 3 h 54 min 16.9 s, 13.5 s, 6.7 s, 0.1 s, 0.1 s. The simulation parameters are the Set 2 of Table \ref{table.simu.parameter.haldane} (Haldane growth function) with  ``standard'' scales $K_{1,4}=10^{5}$ and $K_{2,3,5}=10^{7}$. 
We compute the substrate and the biomass concentrations  at  $t=0.2$ (h) and for $j=1\cdots 20000$ independent Monte Carlo trials. The empirical densities are obtained with a kernel approximation procedure. We also compute the empirical mean and standard deviation from the sample of the pure jump process and plot with a thick grey line the corresponding normal densities: the match is very good. The vertical doted line the value of the substrate and biomass concentration at time $t=0.2$ given by the ODE; again it matches with the mean of all the empirical densities. The initial condition $(S_{0},B_{0})$ is the stable equilibrium solution of the ODE \eqref{eq.chemostat} that does not correspond to the washout: hence these empirical densities could be considered as good approximations of the limit distribution of the stochastic chemostat.}
\label{fig.huge.mc.haldane}
\end{figure}

The conclusions are:
\begin{itemize}
\item
The two approximations (Poisson and normal) are very close to the exact simulation of the pure jump process; the approximation with a larger step $0.5$ is slightly different.
\item
The computation times\footnote{CPU time on a 2.13GHz Intel Core 2 Duo with a RAM of 2 GB.} are:
\begin{itemize}
\item for the exact simulation of the pure jump process: 5 h 45 min 32.6 s;
\item for the Poisson approximation: 33.2 s (with the time step 0.05) and 4.6 s (with the time step 0.5);
\item for the normal approximation: 0.7 s (with the time step 0.05) and 0.1 s (with the time step 0.5).
\end{itemize}
In the present situation, where the parameters $K_{i}$ are rather high, and for non-small concentration of the biomass and the substrate, \emph{the exact simulation of the pure jump process (Gillespie method) should be avoided}.
\item
The resulting empirical densities are very close to normal densities and the solution of the ODE coincide with the mean of these normal densities.
\end{itemize}
A second test is proposed in the case of the Haldane growth function (Set 2 of Table \ref{table.simu.parameter.haldane}): see Figure \ref{fig.huge.mc.haldane}. The conclusions are the same as for the Monod case.

\begin{table}
\begin{center}\small
\begin{tabular}{llll}
\hline
\multicolumn{4}{r}{Haldane model}\\
\hline
& set 1 & set 2 \\
\hline
$k$       & $0.1$   & $0.1$   & stoichiometric constant \\
$\mumax$  & $2$     & $8$     & maximal growth rate ($h^{-1}$)\\
$D$       & $0.4$   & $0.4$   & dilution rate ($h^{-1}$)\\
$\ssin$    & $1$     & $1$     & input concentration ($g/l$)\\
$\ks$     & $4$     & $17$    & half saturation constant ($g/l$)\\
$k_{i}$   & $1$     & $1$     & saturation parameter )\\
\\ \hline
\end{tabular}
\\[0.6cm]
\begin{tabular}{cc}
\includegraphics[width=3cm]{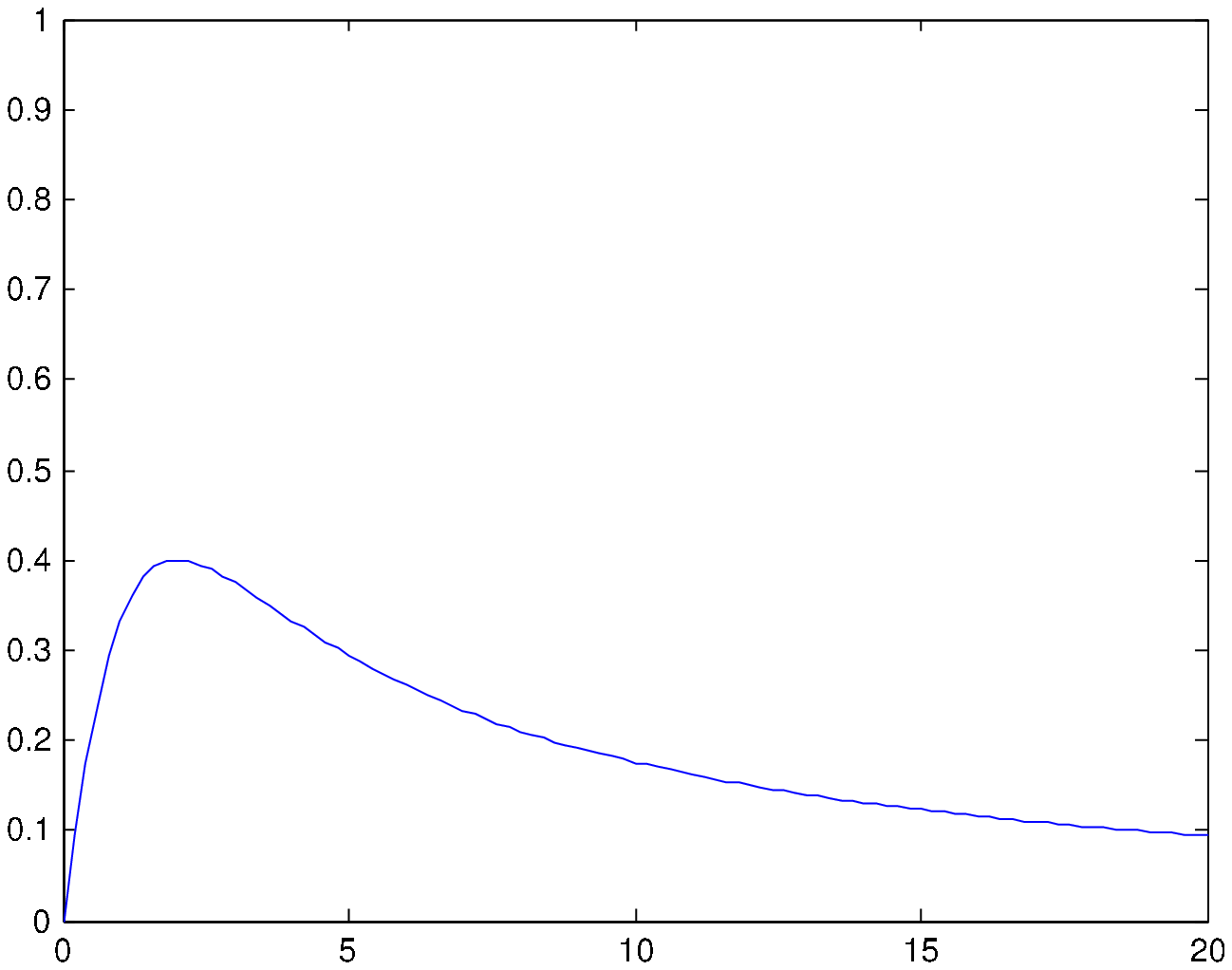}
&
\includegraphics[width=3cm]{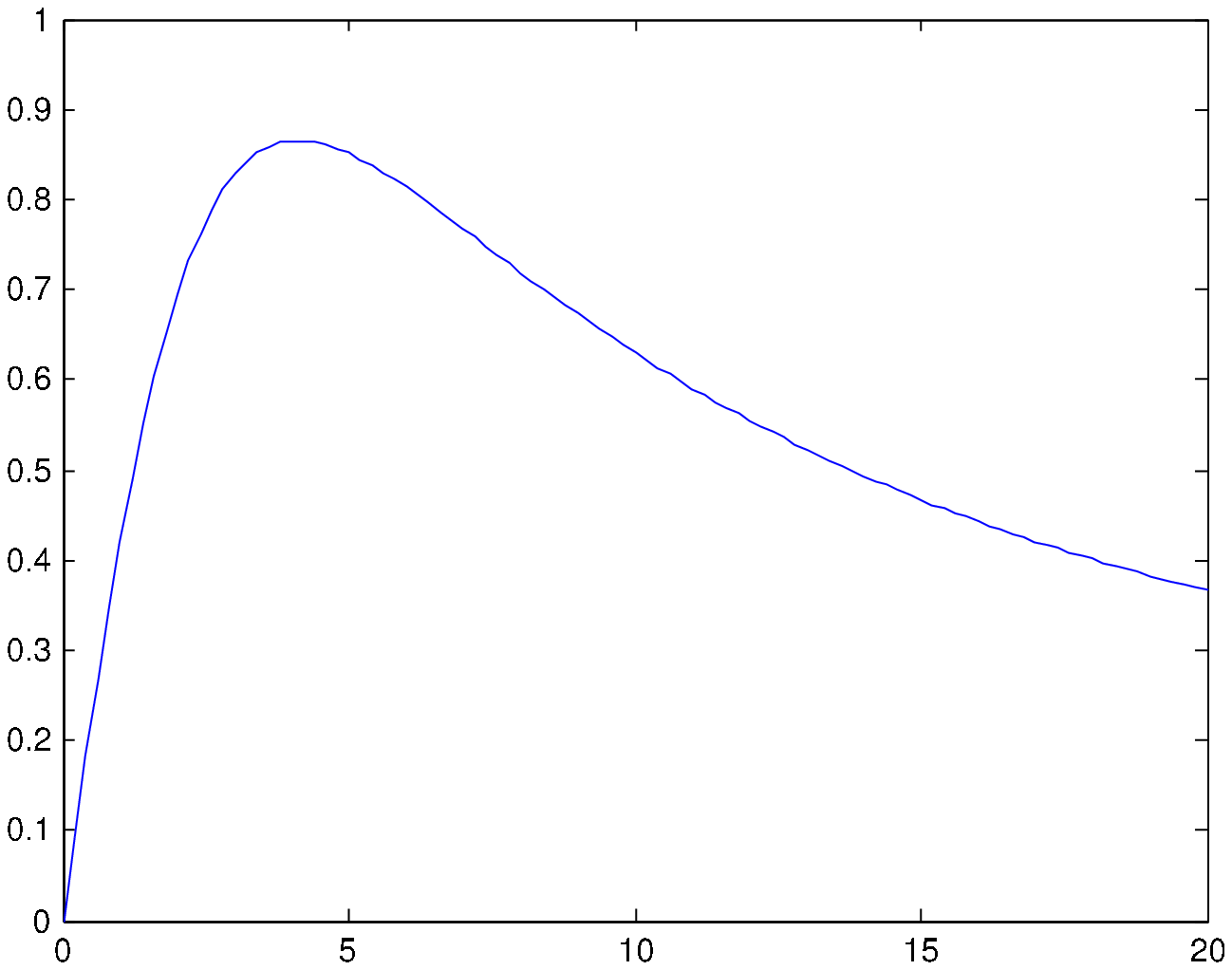}
\\
set 1 & set 2 \end{tabular}
\end{center}
\caption{\it Sets of simulation parameters and the corresponding Haldane growth rate functions $s\to\mu(s)$. The horizontal doted line is the maximum capacity $\mumax$ and the vertical doted line is the asymptotic substrate concentration of the ODE corresponding to the non-washout case.} 
\label{table.simu.parameter.haldane} 
\end{table}

\subsection{About the scales parameters}
\label{sec.simulation.3}

\begin{figure}
\begin{center}
\includegraphics[width=8cm]  {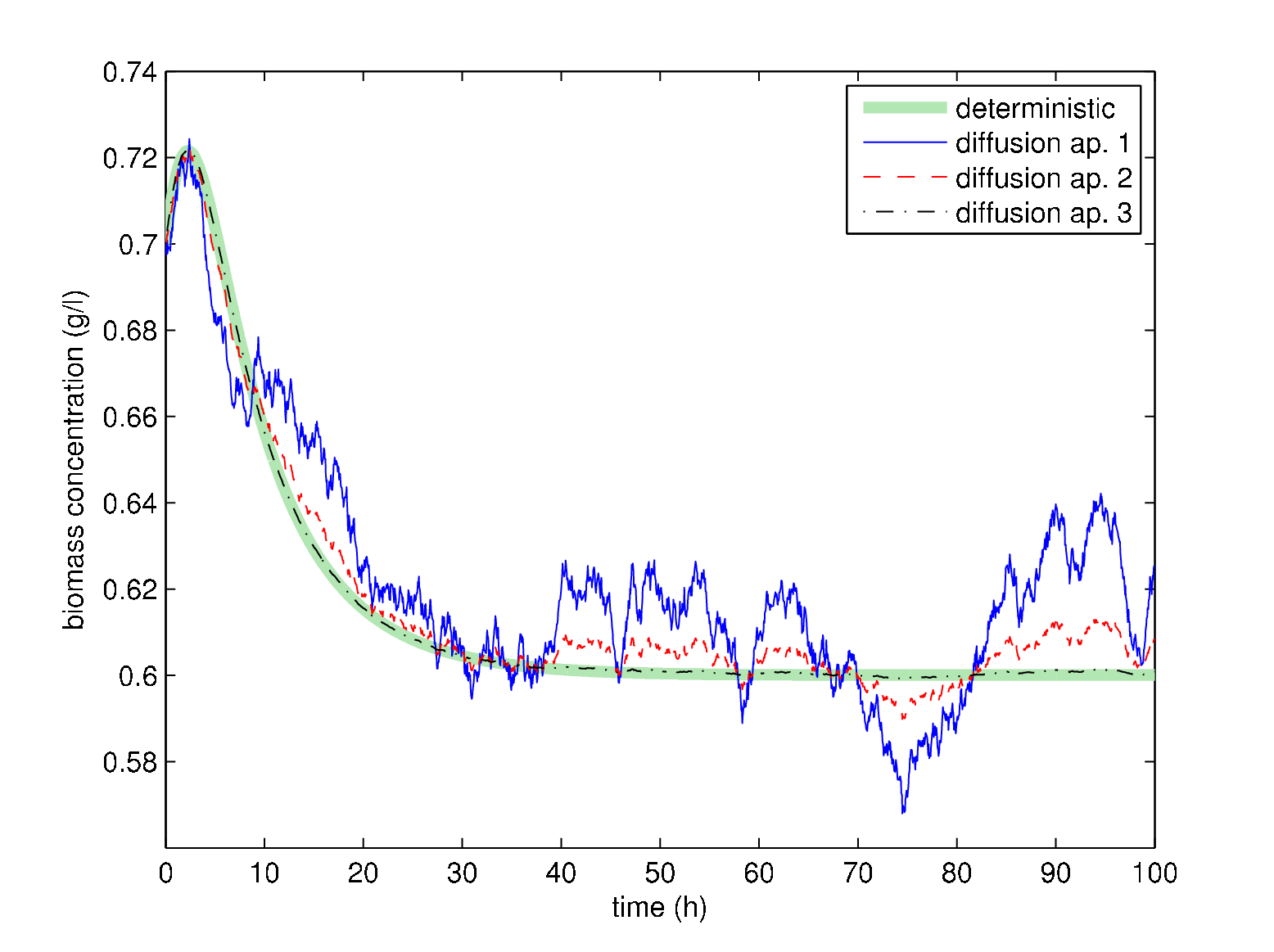}%
\includegraphics[width=8cm]  {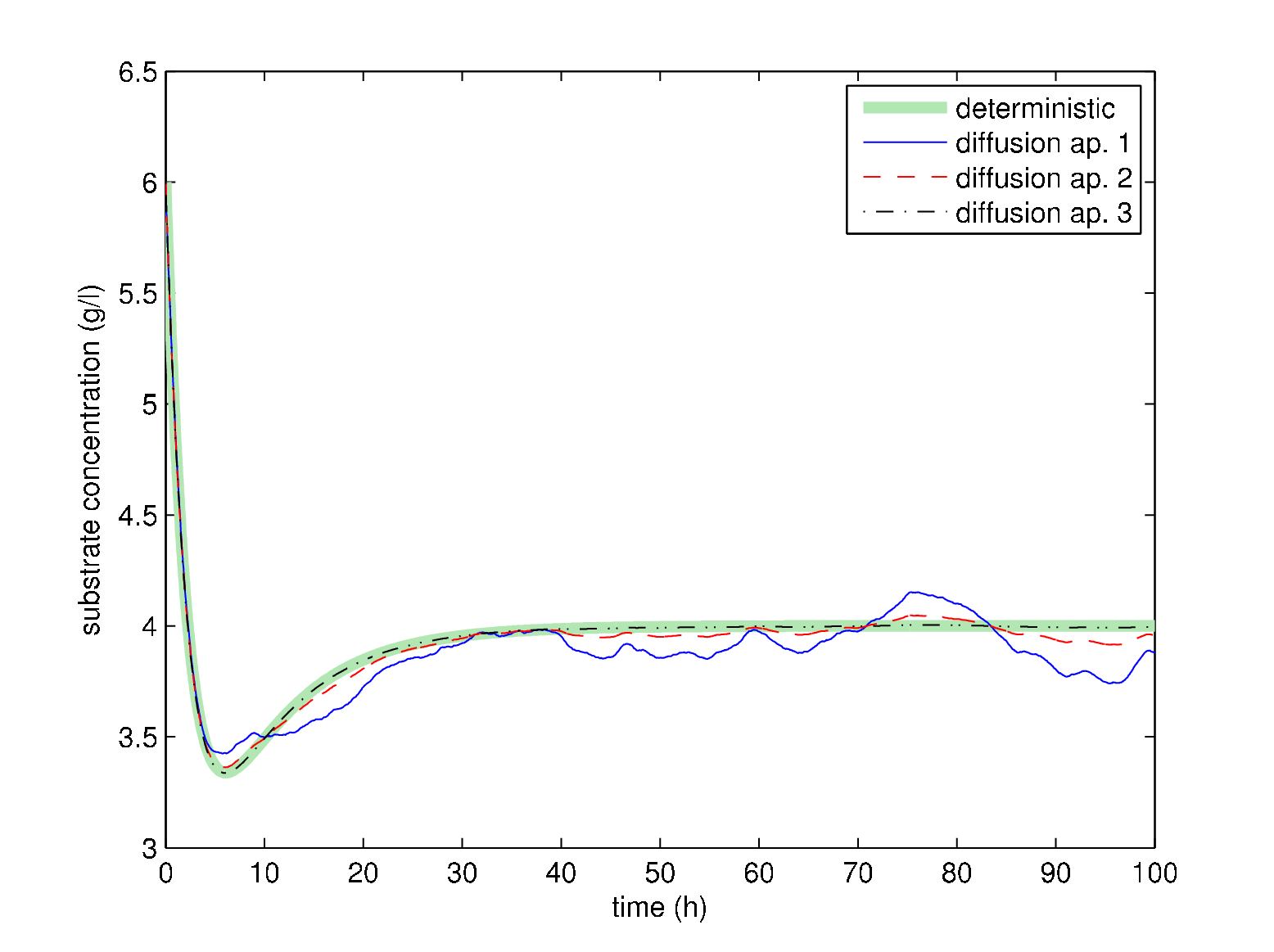}
\includegraphics[width=9.4cm]{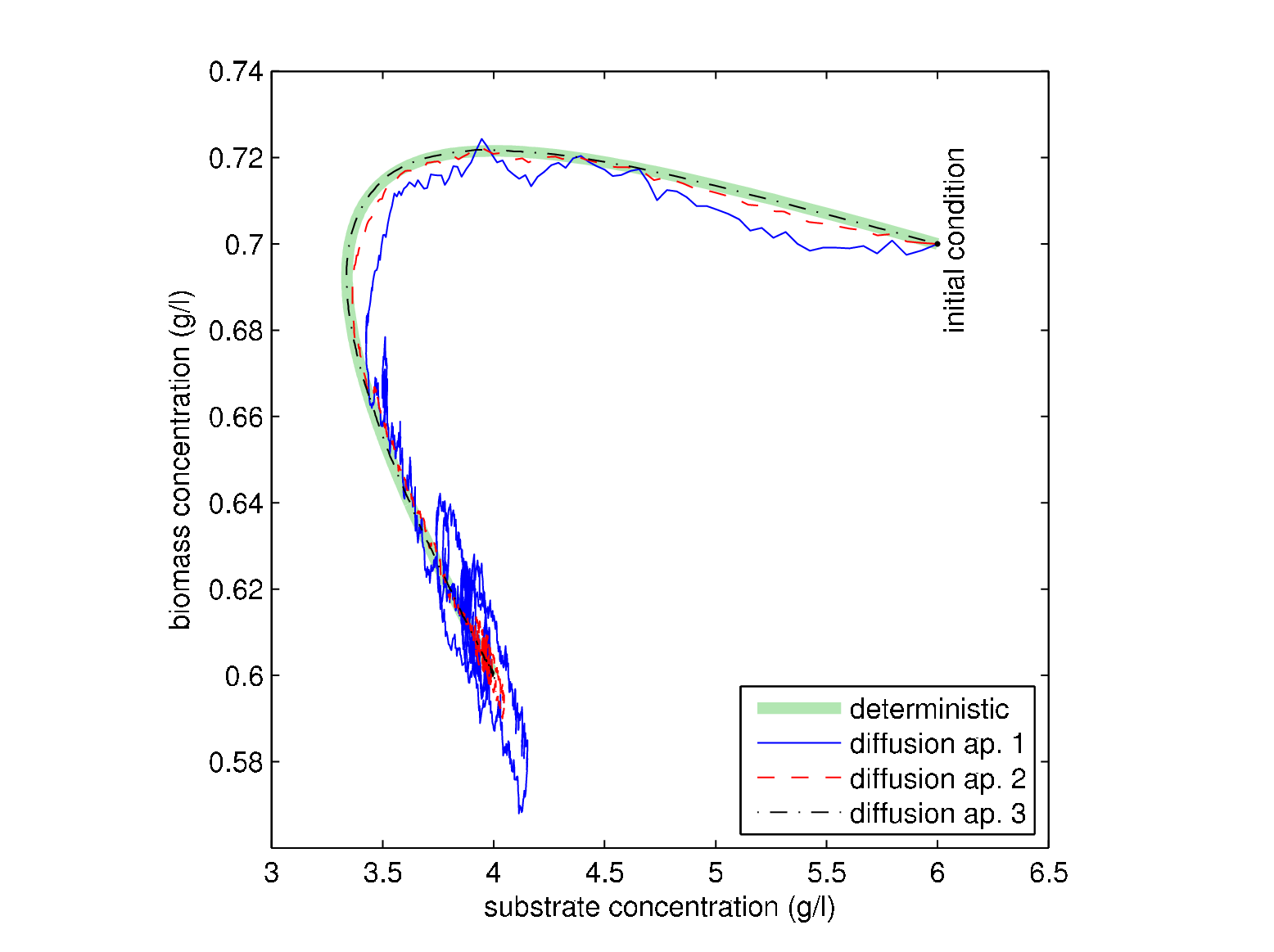}
\end{center}
\caption{\it Diffusion approximation, Case 1, Table \ref{table.simulation.cases} (``standard case'') / Simulation of \eqref{eq.sde.discretization} with Monod specific growth rate \eqref{eq.specific.growth.rate}
with the parameters of Table \ref{table.simu.parameter.monod}
---
Time evolution of the biomass concentration (top left),
time evolution of the substrate concentration (top right),
phase portrait biomass/substrate concentrations (bottom) according to 4 cases:
case 0, case 1.1, case 1.2, case 1.3
(see Table \ref{table.simulation.cases}).
Cases 0 (deterministic) and 1.3 are identical.}
\label{fig.simu.case1}
\end{figure}

\begin{figure}
\begin{center}
\includegraphics[width=8cm]  {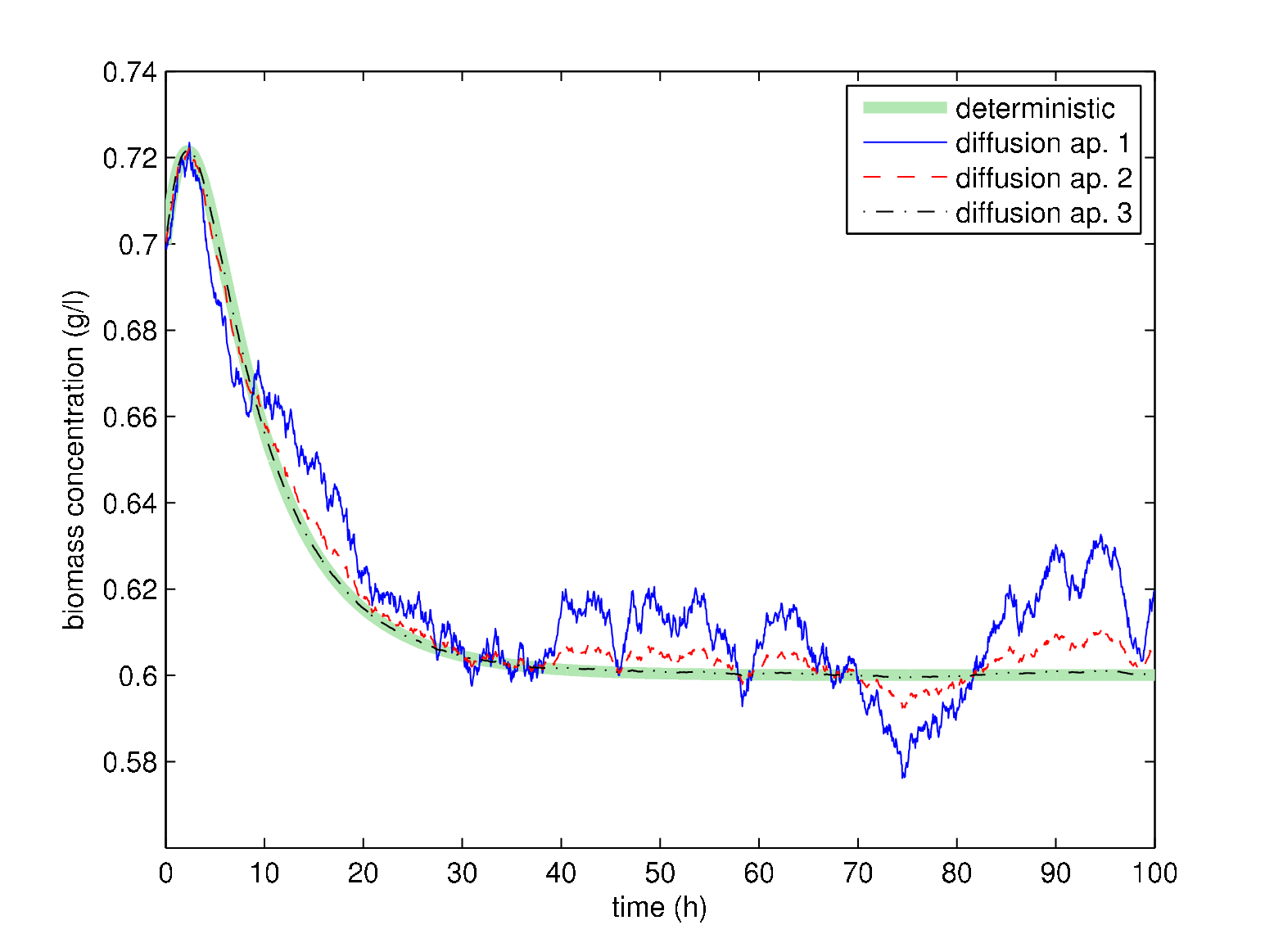}%
\includegraphics[width=8cm]  {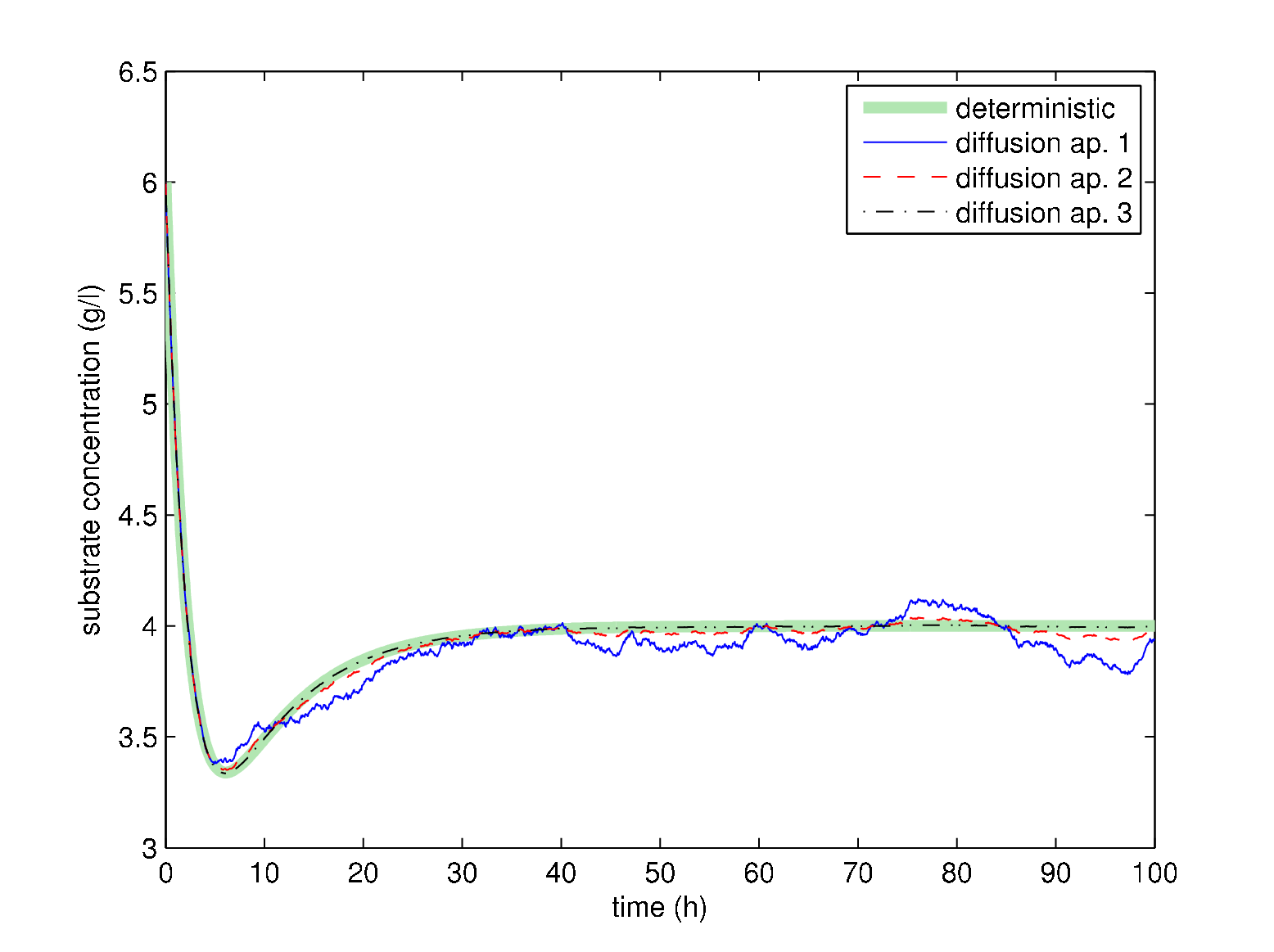}
\includegraphics[width=9.4cm]{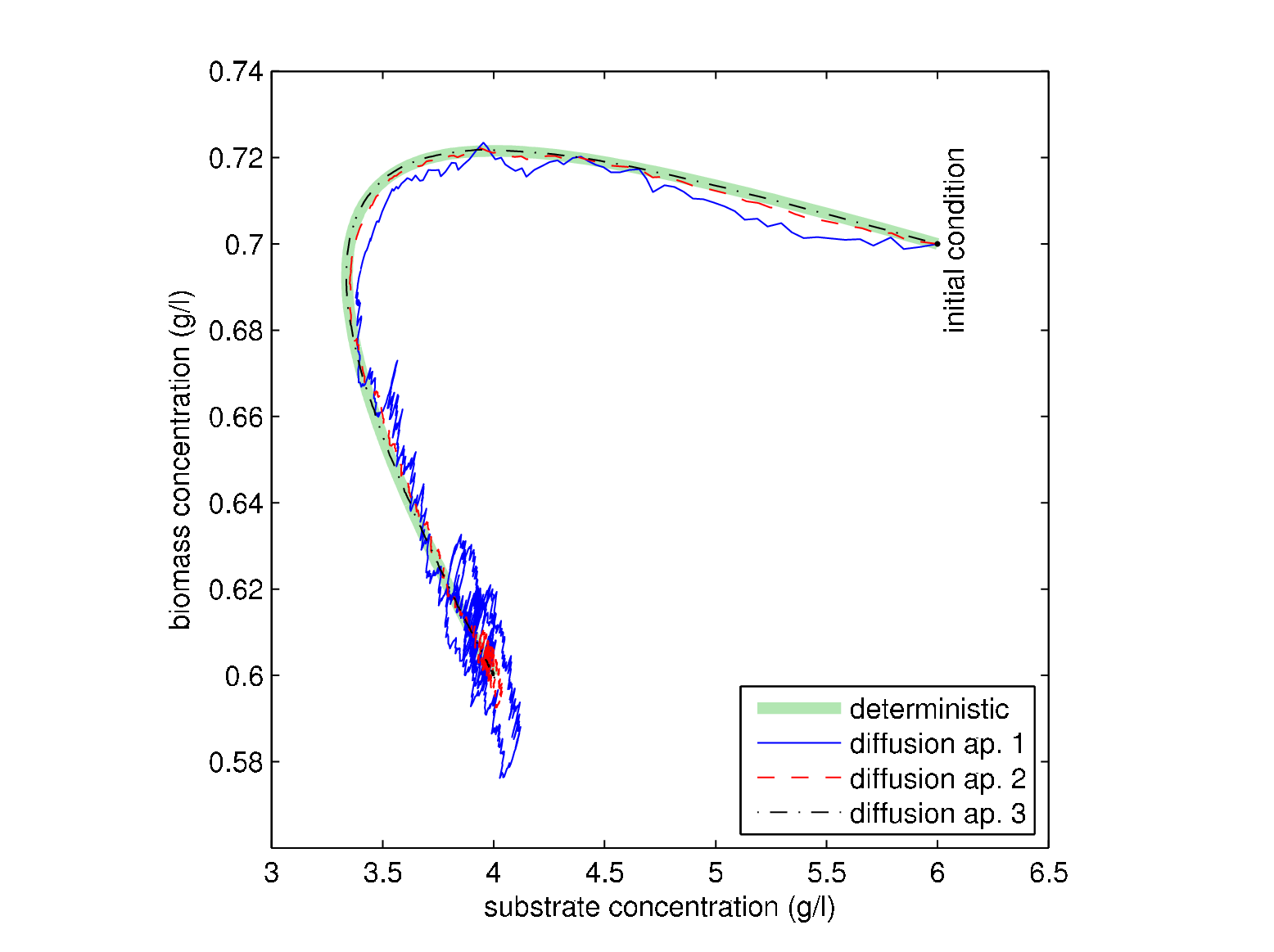}
\end{center}
\caption{\it Diffusion approximation, Case 2, Table \ref{table.simulation.cases} (``unstirred inflow and outflows'') / Simulation of \eqref{eq.sde.discretization} with Monod specific growth rate \eqref{eq.specific.growth.rate}
with the parameters of Table \ref{table.simu.parameter.monod}
---
Time evolution of the biomass concentration (top left),
time evolution of the substrate concentration (top right),
phase portrait biomass/substrate concentrations (bottom) according to 4 cases:
case 0, case 2.1, case 2.2, case 2.3
(see Table \ref{table.simulation.cases}).
Cases 0 (deterministic) and 2.3 are identical.}
\label{fig.simu.case2}
\end{figure}

\begin{figure}
\begin{center}
\includegraphics[width=8cm]  {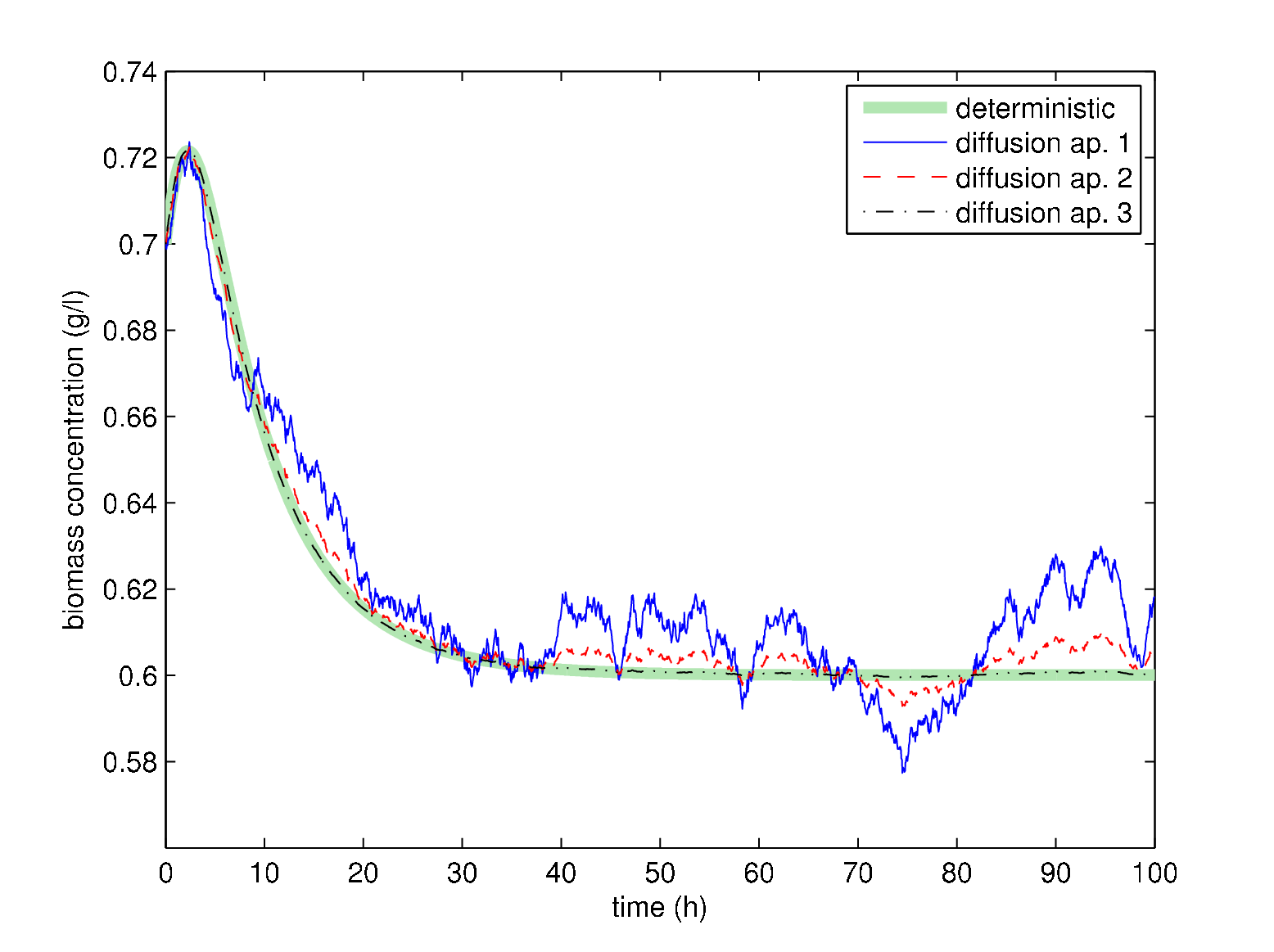}%
\includegraphics[width=8cm]  {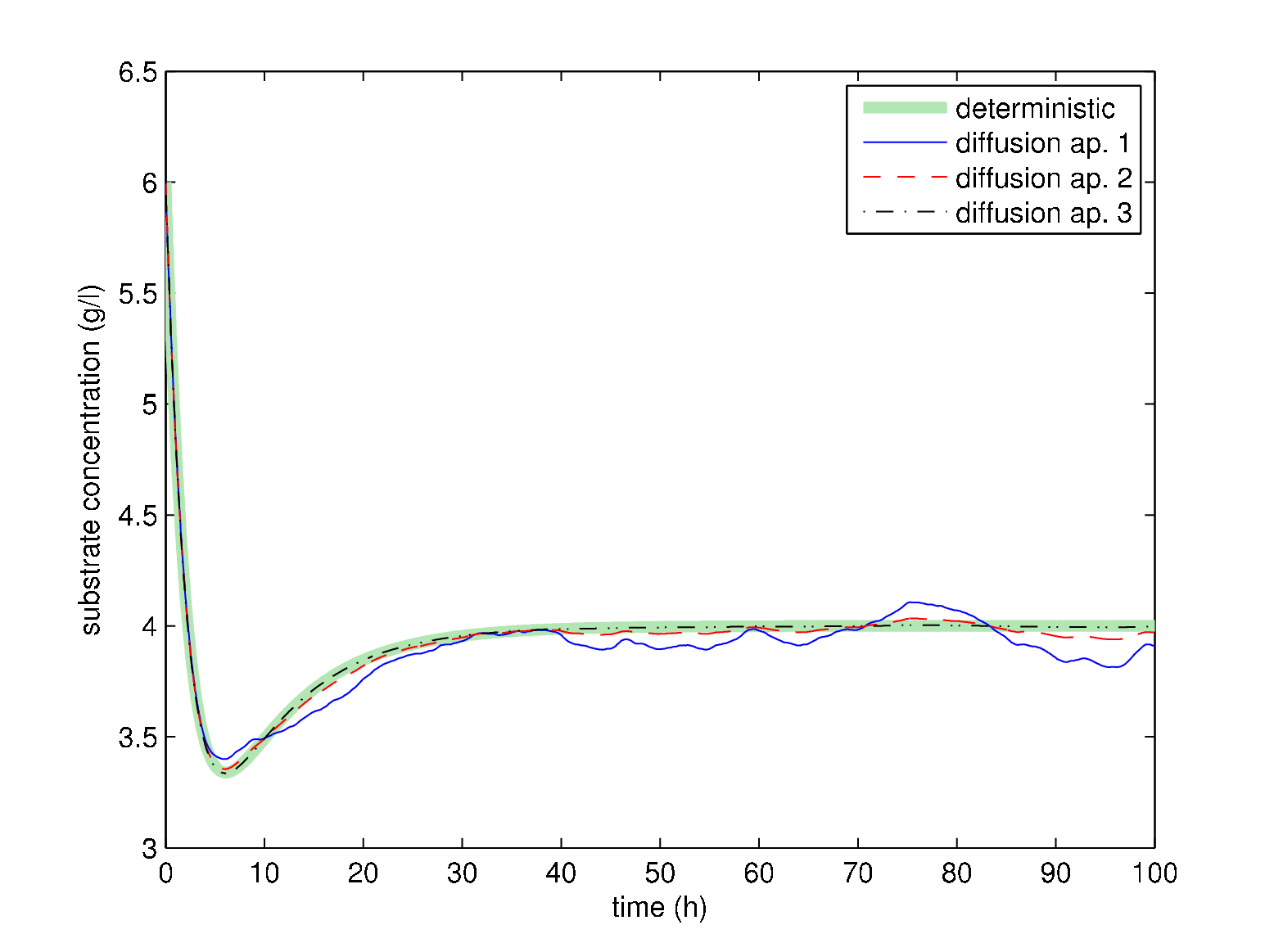}
\includegraphics[width=9.4cm]{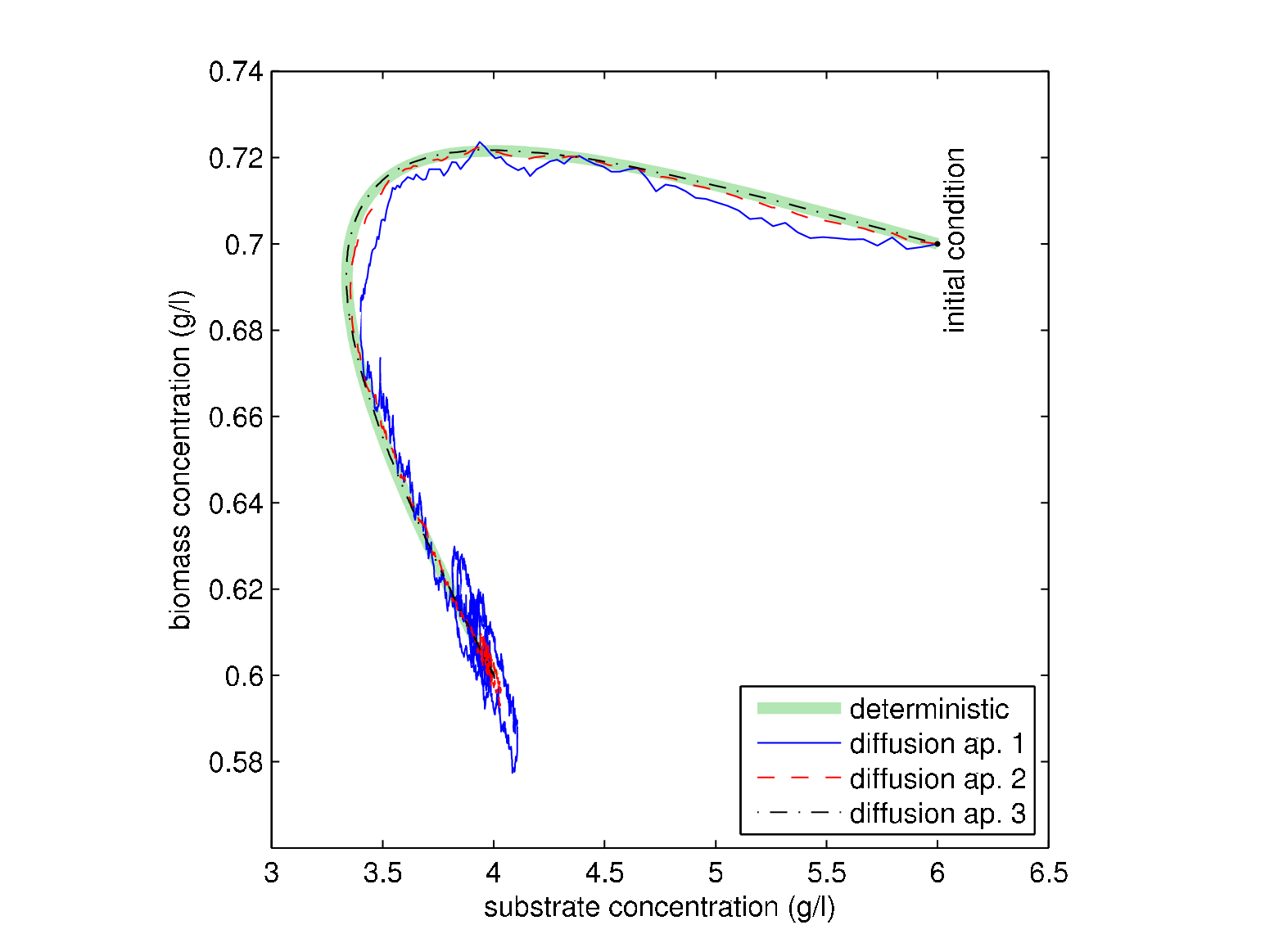}
\end{center}
\caption{\it Case 3, Table \ref{table.simulation.cases}  (``substrate fluid limit case'') / Simulation of \eqref{eq.sde.discretization} with Monod specific growth rate \eqref{eq.specific.growth.rate}
with the parameters of Table \ref{table.simu.parameter.monod}
---
Time evolution of the biomass concentration (top left),
time evolution of the substrate concentration (top right),
phase portrait biomass/substrate concentrations (bottom) according to 4 cases:
case 0, case 3.1, case 3.2, case 3.3
(see Table \ref{table.simulation.cases}).
Cases 0 (deterministic) and 3.3 are identical.}
\label{fig.simu.case3}
\end{figure}

\begin{figure}
\begin{center}
\includegraphics[width=8cm]  {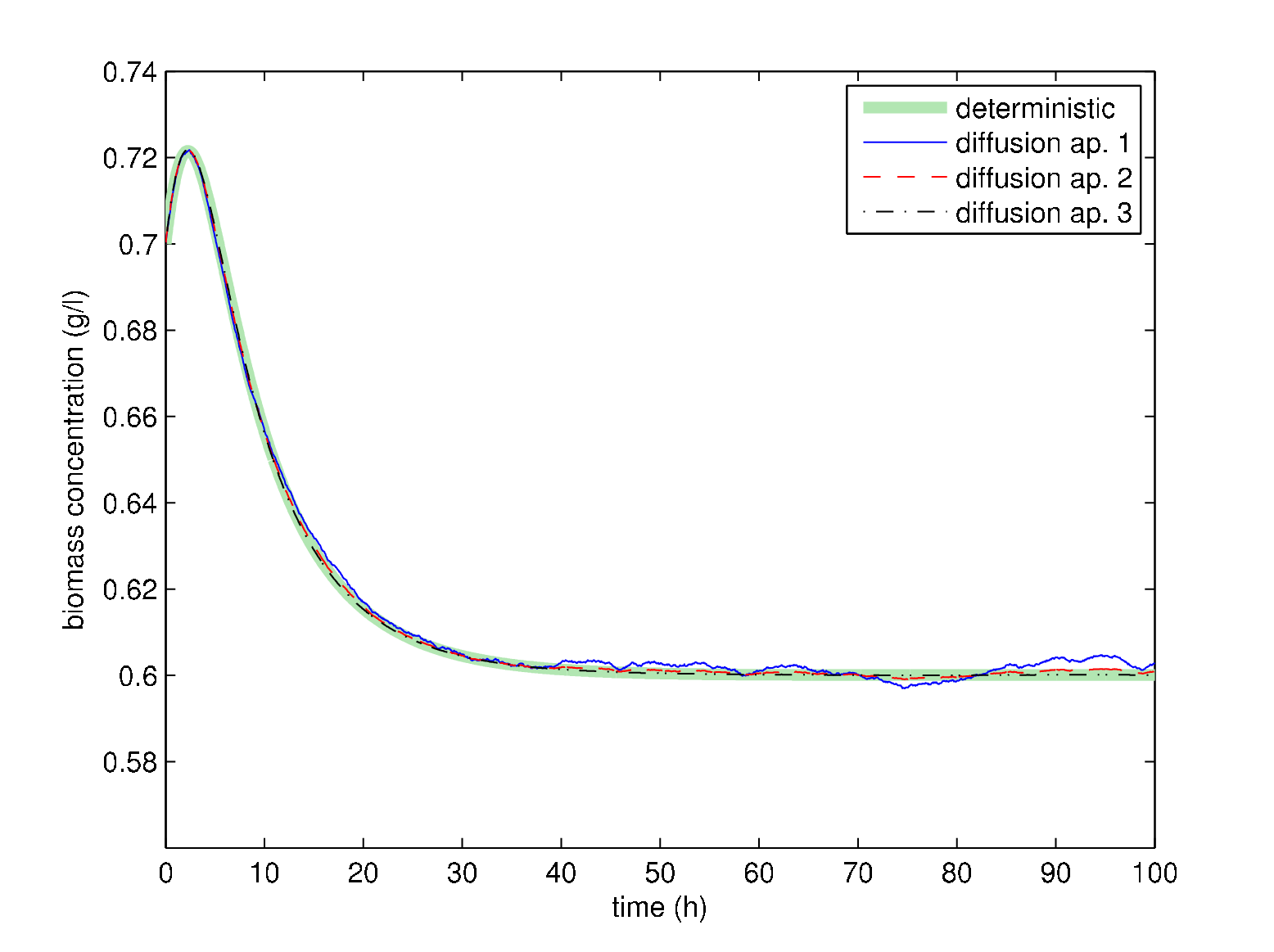}%
\includegraphics[width=8cm]  {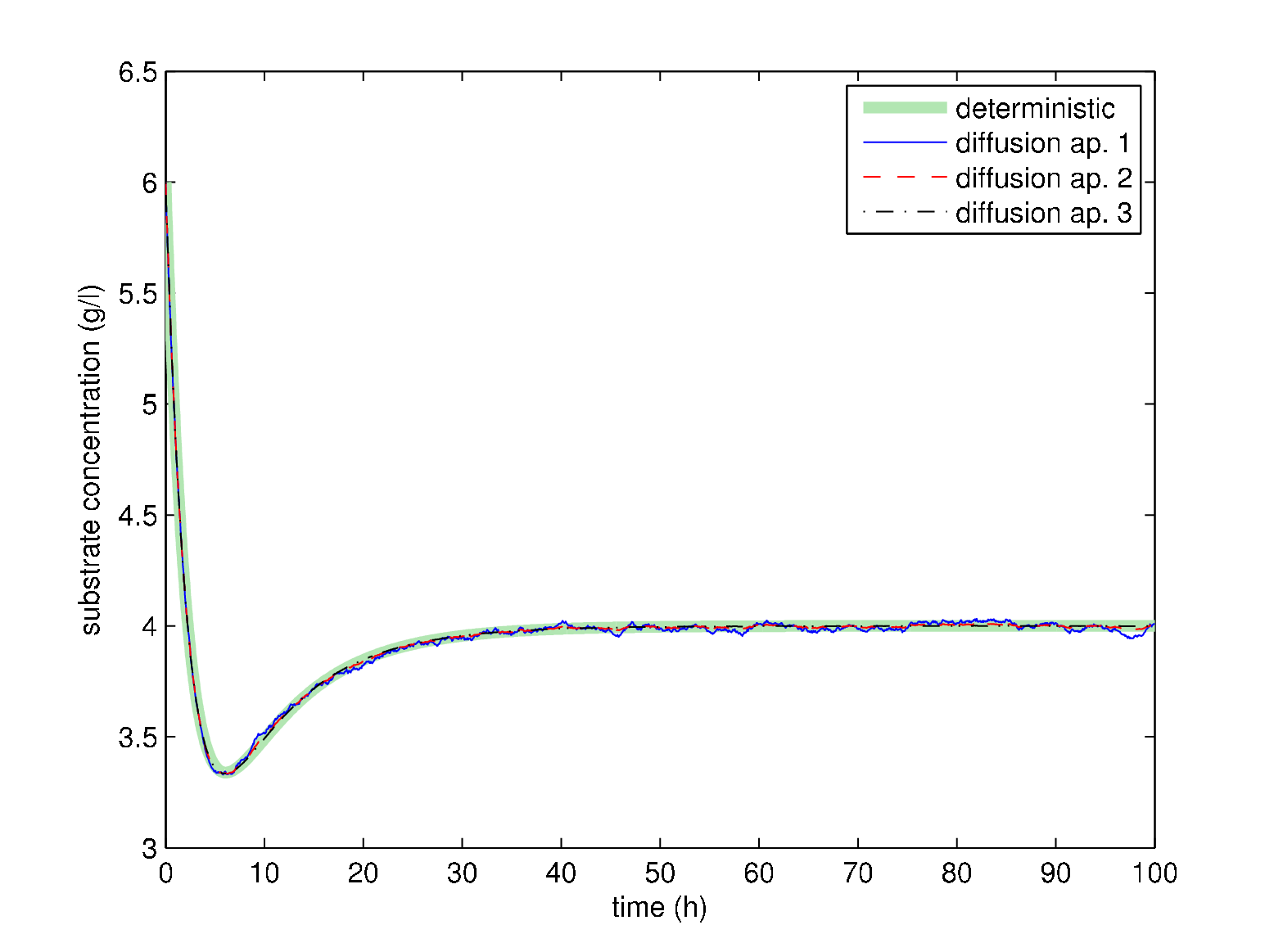}
\includegraphics[width=9.4cm]{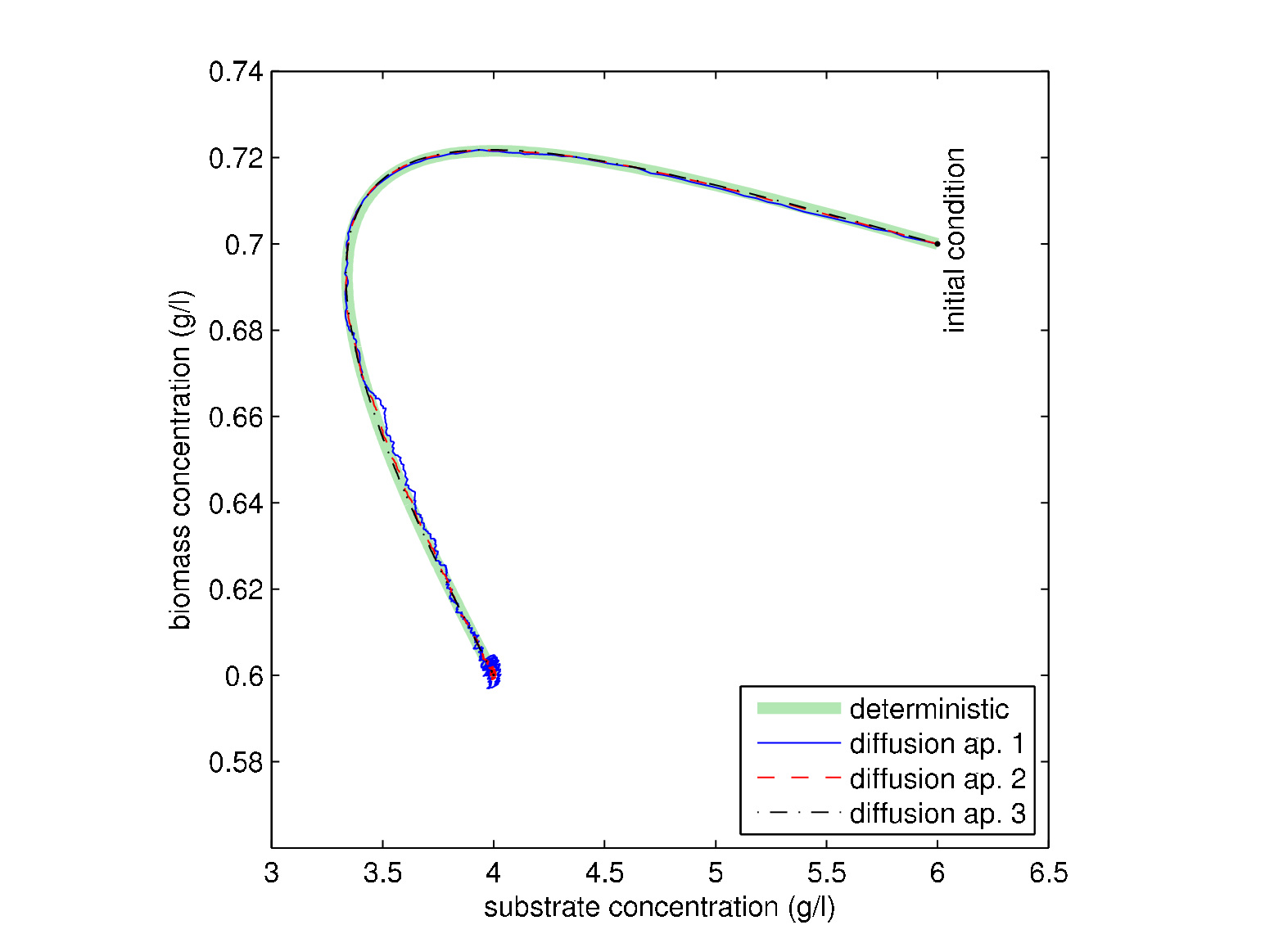}
\end{center}
\caption{\it Case 4, Table \ref{table.simulation.cases} (``biological case'') / Simulation of \eqref{eq.sde.discretization} with Monod specific growth rate \eqref{eq.specific.growth.rate}
with the parameters of Table \ref{table.simu.parameter.monod}
---
Time evolution of the biomass concentration (top left),
time evolution of the substrate concentration (top right),
phase portrait biomass/substrate concentrations (bottom) according to 4 cases:
case 0, case 4.1, case 4.2, case 4.3
(see Table \ref{table.simulation.cases}).
Cases 0 (deterministic) and 4.3 are identical.}
\label{fig.simu.case4}
\end{figure}

\begin{figure}
\begin{center}
\includegraphics[width=7cm]{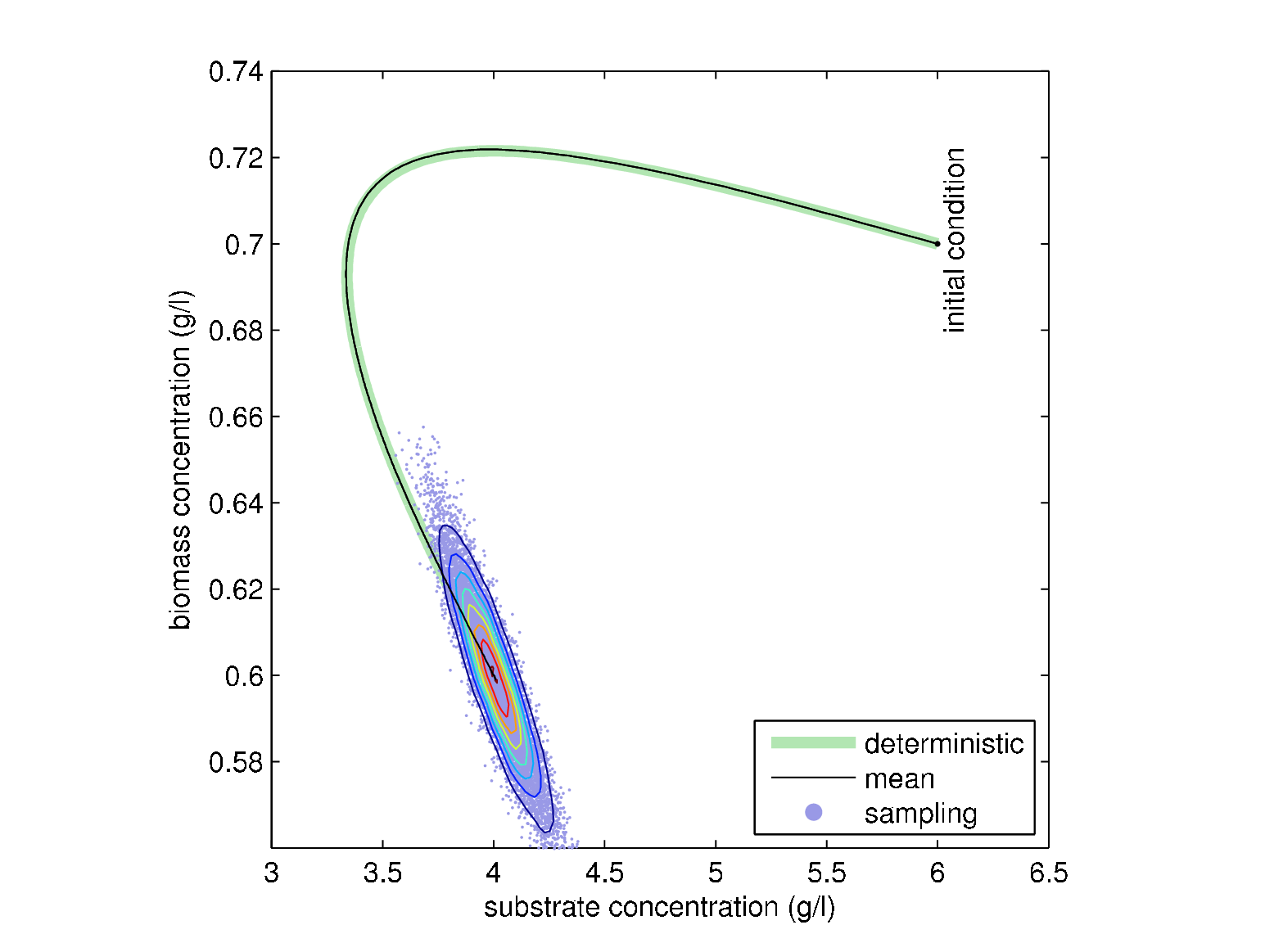}%
\includegraphics[width=7cm]{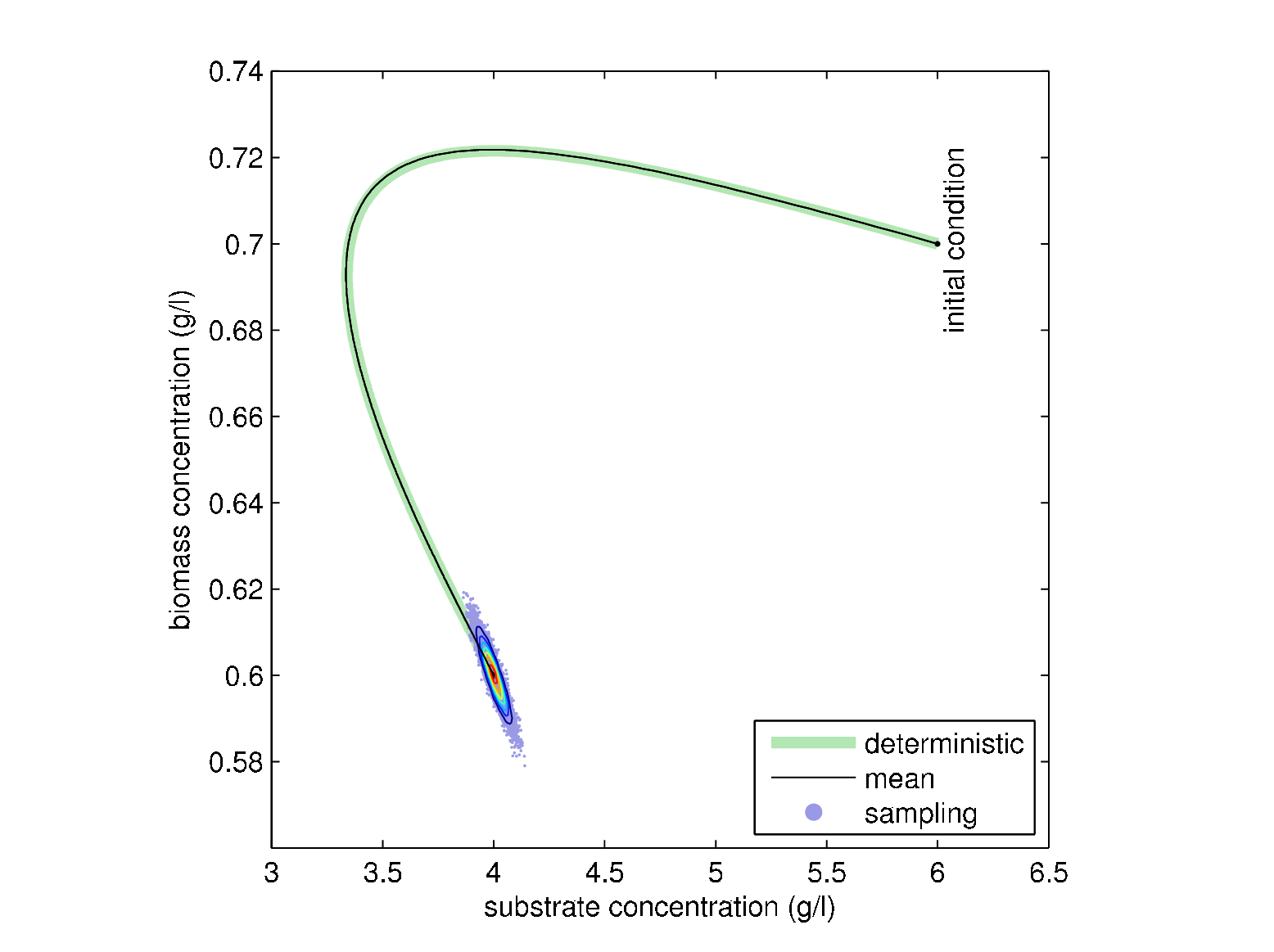}
\end{center}
\caption{\it Cases 1.1 and 1.2 Table \ref{table.simulation.cases} / 
Sampling 10000 Monte Carlo trials  of the law of $(\tilde\beta_{t_{n}},\tilde\sigma_{t_{n}})$ for $t_{n}=100$ --- The deterministic solution and the mean of the sampled trajectories coincide 
--- The final law is represented by the sample and by the contour plot of the corresponding kernel approximation of the p.d.f.}
\label{fig.simu.law.case1}
\end{figure}

\begin{figure}
\begin{center}
\includegraphics[width=7cm]{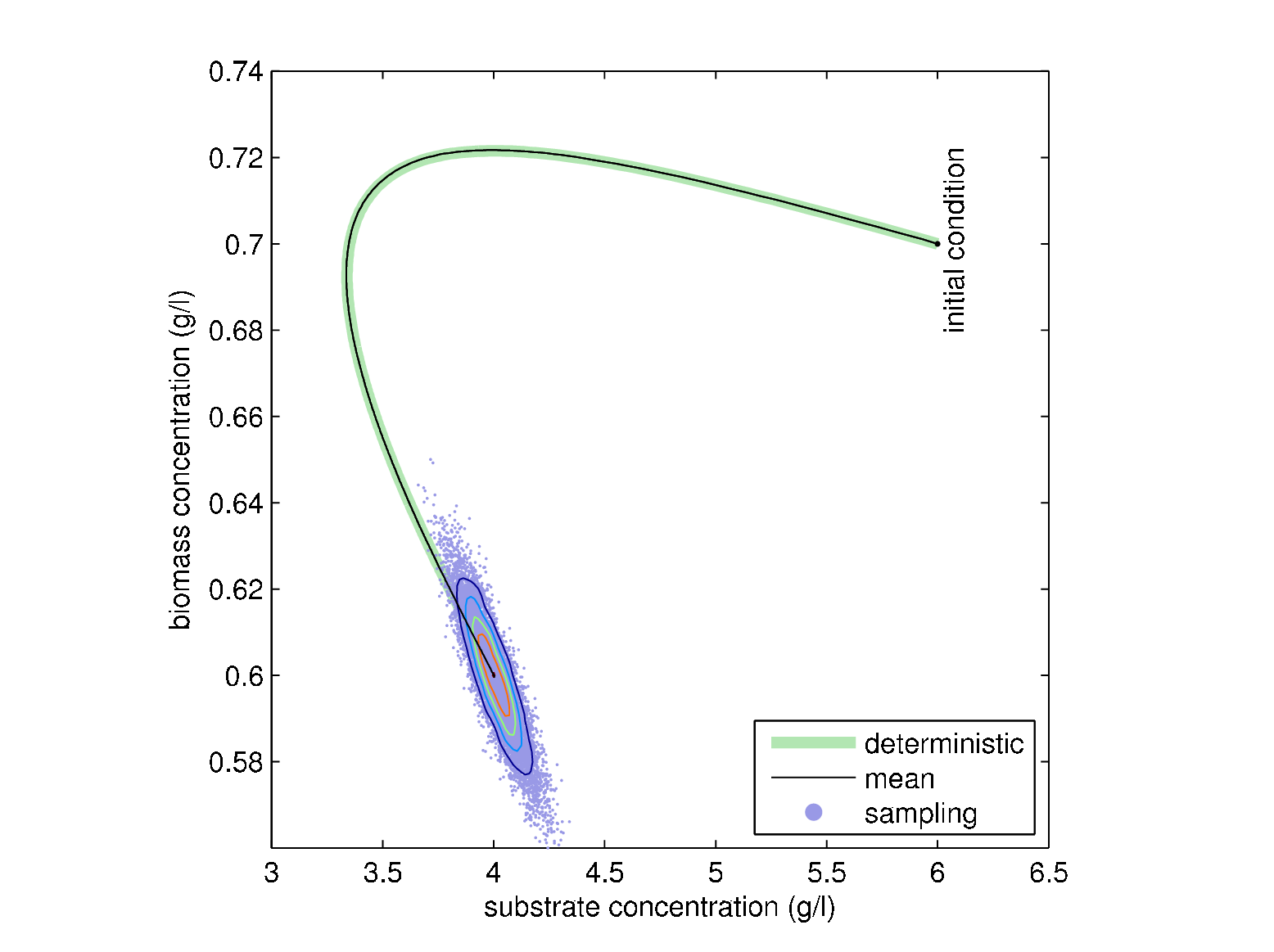}%
\includegraphics[width=7cm]{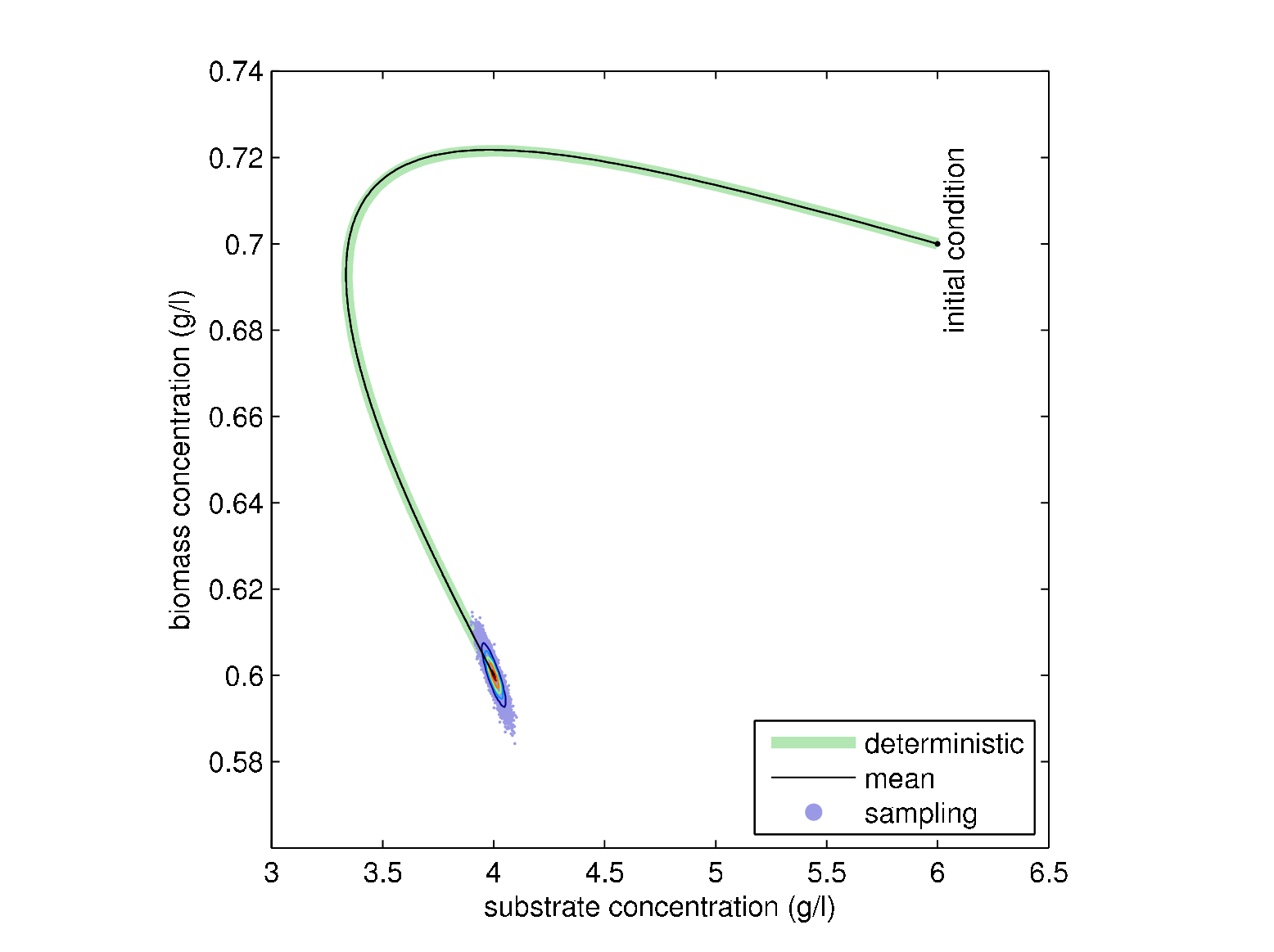}
\end{center}
\caption{\it Cases 2.1 and 2.2 Table \ref{table.simulation.cases} / 
Sampling 10000 Monte Carlo trials  of the law of $(\tilde\beta_{t_{n}},\tilde\sigma_{t_{n}})$ for $t_{n}=100$ --- The deterministic solution and the mean of the sampled trajectories coincide 
--- The final law is represented by the sample and by the contour plot of the corresponding kernel approximation of the p.d.f.}
\label{fig.simu.law.case2}
\end{figure}

\begin{figure}
\begin{center}
\includegraphics[width=7cm]{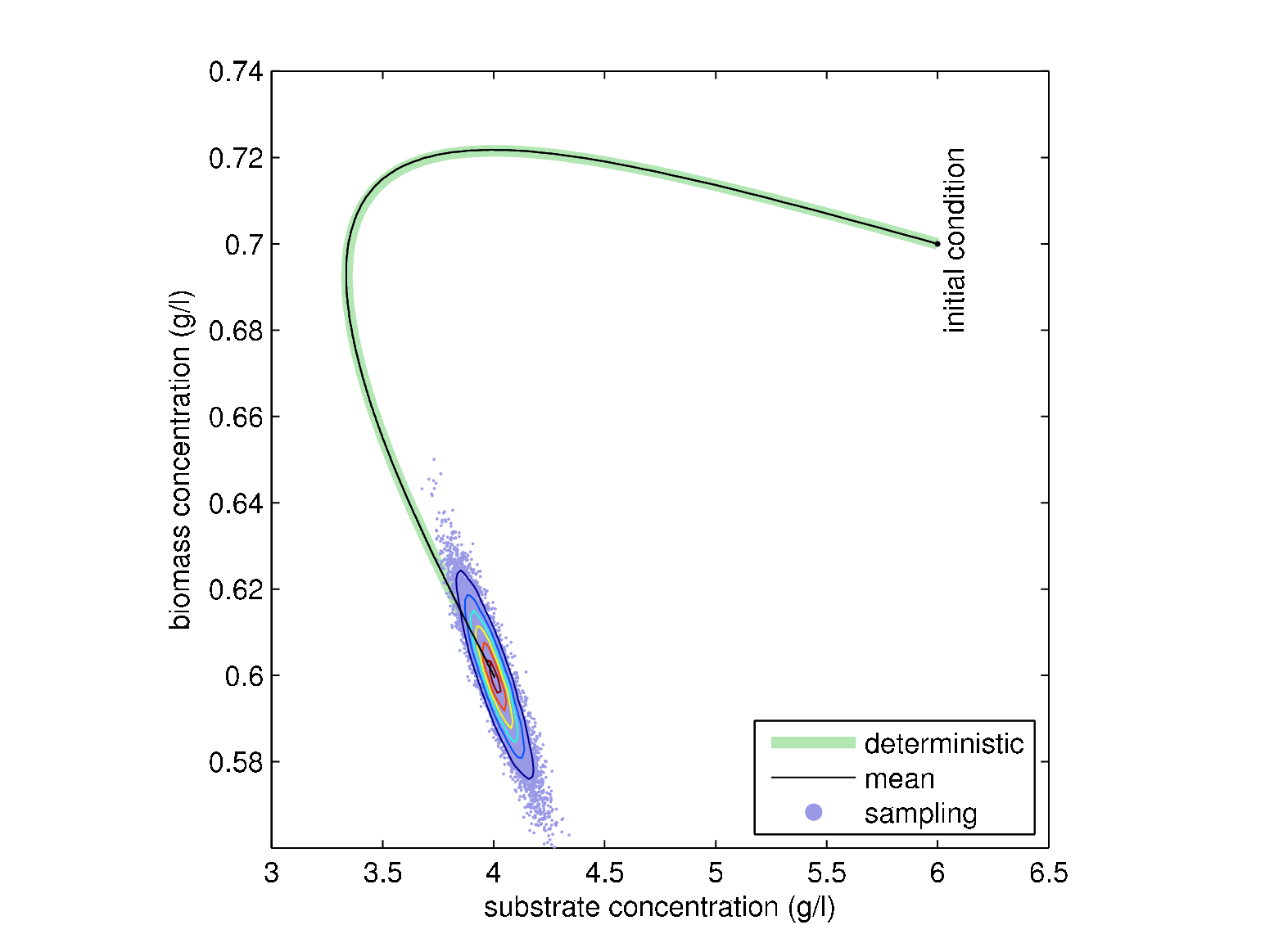}%
\includegraphics[width=7cm]{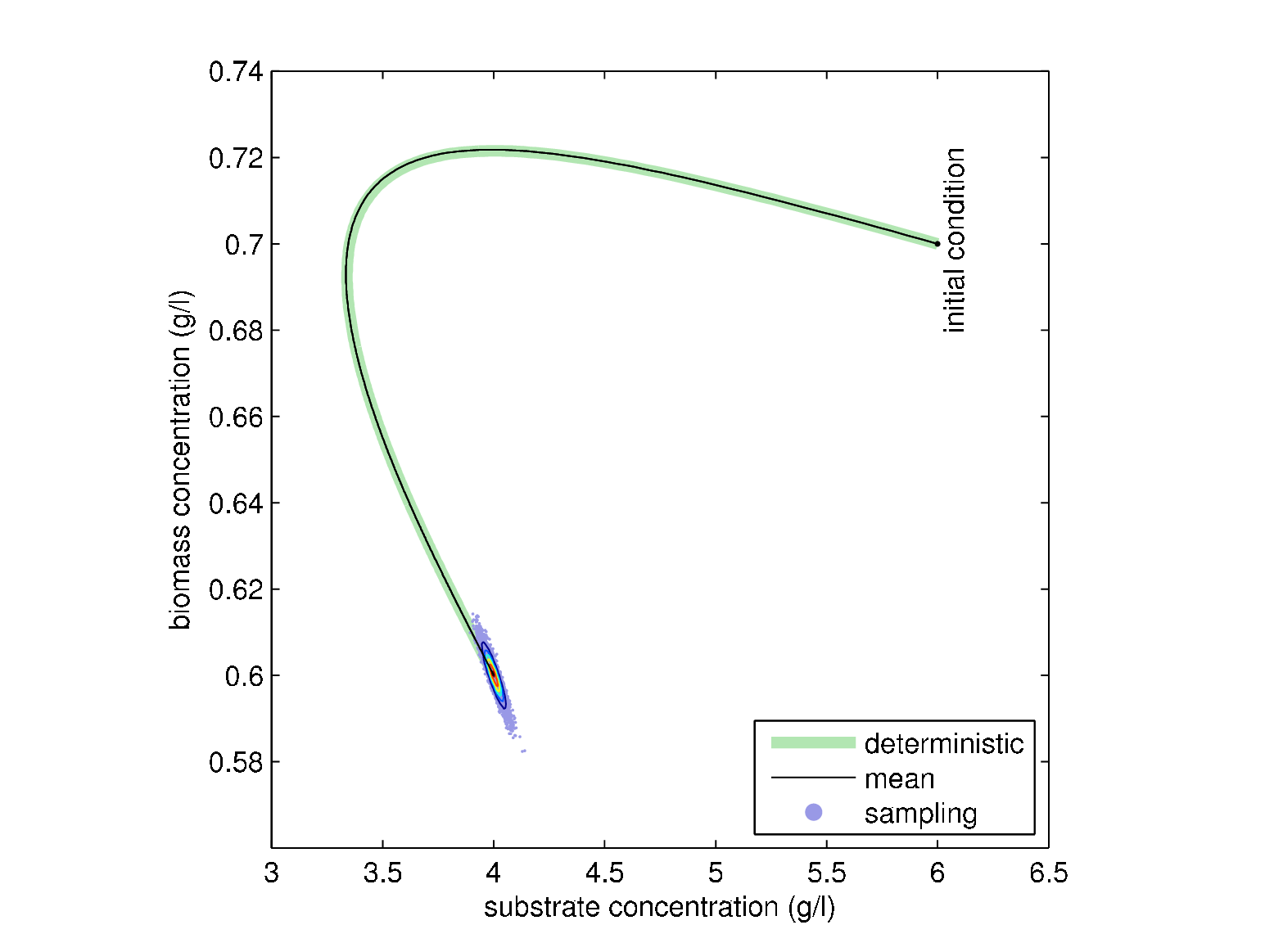}
\end{center}
\caption{\it Cases 3.1 and 3.2 Table \ref{table.simulation.cases} / 
Sampling 10000 Monte Carlo trials  of the law of $(\tilde\beta_{t_{n}},\tilde\sigma_{t_{n}})$ for $t_{n}=100$ --- The deterministic solution and the mean of the sampled trajectories coincide 
--- The final law is represented by the sample and by the contour plot of the corresponding kernel approximation of the p.d.f.}
\label{fig.simu.law.case3}
\end{figure}

\begin{figure}
\begin{center}
\includegraphics[width=7cm]{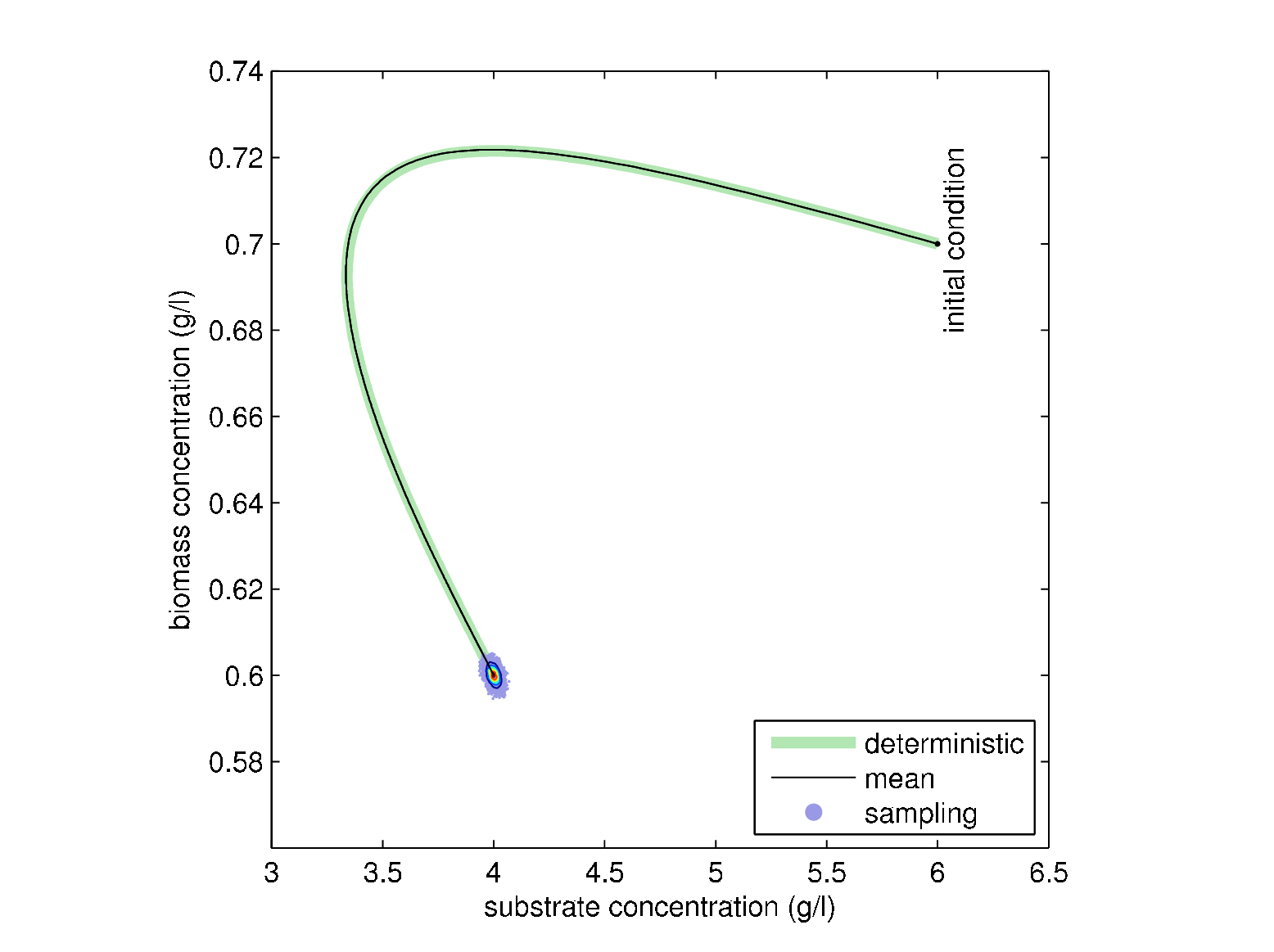}%
\includegraphics[width=7cm]{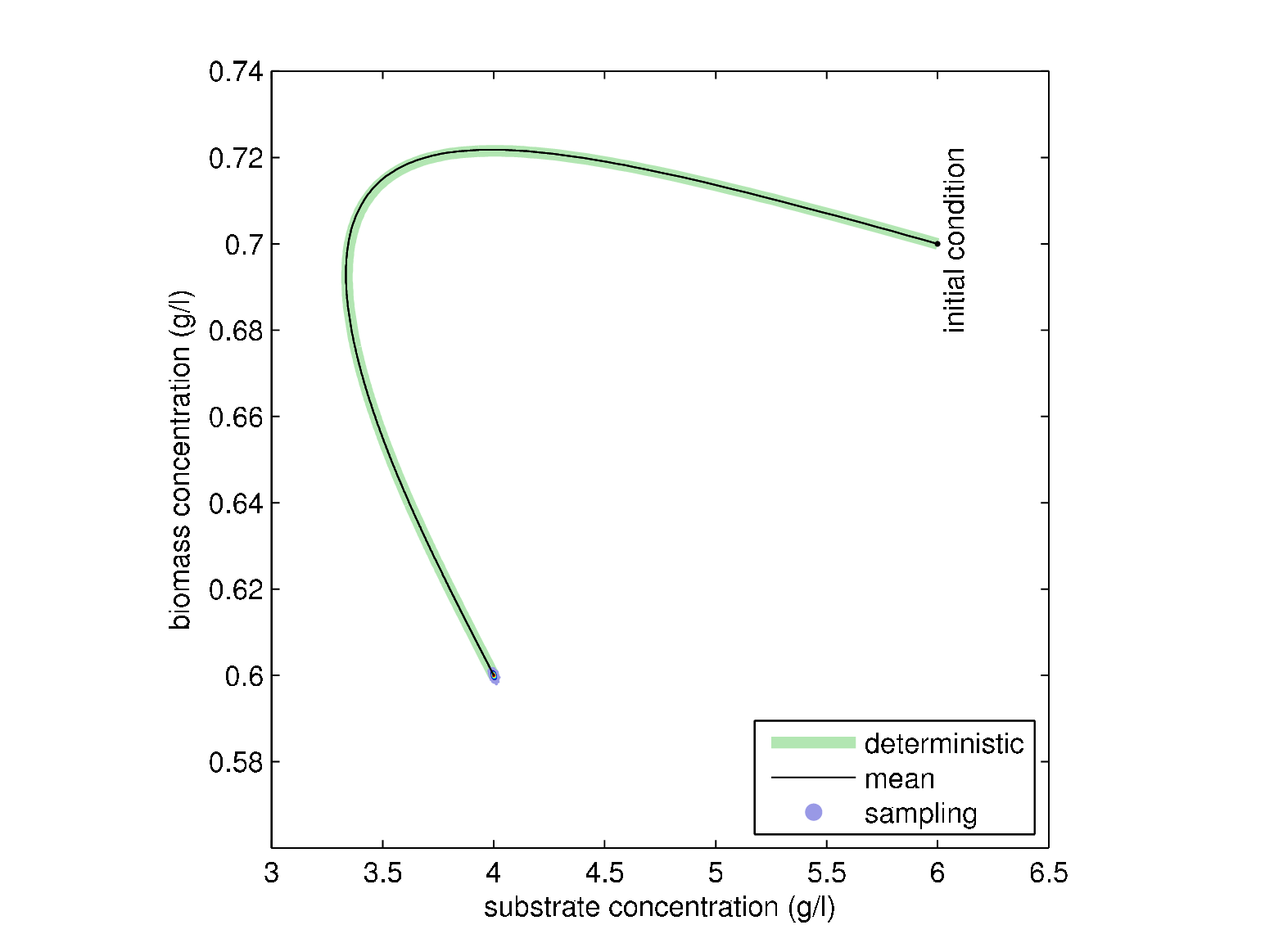}
\end{center}
\caption{\it Cases 4.1 and 4.2 Table \ref{table.simulation.cases} / 
Sampling 10000 Monte Carlo trials  of the law of $(\tilde\beta_{t_{n}},\tilde\sigma_{t_{n}})$ for $t_{n}=100$ --- The deterministic solution and the mean of the sampled trajectories coincide.}
\label{fig.simu.law.case4}
\end{figure}

As we have seen, for large populations, the diffusion approximation $\tilde \xi_{t_{n}}=(\tilde \beta_{t_{n}},\tilde \sigma_{t_{n}})$ given by \eqref{eq.sde.discretization} is very close to the reference pure jump model $X_{t}=(B_{t},S_{t})$. So we now propose simulations of the diffusion approximation in the case of a Monod specific growth rate
according to the scales scenarios of Table \ref{table.simulation.cases}
\begin{itemize}
\item 
Case 1 (``standard''): see Figures \ref{fig.simu.case1} and \ref{fig.simu.law.case1}.
\item 
Case 2 (``unstirred inflowoutflows''): see Figures \ref{fig.simu.case2} and \ref{fig.simu.law.case2}.
\item 
Case 3 (``fluid substrate''): see Figures \ref{fig.simu.case3} and \ref{fig.simu.law.case3}.
\item 
Case 4 (``biological only''): see Figures \ref{fig.simu.case4} and \ref{fig.simu.law.case4}.
\end{itemize}
Figures \ref{fig.simu.case1} to \ref{fig.simu.case4} represent a simulation of a single trajectory in the 3 levels of scale: cases $m,1$ to $m,3$ (for $m=1\cdots4$). Figures \ref{fig.simu.law.case1} to \ref{fig.simu.law.case4} represent the result of 10000 Monte Carlo trials in the 3 levels of scale: cases $m,1$ and $m,2$ (for $m=1\cdots4$). We represent the mean trajectory and the empirical law of $\tilde \xi_{T}$ at final time $T$.

We can conclude that, \emph{at this level of population and scale}:
\begin{itemize}
\item 
The stochasticity is negligible only in the Case 4 (``biological only'') and at the highest scale level (cases $m,3$).
\item 
The ODE solution $x(t)$ matches the (empirical) mean of the stochastic process at these scales (as the stochastic process is solution of a nonlinear equation, there is no reason for the mean of the stochastic process to coincide with the solution of the deterministic equations). 

\end{itemize}
Equivalent results have been obtained for the Haldane case.

\subsection{Comparison with the Imhof-Walcher model \cite{imhof2005a}}
\label{sec.simulation.comp.adhoc}

We compare the processes $\xi_{t}=(\beta_{t},\sigma_{t})^{*}$ given 
by the diffusion approximation model \eqref{eq.diffusion.approximation} with the one given by the ad hoc 
model \eqref{eq.diffusion.approximation.adhoc}. The parameter are: $K=1$, $
\mumax=1$, $D=0.5$, $\Sin=8$, $\ks=0.5$, final time $T=20$, $\Delta t=0.02$, 20000 
Monte Carlo trials; $K_{i}=10^{5}$ for \eqref{eq.diffusion.approximation} and $\cb=
\cs=0.02$ for \eqref{eq.diffusion.approximation.adhoc}. The parameters are chosen 
so that the biomass concentration evolves from 0.5 to about 7.5, and the substrate 
concentration from 5 to about 0.5. Also the limit distribution is lesser than 1 in 
the substrate and greater than 1 in the biomass. Indeed one of the main difference 
between \eqref{eq.diffusion.approximation} and 
\eqref{eq.diffusion.approximation.adhoc} is than for state values less than 1 
(resp. more greater than 1) the noise variance for the first model is greater 
(resp. lesser) than the noise variance for the second model. 

This example illustrates clearly that the two models differ substantially.

\begin{figure}
\begin{center}
\includegraphics[width=7cm]{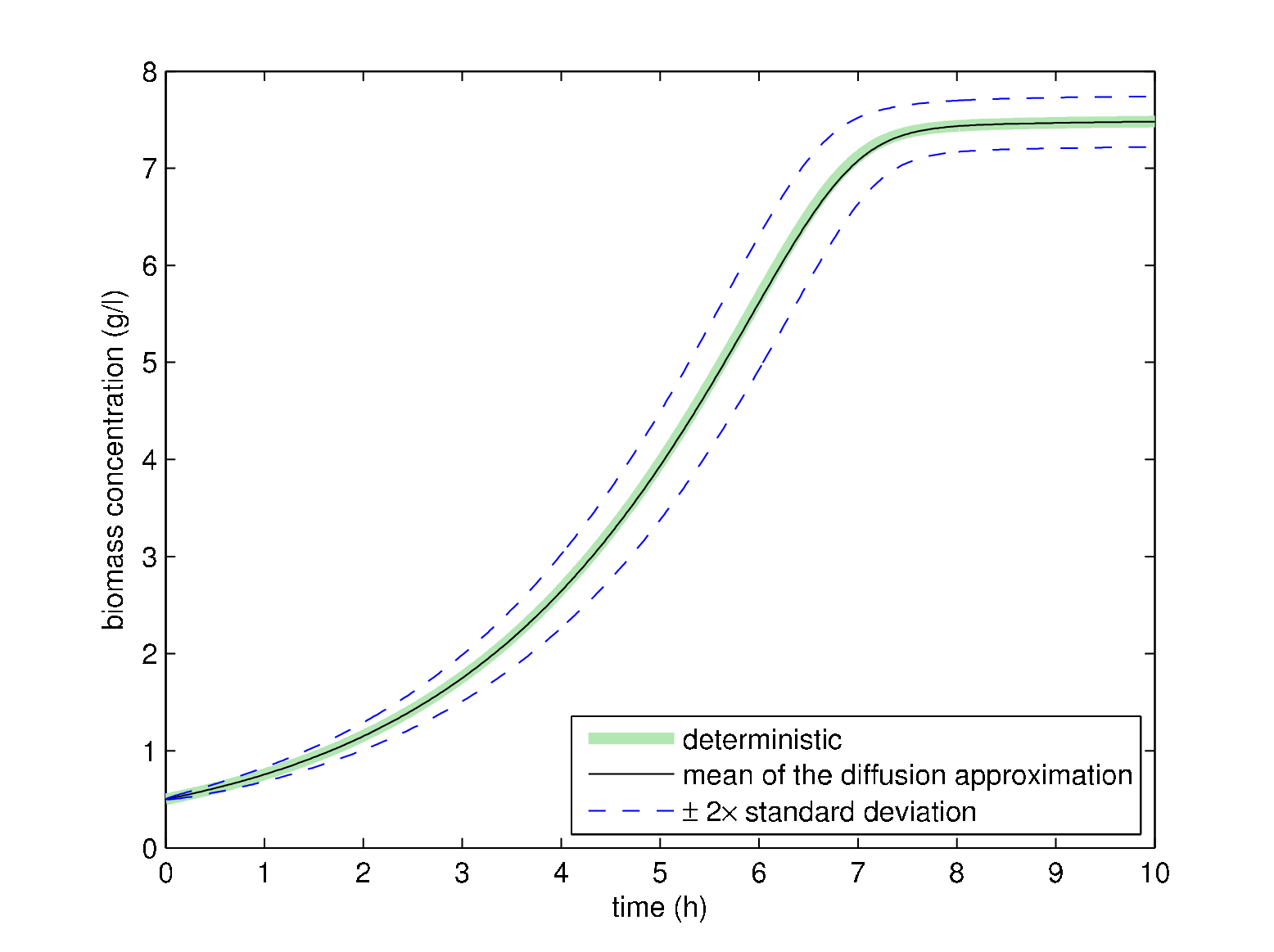}
\includegraphics[width=7cm]{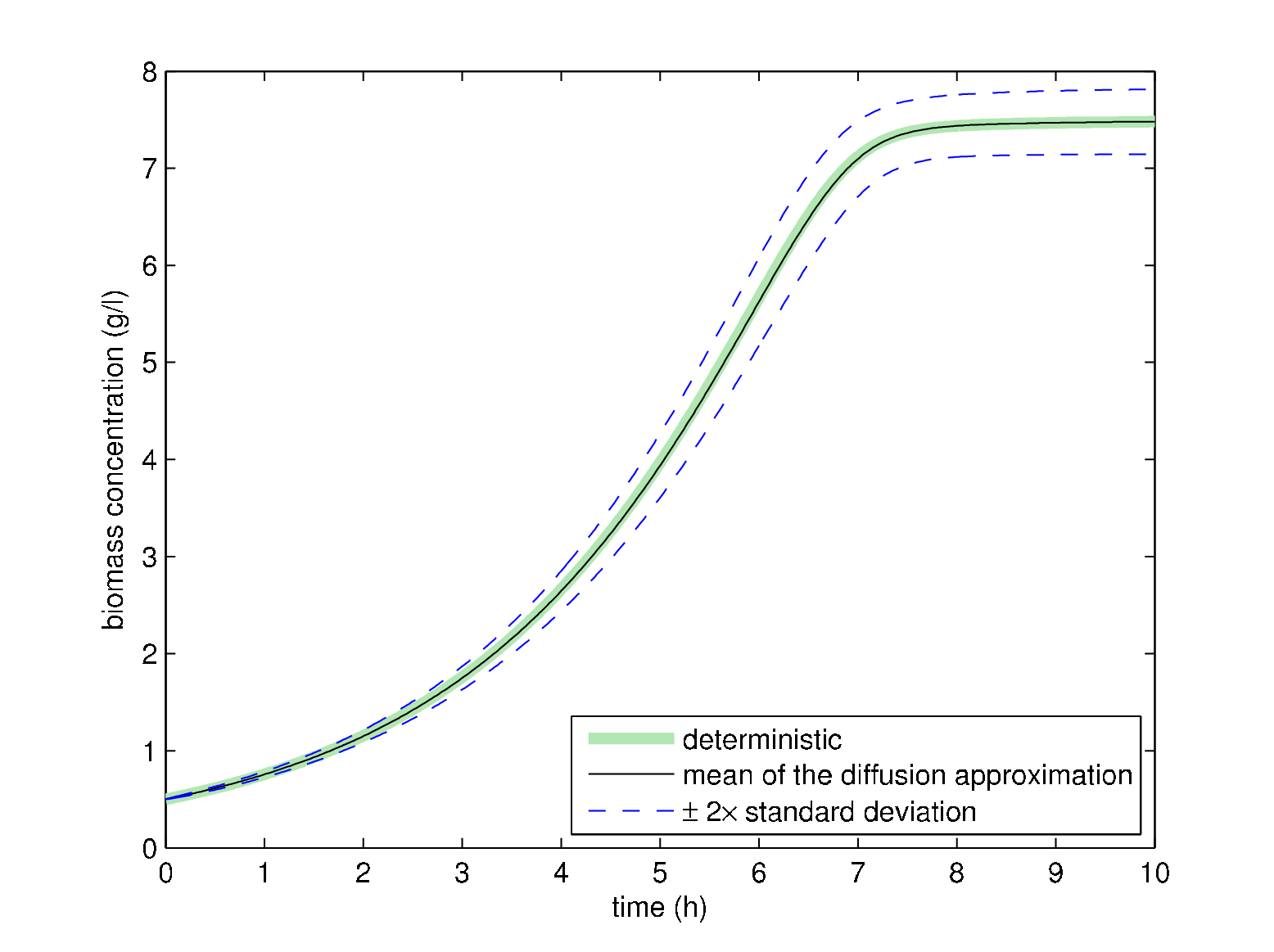}
\\
\includegraphics[width=7cm]{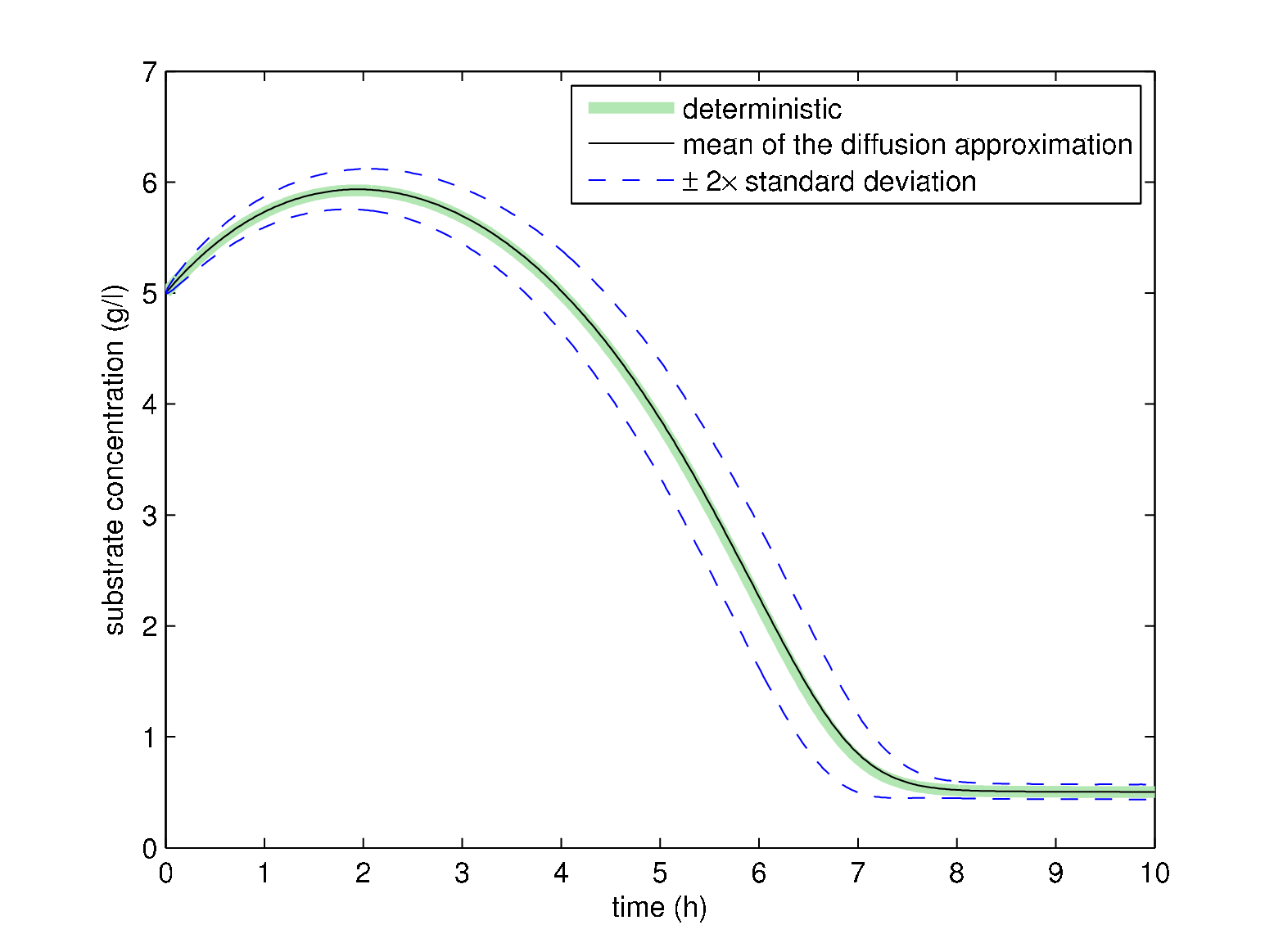}
\includegraphics[width=7cm]{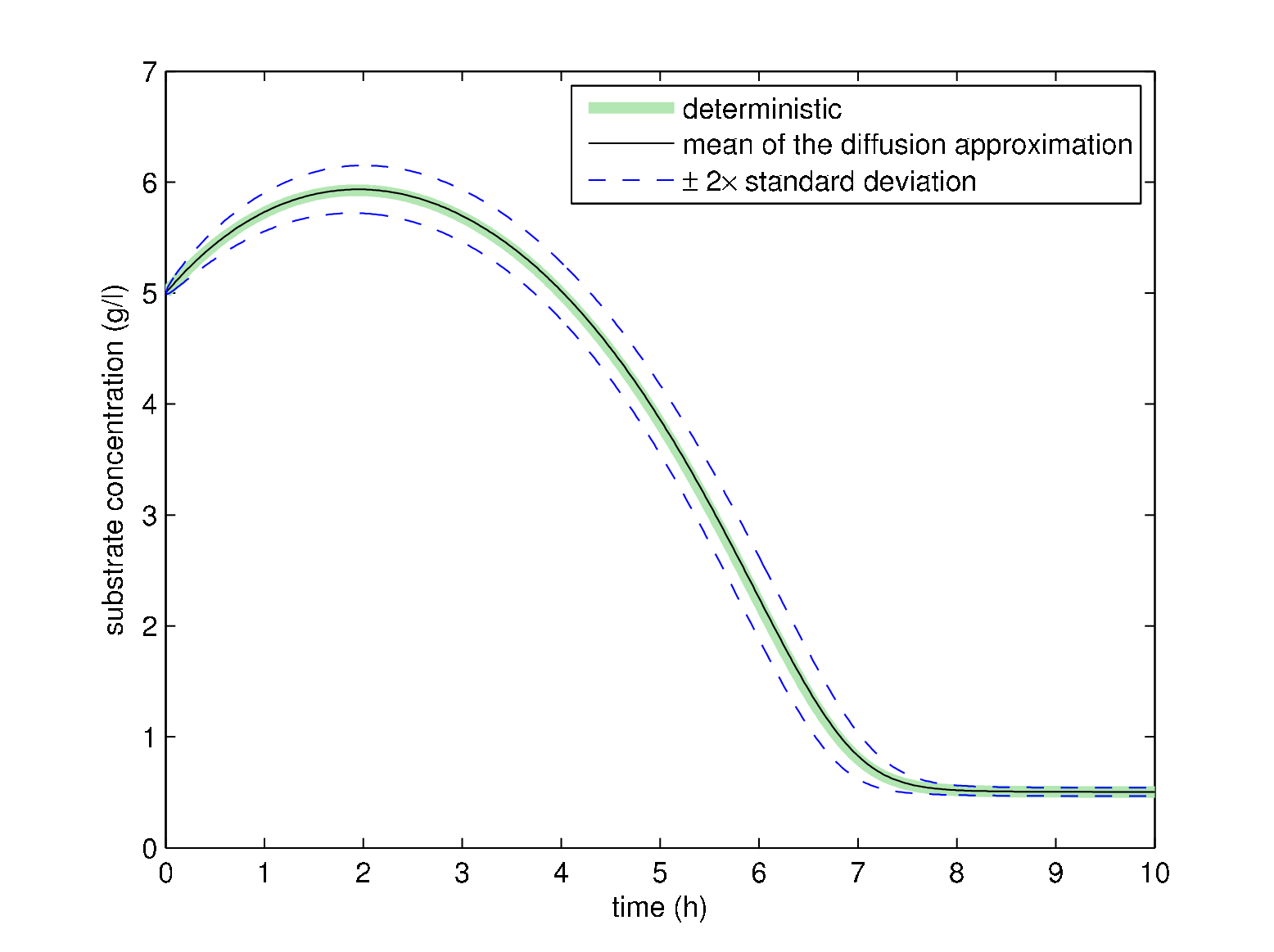}
\\
\includegraphics[width=7cm]{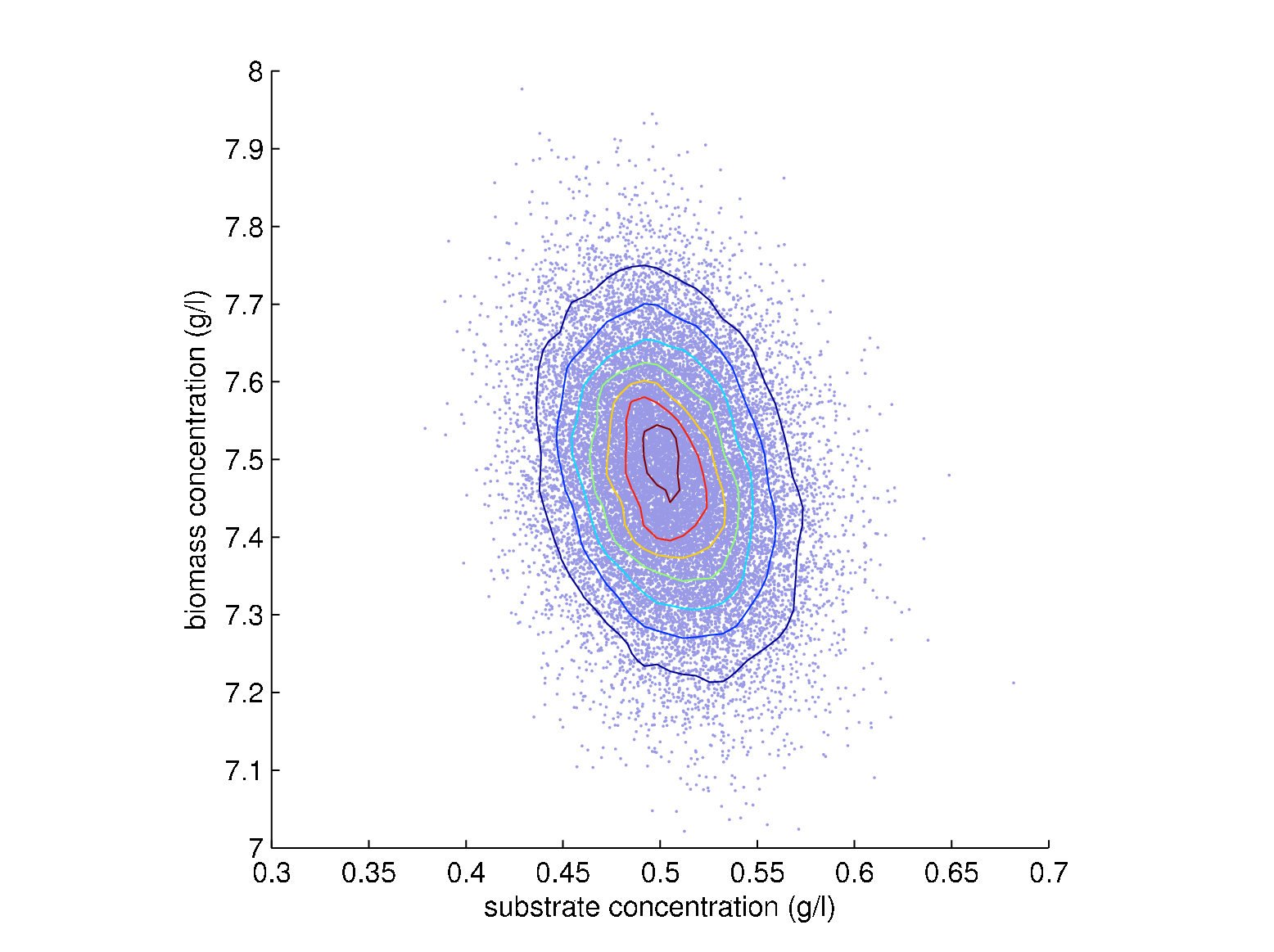}
\includegraphics[width=7cm]{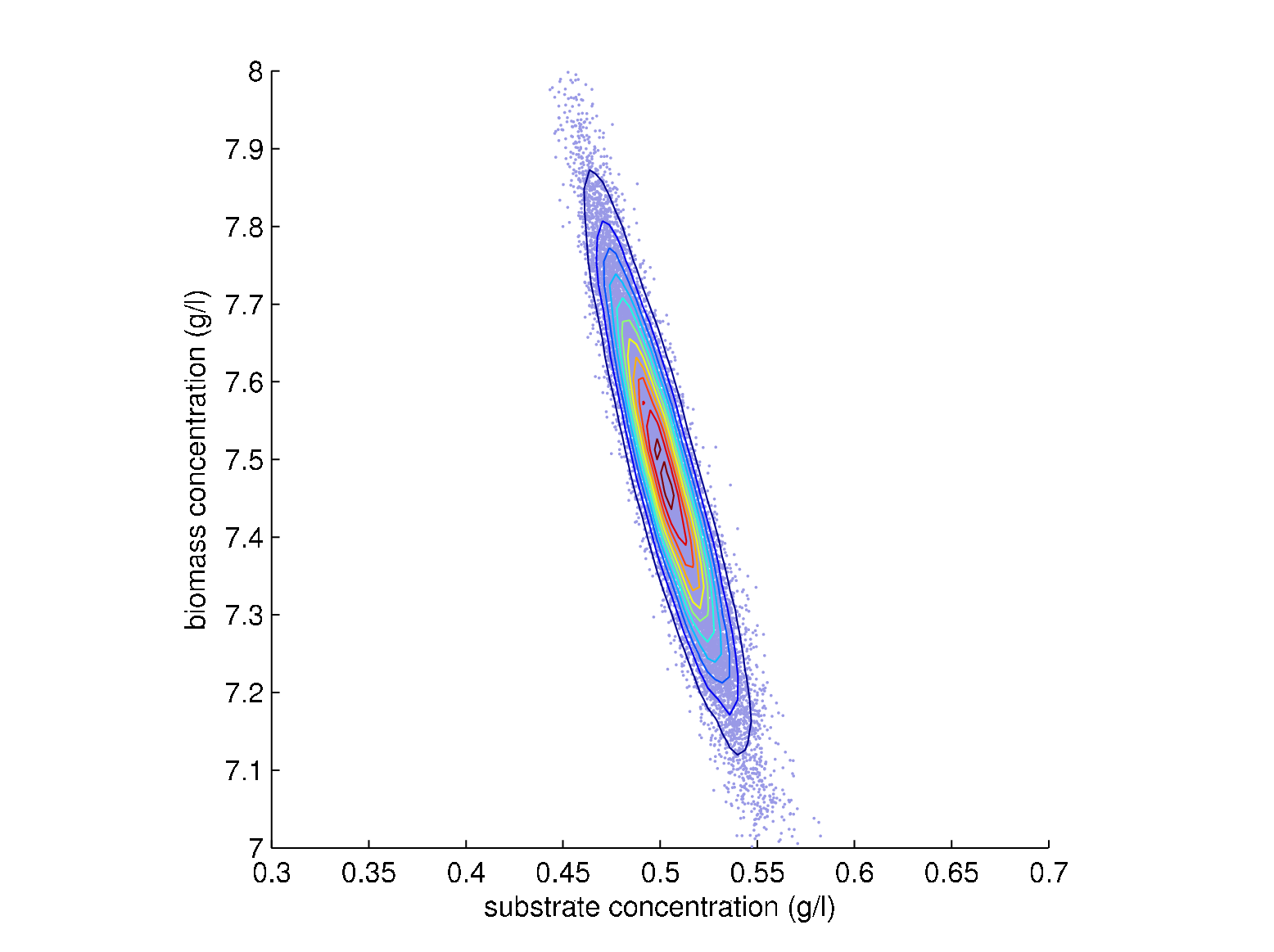}
\end{center}
\caption{\it Comparison of the processes $\xi_{t}=(\beta_{t},\sigma_{t})^{*}$ given by the diffusion approximation model \eqref{eq.diffusion.approximation} and by the ad hoc model \eqref{eq.diffusion.approximation.adhoc}. Evolution of the biomass concentration (top) and substrate concentration (center) during time; final joint density approximation of concentration (bottom). The two models differ substantially: compared to the diffusion approximation, the ad hoc model overestimates (resp underestimates) the noise variance for concentration greater (resp. lesser) than 1. }
\label{fig.simu.comp.hadhoc}
\end{figure}
\section{Discussion}
\label{sec.discussion}

We started from a reference pure jump model $X_{t}$, described by rates/jumps 
structure  of Table \ref{table:rates:jumps} or as a solution of the stochastic 
differential equation \eqref{eq.jump.representation}. The martingale decomposition \eqref{eq.representation.pure.jump} clearly describes that the dynamics of $X_{t}$ is the combination of the classical deterministic dynamics of the chemostat 
\eqref{eq.chemostat} plus martingale terms with coefficients $1/\sqrt{K_{i}}$ and with explicitly known quadratic variations, see \eqref{eq.quadratic.variation}. These quadratic variation terms allow us to assess the difference between the stochastic model and the deterministic one.

We presented the explicit Monte Carlo simulation procedure, called Gillespie method, for the process $X_{t}$. In standard cases, that is for high population levels (i.e. $K_{i}$ large), this procedure is not feasible as it requires us to simulate too many events. In this case, we presented the Poisson approximation \eqref{eq.sde.discretization.2} and the normal approximation \eqref{eq.sde.discretization}, both in discrete-time. These approximations are valid only for large populations, i.e. about the axes, it is necessary to return to the pure jump process representation. In the application discussed here, the Poisson approximation is of little interest: it is more time-consuming than the diffusion approximation and valid only on a very limited scale range between the pure jump model and the normal approximation model.

In contrast with previous stochastic chemostat models \cite{stephanopoulos1979a,gard1999a,imhof2005a} where the stochasticity was introduced according to an ad hoc approach, in the present work we propose a family of models where the structure of the noise emerges from the very dynamics and where the scale parameters can be tuned according to the problem under interest. In particular it allows us to propose hybrid models where the cell population dynamics features stochasticity as the substrate is in fluid dynamics (ODE), corresponding to the Case 3 of Table \ref{table.simulation.cases}. This kind of model has already been proposed in \cite{grasman2005a} in a three trophic levels case where the stochasticity appears only in the top level trophic as a stochastic logistic model and with fluid limit dynamics for the two other levels; it also has been proposed in \cite{crump1979a} with a pure jump process for the biomass dynamics and a fluid limit for the substrate. This approach can also be related to coupled slow/fast reactions in stochastic chemical kinetics \cite{haseltine2002a,ball2006a}.

The approach proposed here can be applied to any model of population dynamics especially in cases of difference of scale between the different dynamics (e.g. cell/substrate). The dynamics of interacting populations cannot be modeled by a single model but rather by a family of models whose domain of validity depends on the scale at which the dynamics are considered. For example the normal approximation model represented as stochastic differential equations \eqref{eq.diffusion.approximation} or the ODE model \eqref{eq.chemostat} are valid in high population levels, hence using such models to infer extinction characteristics like extinction time and extinction probabilities is not valid. This was already noticed by \cite{pollett2001a} and \cite{wilcox2002a}.

In most standard population scales of the chemostat the ODE model is justified. Also, the ODE framework proposes analysis, control and optimization tools that are more accessible than the one of the SDE context. Though, as seen, the stochasticity cannot be neglected in many situations. This stochasticity could be of small intensity in the present single species/single substrate situation but could deeply perturb multiple species/multiple substrates situations. The SDE model could be simulated at a small extra computational cost and offers a more realistic prediction tool. Indeed, as  it can account for the variability of the experiments, the simulation of the SDE offers the possibility to explore in depth the potentialities of the dynamical systems.

The SDE model is also more adapted for the confrontation to the data as it allows us to build a statistical model and the associated likelihood function. One of the next important steps, that we will investigate in coming work, will be to propose an adapted statistical procedure to estimate the scale parameters $K_{i}$, and in a second step to estimate the parameters ($D$, $\ssin$...). In the future we will also investigate the long-term behavior of these models as well as their optimal command.

\section*{Acknowledgements}

The work was partially supported by the French National Research Agency (ANR) within the SYSCOMM project DISCO ANR-09-SYSC-003.

\clearpage
\appendix
\section*{Appendices}

\section{Representation of the process $X_{t}$}
\label{sec.representation.of.X}


\subsubsection*{Infinitesimal generator}

We consider the Markov process $X_{t}$ with infinitesimal generator:
\begin{align}
\label{eq.GI.jump}
   \AAA \phi(x)
   &=
   \sum_{i=1}^{5} \lambda_{i}(x)\,[\phi(x+\nu_{i}(x))-\phi(x)]
\end{align}
for all $\phi:\R^{2}_{+}\mapsto \R$ continuous with compact support \cite[Th. 8-3.1]{ethier1986a}. The infinitesimal generator can also be understood in the following way:
\begin{align*}
\P(X_{t+\Delta t}=x+y|X_{t}=x)
\simeq
\left\{
\begin{array}{ll}
   \lambda_{i}(x)\,\Delta t+o(\Delta t)\,,
   & \textrm{if }y=\nu_{i}(x)\textrm{ for }i=1\cdots 5,
   \\
   1-\sum_{i=1}^{5}\lambda_{i}(x)\,\Delta t+o(\Delta t)\,,
   & \textrm{if }y=0,
   \\
   o(\Delta t)\,,
   & \textrm{otherwise}.
\end{array}
\right.
\end{align*}
or as
\[
   \AAA \phi(x)
   =
   \lim_{t\to 0}
   \frac{\E\phi(X^{x}_{t})-\phi(x)}{t}
\]
where $X^{x}_{t}$ is the process $X_{t}$ starting from $x$. It can be rewritten as:
\begin{align*}
   \AAA \phi(x)
   &=
   \lambda(x)\,\int_{\R^{2}_{+}} [\phi(y)-\phi(x)]\,\rho(x,\rmd y)
\end{align*}
with
\begin{align}
  \lambda(x)
  &\eqdef
  \sum_{i=1}^{5} \lambda_{i}(x)\,,
\\
  \rho(x,\rmd y)
  &\eqdef
  \sum_{i=1}^{5} \bar\lambda_{i}(x) \, \delta_{x+\nu_{i}(x)}(\rmd y)
  &
  \textrm{with }
  \bar\lambda_{i}(x)
  \eqdef
  \frac{\lambda_{i}(x)}{\sum_{i'=1}^{5} \lambda_{i'}(x)}
\end{align}

We define the jump times:  $\tau_{0}=0$ and
\[
  \tau_{n}
  \eqdef
  \inf\big\{t>\tau_{n-1}\,;\,X_{t}\neq X_{\tau_{n-1}}\big\}
\]
and the embedded jump chain
\[ 
   Y_{n} \eqdef X_{\tau_{n}}\,.
\]
It is well know that  \fenumi\ $Y_{n}$ is a Markov chain on $\R^{2}_{+}$ with transition probability $\rho(x,\rmd y)$; \fenumii\ for all $n\geq 1$, conditionally on $Y_{0},\dots,Y_{n-1}$, the holding times $\tau_{1}-\tau_{0},\dots,\tau_{n}-\tau_{n-1}$ are independent and exponentially distributed of intensity parameters $\lambda(Y_{0}),\dots,\lambda(Y_{n-1})$, see \cite{norris1998a}. These properties are at are the basis of the Gillespie simulation algorithm 
(see Algorithm \ref{algo.gillespie}).

\subsubsection*{Non-explosion and existence of moments}

To study the non-explosion and existence of moments, we define the mean jump size function:
\begin{align}
\label{eq.mK}
  m_{K}(x)
  \eqdef
  \int_{\R^{2}_{+}} (y-x)\,\rho(x,\rmd y)
  =
  \sum_{i=1}^{5} \nu_{i}(x)\,\bar\lambda_{i}(x)\,.
\end{align}
and, for $p\geq 1$
\begin{align}
\label{eq.|mK|}
  |m_{K}|_{p}(x)
  \eqdef
  \int_{\R^{2}_{+}} |y-x|^{p}\,\rho(x,\rmd y)
  =
  \sum_{i=1}^{5} |\nu_{i}(x)|^{p}\,\bar\lambda_{i}(x)\,.
\end{align}
Note that from \eqref{eq.fK1}:
\begin{align}
\label{eq.fK}
  f_{K}(x)
  =
  \lambda(x)\,m_{K}(x)
\end{align}
and if we replace $\nu_{i}(x)$ by $\nu_{i}$ in \eqref{eq.fK} we get:
\begin{align*}
  \sum_{i=1}^{5} \nu_{i}\,\lambda_{i}(x)
  =
  \begin{pmatrix}
    \mu(s)\,b - D\,b
    \\
    -k\,\mu(s)\,b + D\,\ssin-D\,s
  \end{pmatrix}
  =
  f(x)
\end{align*}
where $f(x)$ is the right-hand-side function of the ODE \eqref{eq.chemostat}. 

$f_{K}(x)$ is the instantaneous mean of the process $X_{t}$, more precisely we will show in Proposition \ref{proposition.jump.representation}, that $\E(X_{t+\Delta t}|X_{t}=x)\simeq f_{K}(x)\,\Delta t$. As proved in the next lemma $f_{K}(x)$ is essentially $f(x)$, so locally in time the mean of the process $X_{t}$ behaves like $x(t)$.
 
\begin{lemma}
\label{lemma.f.fK}
Consider $\nu_{i}(x)$ defined in Table \ref{table:rates:jumps}, $\nu_{i}$ defined by  \eqref{eq.nui} and $f_{K}(x)$ defined by \eqref{eq.fK1}. First $|\nu_{i}(x)|\leq|\nu_{i}|$. Then let
\[
  \RR_{K}
  \eqdef
  \Big\{
    x=\left(\begin{smallmatrix}b\\ s\end{smallmatrix}\right)\in\R^{2}_{+}
    \,;\,
    \textstyle
    b\leq \frac{1}{K_{4}}
    \textrm{ or }
    s\leq \frac{1}{K_{2}}
    \textrm{ or }
    s\leq \frac{1}{K_{5}}
  \Big\}\,.
\]
For $x\not\in \RR_{K}$, $\nu_{i}(x)=\nu_{i}$, $m_{K}(x)=m(x)$, $f_{K}(x)=f(x)$. For all $x\in \R^{2}_{+}$:
\begin{align}
\label{lemma.f.fK.1}
  |\nu_{i}(x)-\nu_{i}| 
  &\leq \frac{1}{\min_{j=1\cdots 5}K_{j}}
  \,,\quad i=1,\dots,5\,,
\\
\label{lemma.f.fK.2}
  |m_{K}(x)-m(x)| 
  &\leq \frac{1}{\min_{j=1\cdots 5}K_{j}}\,,
\\
\label{lemma.f.fK.3}
  |f_{K}(x)-f(x)| 
  &\leq 
  (1-K_{2}\,s)^{+}\,k\,\mu(s)\,b
  +
  (1-K_{4}\,b)^{+}\,D\,b
  +
  (1-K_{5}\,s)^{+}\,D\,s
\end{align}
so that $f_{K}(x)\to f(x)$ when $K_{i}\to\infty$ for all $i=1,\dots,5$ and this convergence is uniform on any compact set of $(0,\infty)^{2}$.
\end{lemma}

\proof

For $x\not\in \RR_{K}$ it is clear that $\nu_{i}(x)=\nu_{i}$ so $f_{K}(x)=f(x)$.
For all $x$, we have $\nu_{1}(x)\equiv \nu_{1}$ and $\nu_{3}(x)\equiv \nu_{3}$ and:
\begin{align*}
  \nu_{2}(x)-\nu_{2}
  &=
  \textstyle 
  \frac{1}{K_{2}}
  \left(\begin{smallmatrix}
    0
    \\
    (1-K_{2}\,s)^{+}
  \end{smallmatrix}\right)
  \,,
\\
  \nu_{4}(x)-\nu_{4}
  &=
  \textstyle 
  \frac{1}{K_{4}}
  \left(\begin{smallmatrix}
    (1-K_{4}\,b)^{+}
    \\
    0
  \end{smallmatrix}\right)
  \,,
  \\
  \nu_{5}(x)-\nu_{5}
  &=
  \textstyle
  \frac{1}{K_{5}}
  \left(\begin{smallmatrix}
    0
    \\
    (1-K_{5}\,s)^{+}
  \end{smallmatrix}\right)
\end{align*}
so we get \eqref{lemma.f.fK.1}. The following assertions of the lemma are straightforward.\carre

\bigskip

Non-explosion and existence of moments are given by the following result:
\begin{theorem}[Hamza and Klebaner \cite{hamza1995a}]
\label{theorem.hamza1995a}
Suppose that $\lambda(x)>0$ for all $x\in\R^{2}_{+}$ and that there exist $C>0$ and $p\geq 1$ such that $\E(|X_{0}|^{p})<\infty$ and
\begin{align}
\label{cond.theorem.hamza1995a}
  \lambda(x)\,|f_{K}|_{p}(x)
  \leq
  C\,(1+|x|^{p})\,,\quad
  \forall x\in\R^{2}_{+}
\end{align}
then the Markov process $(X_{t})_{t\geq 0}$ is non-explosive, that is $\tau_{n}\to\infty$ a.s., and for all $T>0$ there exists $C>0$ s.t. $\E(|X_{t}|^{p})\leq C$ for all $t\leq T$.
\end{theorem}
Indeed $\lambda(x)=\sum_{i}\lambda_{i}(x)\geq K_{3}\,D\,\ssin>0$ and
\begin{align*}
  |f_{K}|_{p}(x)
  &=
  \sum_{i=1}^{5} |\nu_{i}(x)|^{p}\,\bar\lambda_{i}(x)
\\  
  &\leq
  \frac{\mu(s)\,b}{K_{1}^{p-1}}
  +
  \frac{k\,\mu(s)\,b}{K_{2}^{p-1}}  
  +
  \frac{D\,\ssin}{K_{3}^{p-1}} 
  +
  \frac{D\,b}{K_{4}^{p-1}} 
  +
  \frac{D\,s}{K_{5}^{p-1}} 
\\  
  &\leq
  \Big(
  \frac{1}{K_{1}^{p-1}}
  +
  \frac{1}{K_{2}^{p-1}}  
  +
  \frac{1}{K_{3}^{p-1}} 
  +
  \frac{1}{K_{4}^{p-1}} 
  +
  \frac{1}{K_{5}^{p-1}} 
  \Big)\,(1+|x|)
\end{align*}
so that \eqref{cond.theorem.hamza1995a} is fulfilled for all $p\geq 1$. Hence the Markov process $X_{t}$ with infinitesimal generator $\AAA$ defined by \eqref{eq.GI.jump} is non-explosive and $X_{t}$ admits moments of all order for all $t\geq 0$.

\subsubsection*{Representation for the process $X_{t}$}

We now give a representation for the process $X_{t}$ as a solution of a stochastic differential equation driven by random Poisson measures:
\begin{proposition}
\label{proposition.jump.representation}
The process $X_{t}$ is defined for all $t\geq 0$ and it is solution of the jump SDE \eqref{eq.jump.representation}
where $N^{i}$ are independent random Poisson measures with intensity measure $\rmd u\times \rmd v$ (Lebesgue measure). 
\end{proposition}

\proof
We first verify that the stochastic integral in \eqref{eq.jump.representation} is defined. According to \cite[\S\,II-3]{ikeda1981a}, this integral is defined if:
\[
  \E 
  \int_{0}^{t}\int_{0}^{\infty} 
          |\nu_{i}(X_u)|\,\indic_{\{v\leq \lambda_{i}(X_u)\}}
          \, \rmd v \,\rmd u 
  <\infty\,.
\]
We have
\begin{align*}
  &
  \E 
  \int_{0}^{t}\int_{0}^{\infty} 
          |\nu_{i}(X_u)|\,\indic_{\{v\leq \lambda_{i}(X_u)\}}
           \, \rmd v\,\rmd u
\\
  & \qquad \qquad \qquad
  \leq
  \frac{1}{K_{i}}\,\E 
  \int_{0}^{t}  \lambda_{i}(X_u)\,\rmd u
  \leq
  C\, 
  \int_{0}^{t}  (1+\E|X_u|)\,\rmd u
\end{align*}
which is finite according to Theorem \ref{theorem.hamza1995a}.
\carre

\bigskip

Consider the centered random Poisson measure:
\[
   \tilde N^{i}(\rmd u\times \rmd v)
   \eqdef
   N^{i}(\rmd u\times \rmd v)
   -
   \rmd u\times \rmd v\,.
\]
According to \cite[\S\,II-3]{ikeda1981a}
\begin{align}
\label{eq.Mi}
  M^{i}_{t}
  \eqdef
  \int_{0}^{t}\int_{0}^{\infty} 
          \nu_{i}(X_{u^{-}})\,\indic_{\{v\leq \lambda_{i}(X_{u^{-}})\}}
          \,\tilde N^{i}(\rmd u\times\rmd v)
  \,,\qquad i=1,\dots,5
\end{align}
are five independent square-integrable martingales with finite moments of all orders. Let:
\begin{align}
\label{eq.M}
  M_{t} \eqdef \sum_{i=1}^{5}M^{i}_{t}\,.
\end{align}
From \eqref{eq.jump.representation}:
\begin{align}
\label{eq.jump.representation2}
  X_{t}
  &=
  X_{0}
  +
  \int_{0}^{t}f_{K}(X_u)\,\rmd u
  +
  M_{t}
  \,.
\end{align}

We want to study the behavior of the martingales $M^{i}_{t}$ as the $K_{i}\to\infty$. First note that
  $M^{i}_{t}
  =
  \left(\begin{smallmatrix}
    m^{i}_{t}
    \\
    0
  \end{smallmatrix}\right)$ for $i=1,4$ and
  $M^{i}_{t}
  =
  \left(\begin{smallmatrix}
    0
    \\
    m^{i}_{t}
  \end{smallmatrix}\right)$ for $i=2,3,5$ with
\begin{subequations}
\label{eq.mi}
\begin{align}
\label{eq.m1}
  m^{1}_{t}
  &\eqdef
  \hphantom{-}
  \frac{1}{K_{1}}\,
  \int_{0}^{t}\int_{0}^{\infty} 
          \indic_{\{v\leq \lambda_{1}(X_{u^{-}})\}}
          \,\tilde N^{1}(\rmd u\times\rmd v)\,,
\\
\label{eq.m2}
  m^{2}_{t}
  &\eqdef
  -
  \frac{1}{K_{2}}\,
  \int_{0}^{t}\int_{0}^{\infty} 
          (1\wedge K_{2}\,S_{u^{-}})\,
          \indic_{\{v\leq \lambda_{2}(X_{u^{-}})\}}
          \,\tilde N^{2}(\rmd u\times\rmd v)\,,
\\
\label{eq.m3}
  m^{3}_{t}
  &\eqdef
  \hphantom{-}\frac{1}{K_{3}}\,
  \int_{0}^{t}\int_{0}^{\infty} 
          \indic_{\{v\leq \lambda_{3}(X_{u^{-}})\}}
          \,\tilde N^{3}(\rmd u\times\rmd v)\,,
\\
\label{eq.m4}
  m^{4}_{t}
  &\eqdef
  -
  \frac{1}{K_{4}}\,
  \int_{0}^{t}\int_{0}^{\infty} 
          (1\wedge K_{4}\,B_{u^{-}})\,
          \indic_{\{v\leq \lambda_{4}(X_{u^{-}})\}}
          \,\tilde N^{4}(\rmd u\times\rmd v)\,,
\\
\label{eq.m5}
  m^{5}_{t}
  &\eqdef
  -
  \frac{1}{K_{5}}\,
  \int_{0}^{t}\int_{0}^{\infty} 
          (1\wedge K_{5}\,S_{u^{-}})\,
          \indic_{\{v\leq \lambda_{5}(X_{u^{-}})\}}
          \,\tilde N^{5}(\rmd u\times\rmd v)\,.
\end{align}
\end{subequations}
As $m^{i}_{t}$ is of the form $m^{i}_{t} = 
  \int_{0}^{t}\int_{0}^{\infty} \gamma^{i}(u^-,v)\,\tilde N^{i}(\rmd u\times\rmd v)$
then the associated predictable quadratic variation is $\crochet{m^{i}}_{t} = 
  \int_{0}^{t}\int_{0}^{\infty} [\gamma^{i}(u,v)]^{2}\,\rmd v\,\rmd u$
so we can easily check that
\begin{align*}
  \crochet{m^{1}}_{t}
  &
  =
  \frac{1}{K_{1}}\, \int_{0}^{t} \mu(S_u)\,B_{u}\,\rmd u\,,
\\
  \crochet{m^{2}}_{t}
  &
  =
  \frac{1}{K_{2}}\,
  \int_{0}^{t} (1\wedge K_{2}\,S_u)^{2}\, k\,\mu(S_u)\,B_u \,\rmd u\,,
\\
  \crochet{m^{3}}_{t}
  &
  =
  \frac{1}{K_{3}}\, D\,\ssin\,t\,,
\\
  \crochet{m^{4}}_{t}
  &
  =
  \frac{1}{K_{4}}\,
  \int_{0}^{t} (1\wedge K_{4}\,B_u)^{2}\, D\,B_u \,\rmd u\,,
\\
  \crochet{m^{5}}_{t}
  &
  =
  \frac{1}{K_{5}}\,
  \int_{0}^{t}  (1\wedge K_{5}\,S_u)^{2}\, D\,S_u \,\rmd u\,.
\end{align*}
Define
\begin{align}
\label{eq.bar.mi}
  \bar m^{i}_{t}
  \eqdef
  \sqrt{K_{i}}\,m^{i}_{t}
  =
  \sqrt{K_{i}}\,
  \int_{[0,t]\times[0,\infty)} \gamma^{i}(u^{-},v)\,\tilde N^{i}(\rmd u\times \rmd v)
  \,.
\end{align}
So we obtained the following representation of the process $X_{t}$ that emphases the dependence on the $K_{i}$:
\begin{subequations}
\label{eq.representation.pure.jump}
\begin{align}
\label{eq.representation.pure.jump.B}
  \rmd B_{t}
  &=
  \big(
    \mu(S_{t})\,B_{t}
    -
    (1\wedge K_{4}\,B_{t})\,D\,B_{t}
  \big)\,\rmd t 
  \textstyle
  +
    \frac{1}{\sqrt{K_{1}}}\,\rmd\bar m^{1}_{t}
    +
    \frac{1}{\sqrt{K_{4}}}\,\rmd\bar m^{4}_{t}
\\
\nonumber
  \rmd S_{t}
  &=
  \big(
    -
    (1\wedge K_{2} \,S_{t})\,k\,\mu(S_{t})\,B_{t}
    + D\,\ssin
    - (1\wedge K_{5} \,S_{t})\,D\,S_{t}
  \big)\,\rmd t
\\
\label{eq.representation.pure.jump.S}
  & \qquad \qquad \qquad \qquad\qquad\qquad
  \textstyle
  +
    \frac{1}{\sqrt{K_{2}}}\,\rmd \bar m^{2}_{t}
    +
    \frac{1}{\sqrt{K_{3}}}\,\rmd \bar m^{3}_{t} 
    +
    \frac{1}{\sqrt{K_{5}}}\,\rmd \bar m^{5}_{t}
\end{align}
\end{subequations}
where $\bar m^{i}_{t}$ are independent square integrable martingales with the following quadratic variations:
\begin{subequations}
\label{eq.quadratic.variation}
\begin{align}
\label{eq.quadratic.variation.1}
  \crochet{\bar m^{1}}_{t}
  &
  =
  \int_{0}^{t} \mu(S_u)\,B_{u}\,\rmd u\,,
\\
\label{eq.quadratic.variation.2}
  \crochet{\bar m^{2}}_{t}
  &
  =
  \int_{0}^{t} (1\wedge K_{2}\,S_u)^{2}\, k\,\mu(S_u)\,B_u \,\rmd u\,,
\\
\label{eq.quadratic.variation.3}
  \crochet{\bar m^{3}}_{t}
  &
  =
  D\,\ssin\,t \vphantom{\int_{0}^{t}}\,,
\\
\label{eq.quadratic.variation.4}
  \crochet{\bar m^{4}}_{t}
  &
  =
  \int_{0}^{t} (1\wedge K_{4}\,B_u)^{2}\, D\,B_u \,\rmd u\,,
\\
\label{eq.quadratic.variation.5}
  \crochet{\bar m^{5}}_{t}
  &
  =
  \int_{0}^{t}  (1\wedge K_{5}\,S_u)^{2}\, D\,S_u \,\rmd u\,.
\end{align}
\end{subequations}

\section{Existence and uniqueness for a solution of the SDE \eqref{eq.diffusion.approximation}}
\label{sec.existence.uniqueness}

To prove that the system \eqref{eq.diffusion.approximation} admits a strong solution and pathwise uniqueness holds we use the results of \cite[p. 134]{rogers2000b} or \cite{durrett1996a}. Let $\DD_N=[\frac1N,N]\times [-\frac{K_{5}}{K_{3}}\,\ssin+\frac1N,N]$
and rewrite \eqref{eq.diffusion.approximation} as:
\begin{align}
\label{eq.diffusion.approximation.vec}
  \rmd \xi_{t}
  =
  f(\xi_{t}) \,\rmd t + g(\xi_{t}) \,\rmd  W_{t}  
\end{align}
where  $\xi_{t}=(\beta_{t}, \sigma_{t})^*$ and $W_{t}=(\Wb_t,\Ws_t)^*$. The coefficient $g$ is not Lipschitz continuous, it is globally Lipschitz on $\DD_{N}$ for all $N$, but we can find Lipschitz continuous coefficients $f_{N}$ and $g_{N}$ such that:
\begin{align*}
  f_{N}(\xi)
  &=f(\xi) \textrm{ for }\xi\in\DD_{N}\,,
  &
  f_{N}(\xi)
  &=0\textrm{ for }\xi\not\in\DD_{2N}\,,
\\
  g_{N}(\xi)
  &=g(\xi) \textrm{ for }\xi\in\DD_{N}\,,
  &
  g_{N}(\xi)
  &=0\textrm{ for }\xi\not\in\DD_{2N}\,.
\end{align*}
The the system
\begin{align*}
  \rmd \xi_{t}^N
  =
  f_{N}(\xi_{t}^N) \,\rmd t + g_{N}(\xi_{t}^N) \,\rmd  W_{t}  
\end{align*}
admits a unique strong solution $\xi ^{N}$. Let:
\[
   T_{N}=\inf\{t\geq0\,;\,\xi_t^{N}\not\in\DD_N\}\,.
\]
Hence if $N\geq M$ then $\xi^N=\xi^M$ for all $t\leq T_M$, so we can define a process $\xi^{\infty}$ such that:
\[
  \xi^{\infty}_{t}=\xi^{N}_{t}
  \quad
  \forall t\leq T_{N}
\]
and will be solution of \eqref{eq.diffusion.approximation.vec} up to the explosion time $T_{\infty}=\lim_{N\to\infty} T_{N}$. Then by stability property and by the fact that the solution cannot cross the boundary of $\DD$ we get $T_{\infty}=\infty$ a.s. which proves the strong existence and pathwise uniqueness defined for all $t\geq 0$.

\cleardoublepage
\addcontentsline{toc}{section}{Reference}
\bibliographystyle{plain}


\end{document}